\documentclass[aps,prx,reprint,superscriptaddress,floatfix]{revtex4-2}
\usepackage[T1]{fontenc}
\usepackage{graphicx}
\usepackage{bibunits}

\begin{document}

\title{Probing Residual Noise at a Decoherence Sweet Spot in a $^{28}$Si/SiGe Spin Qubit}

\author{Shinwoo Lee}
\thanks{These authors contributed equally to this work.}
\affiliation{
NextQuantum Center, Department of Physics and Astronomy,
and Institute of Applied Physics,
Seoul National University, Seoul 08826, Korea
}

\author{Hanseo Sohn}
\thanks{These authors contributed equally to this work.}
\affiliation{
NextQuantum Center, Department of Physics and Astronomy,
and Institute of Applied Physics,
Seoul National University, Seoul 08826, Korea
}

\author{Jaemin Park}
\affiliation{
NextQuantum Center, Department of Physics and Astronomy,
and Institute of Applied Physics,
Seoul National University, Seoul 08826, Korea
}

\author{Hyeongyu Jang}
\affiliation{
NextQuantum Center, Department of Physics and Astronomy,
and Institute of Applied Physics,
Seoul National University, Seoul 08826, Korea
}

\author{Jonginn Yun}
\affiliation{
NextQuantum Center, Department of Physics and Astronomy,
and Institute of Applied Physics,
Seoul National University, Seoul 08826, Korea
}

\author{Jun Yoneda}
\affiliation{
Department of Advanced Materials Science,
Graduate School of Frontier Sciences,
The University of Tokyo,
Kashiwa, Chiba 277-8561, Japan
}

\author{Lucas E. A. Stehouwer}
\affiliation{
QuTech and Kavli Institute of Nanoscience,
Delft University of Technology,
PO Box 5046, 2600 GA Delft, The Netherlands
}

\author{Davide Degli Esposti}
\affiliation{
QuTech and Kavli Institute of Nanoscience,
Delft University of Technology,
PO Box 5046, 2600 GA Delft, The Netherlands
}

\author{Giordano Scappucci}
\affiliation{
QuTech and Kavli Institute of Nanoscience,
Delft University of Technology,
PO Box 5046, 2600 GA Delft, The Netherlands
}

\author{Dohun Kim}
\email{dohunkim@snu.ac.kr}
\affiliation{
NextQuantum Center, Department of Physics and Astronomy,
and Institute of Applied Physics,
Seoul National University, Seoul 08826, Korea
}

\date{September 14, 2026}

\begin{abstract}

In $^{28}$Si/SiGe spin qubits with micromagnets, the longitudinal stray field gradient transduces charge noise into qubit frequency noise and limits coherence. We compare two neighboring qubits in the same device with 800~ppm residual $^{29}$Si, one near a decoherence sweet spot where the gradient is locally minimized and the other at a position with a larger gradient. At the sweet spot, $T_2^*$ reaches 67~$\mu\mathrm{s}$ and the Carr--Purcell--Meiboom--Gill coherence time reaches 4.6~ms, whereas $T_2^*$ is 5.2~$\mu\mathrm{s}$ at the neighboring qubit. Near 1~Hz, the frequency noise power spectral density at the sweet spot is nearly two orders of magnitude lower than at the neighboring qubit. The magnitude and low-frequency decay of the residual spectrum are compatible with the prediction for $^{29}$Si nuclear spin noise, and the weak interqubit correlation indicates that local noise becomes important for dephasing at the sweet spot. Despite a finite correlation with the charge sensor, sweet-spot operation strongly reduced the transduction of charge noise, bringing the residual frequency noise close to the level predicted for $^{29}$Si nuclear spin fluctuations.
\end{abstract}


\maketitle

\section{Introduction}
Semiconductor spin qubits in silicon offer a promising route to scalable quantum information processing \cite{loss_quantum_1998}, combining compatibility with industrial fabrication \cite{vandersypen_quantum_2019,maurand_cmos_2016,steinacker_industry-compatible_2025,george_12-spin-qubit_2025} and long coherence times achievable through isotopic enrichment of $^{28}$Si \cite{zwanenburg_silicon_2013,veldhorst_addressable_2014}. Among silicon spin-qubit architectures, $^{28}$Si/SiGe heterostructures are widely studied as hosts for scalable spin-qubit arrays \cite{vandersypen_quantum_2019}. In these devices, fast and individually addressable spin control is often achieved by integrating a micromagnet on top of the device \cite{obata_coherent_2010,yoneda_robust_2015,dumoulin_stuyck_low_2021,philips_universal_2022}, which generates a spatially varying stray field with transverse and longitudinal components at each qubit location. The transverse component enables all-electrical spin manipulation through electric-dipole spin resonance (EDSR), while the spatial variation of the longitudinal component separates the resonance frequencies of neighboring qubits by tens to hundreds of megahertz, enabling frequency-selective addressing \cite{yoneda_robust_2015,pioro-ladriere_electrically_2008,yoneda_quantum-dot_2018}. However, the local gradient of the longitudinal field, referred to as the decoherence gradient, also dephases the qubit by transducing positional fluctuations of the electron driven by charge noise into qubit frequency noise. The micromagnet therefore introduces a tradeoff between fast control and long coherence because field profiles favorable for EDSR and frequency-selective addressability can also increase the decoherence gradient and the resulting sensitivity to charge noise \cite{dumoulin_stuyck_low_2021,yoneda_quantum-dot_2018,kha_micromagnets_2015}. As a result, reported values of the inhomogeneous dephasing time $T_2^*$ in micromagnet-based $^{28}$Si/SiGe devices range from a few microseconds to approximately 20~$\mu$s, even with field gradient optimization \cite{philips_universal_2022,yoneda_quantum-dot_2018} and isotopic purification to a residual $^{29}$Si concentration of 60 ppm \cite{struck_low-frequency_2020}. Charge noise has generally been regarded as the dominant limitation on $T_2^*$ in these devices \cite{yoneda_quantum-dot_2018,paquelet_wuetz_reducing_2023}, but how far $T_2^*$ can be extended by operating at a minimum of the decoherence gradient remains an open question.

Hyperfine interactions with nuclear spins provide another well-established source of dephasing in semiconductor spin qubits, as observed in GaAs \cite{hanson_spins_2007,cao_tunable_2016} and natural-silicon devices \cite{kawakami_electrical_2014}. In isotopically purified $^{28}$Si/SiGe devices with micromagnets, nuclear spin noise has generally been considered secondary to charge noise in limiting $T_2^*$; however, a theoretical study predicts that residual $^{29}$Si alone sets a limit of a few to a few tens of microseconds at 800 ppm \cite{witzel_nuclear_2012}. In addition, $^{73}$Ge nuclei in the SiGe barrier are predicted to provide a nuclear dephasing channel of comparable magnitude, which becomes a significant nuclear noise source at $^{29}$Si concentrations below approximately 1000 ppm \cite{witzel_nuclear_2012,kerckhoff_magnetic_2021,cvitkovich_coherence_2024}. Because the predicted nuclear limits are comparable to the $T_2^*$ values commonly observed in these devices (approximately 20~$\mu$s), isolating the individual contribution of each noise source from coherence measurements remains difficult. Therefore, two open challenges remain. The first is to realize an operating environment that maintains fast spin control while minimizing the decoherence gradient, so that the transduction of charge noise is strongly reduced. The second is to experimentally determine, in such an environment, the extent to which intrinsic nuclear spin noise limits the coherence of $^{28}$Si/SiGe spin qubits.

In this study, we examine sweet-spot operation of a spin qubit in an isotopically purified $^{28}$Si/SiGe device with an integrated micromagnet, and characterize the residual noise sources. We quantify the qubit coherence with $T_2^*$ obtained from Ramsey measurements, and because this value depends on the data acquisition time $t_{\mathrm{acq}}$, we denote it by $T_2^*(t_{\mathrm{acq}})$ throughout this work. By comparing two qubits located at different positions in the spatially varying micromagnet stray field, we show that the qubit at the sweet spot, where the decoherence gradient is locally minimized, exhibits a $T_2^*$(31~s) of up to 67~$\mu$s, more than an order of magnitude longer than the $T_2^*$(30~s) value of 5.2~$\mu$s measured for a neighboring qubit approximately 100 nm away. Noise spectroscopy and cross-correlation analysis show that the transduction of charge noise is strongly reduced at the sweet spot and that the residual frequency noise exhibits weak correlation between neighboring qubits. Although a finite low-frequency correlation with the charge sensor points to a residual local charge contribution, the residual noise spectrum at the sweet spot lies close to the $^{29}$Si nuclear spin noise prediction of this material.

\section{Device Design}
\begin{figure}[t]
    \centering
    \includegraphics[width=\linewidth]{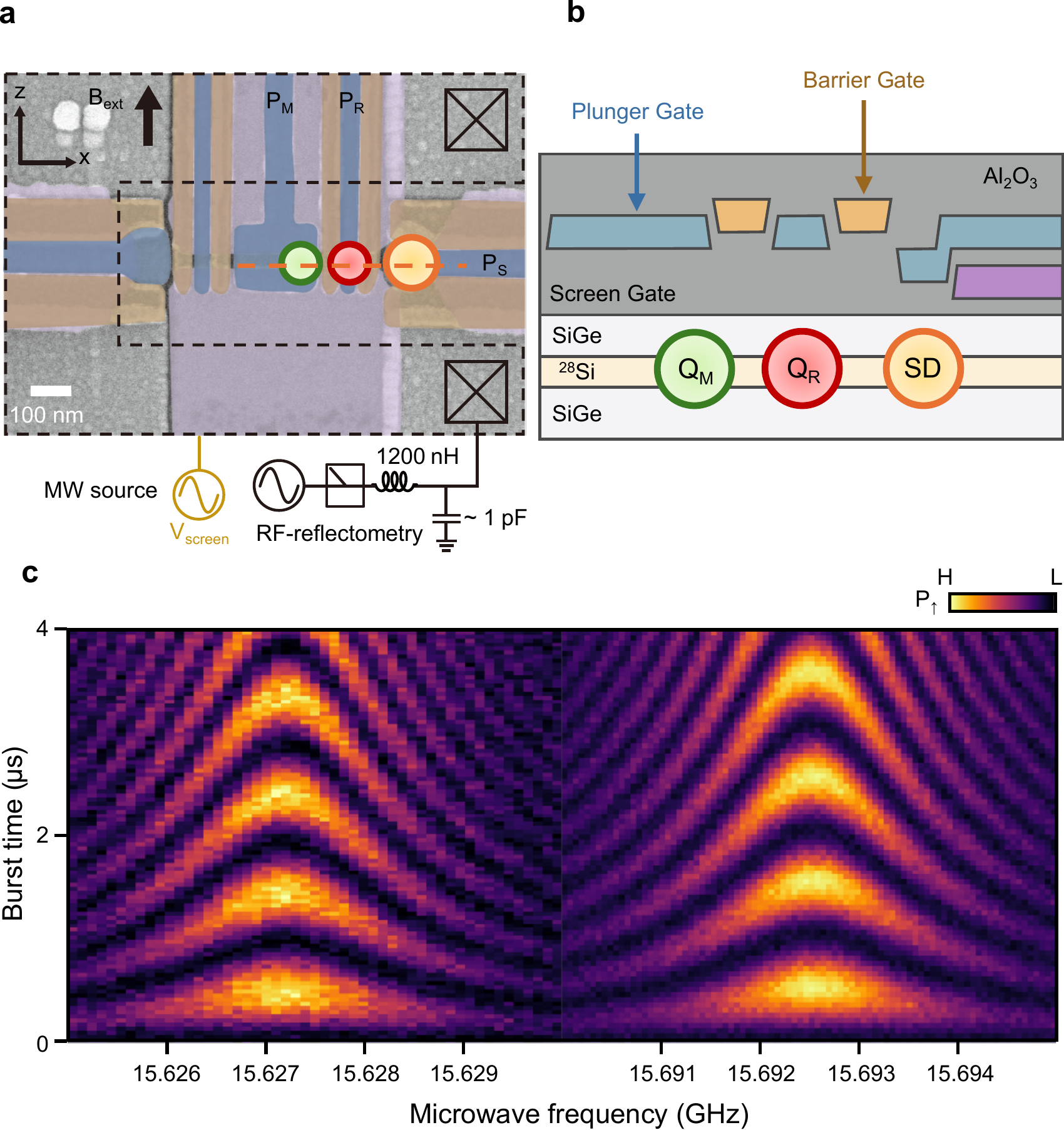}
    \caption{
        Schematic of the $^{28}$Si quantum dot device used in this experiment. False-color scanning electron micrograph (a) and cross-sectional view of the device (b). Three overlapping gate layers are fabricated on an isotopically enriched $^{28}$Si/SiGe heterostructure. The first (purple), second (blue), and third (orange) layers define the screen, plunger, and barrier gates for the qubit array and the sensor dot (SD), respectively. We investigate the rightmost quantum dot ($Q_R$; red circle beneath the right plunger gate, $P_R$) and the central quantum dot ($Q_M$; green circle beneath the middle plunger gate, $P_M$). The larger orange circle represents the SD, used for charge detection via RF reflectometry with an LC tank circuit. For qubit control, a microwave voltage $V_{\mathrm{screen}}$ is applied to a screen gate to drive spin resonance. The thin black arrows indicate the x and z axes, and the bold arrow labeled $B_{\mathrm{ext}}$ indicates the direction of the external magnetic field. In (a), the black dashed line indicates the micromagnet, and the orange dashed line indicates the position of the cross section shown in (b). (c) Rabi chevron patterns of $Q_M$ (left panel) and $Q_R$ (right panel). Both qubits are driven at similar microwave powers. The resonance frequency of $Q_R$ is approximately 70 MHz higher than that of $Q_M$, owing to the micromagnet-induced field gradient.
    }
    \label{fig:device}
\end{figure}

The device is fabricated on an isotopically purified $^{28}$Si/SiGe heterostructure wafer featuring a 9-nm silicon quantum well with a residual $^{29}$Si concentration of approximately 800 ppm \cite{degli_esposti_wafer-scale_2022,park_highly_2026}. A two-dimensional electron gas is accumulated in the quantum well by applying voltages to overlapping gate electrodes, which include screen, plunger, and barrier gates [Figs.~\ref{fig:device}(a) and \ref{fig:device}(b)]. Two quantum dots, labeled $Q_R$ and $Q_M$, are defined under the plunger gates $P_R$ and $P_M$, and a nearby sensor dot (SD) under $P_S$ is used for charge detection via radiofrequency (RF) reflectometry (Appendix A).

An external magnetic field $B_{\mathrm{ext}}$ of 0.44 T is applied along the in-plane z-direction to set the quantization axis, with the stray field adding at the dot positions [Fig.~\ref{fig:device}(a)]. A cobalt micromagnet—150 nm thick and deposited approximately 200 nm above the quantum well—generates a spatially varying stray field $B^{\mathrm{MM}}$ whose components serve distinct roles in qubit operation. The gradients of the transverse stray field components ($\partial B_x^{\mathrm{MM}}/\partial z$ and $\partial B_y^{\mathrm{MM}}/\partial z$), which couple the in-plane motion of the electron to the transverse spin direction, provide the driving field for EDSR, and a microwave tone near 15.7 GHz applied to the screen gate drives coherent spin rotations with Rabi frequencies on the order of 1 MHz [Fig.~\ref{fig:device}(c)]. The spatial variation of the longitudinal field separates the resonance frequencies of the two qubits by approximately 70 MHz, enabling individual qubit control, while its local gradients ($\partial B_z^{\mathrm{MM}}/\partial x$ and $\partial B_z^{\mathrm{MM}}/\partial z$), collectively referred to as the decoherence gradient, convert dot displacement into qubit frequency noise. We align the micromagnet to the dot array so that a local maximum of $B_z^{\mathrm{MM}}$, where the qubit frequency is least sensitive to dot displacement, lies near one of the dot positions while a strong transverse driving gradient is retained. Simulations of the stray field show that such a maximum exists near $Q_R$, with rapidly varying gradients across the array and in-plane derivatives that nearly vanish at $Q_R$, as verified in Sec. III A.

\section{Experimental Results}

\subsection{Ramsey and dynamical decoupling measurements}
\begin{figure*}[t]
    \centering
    \includegraphics[width=\linewidth]{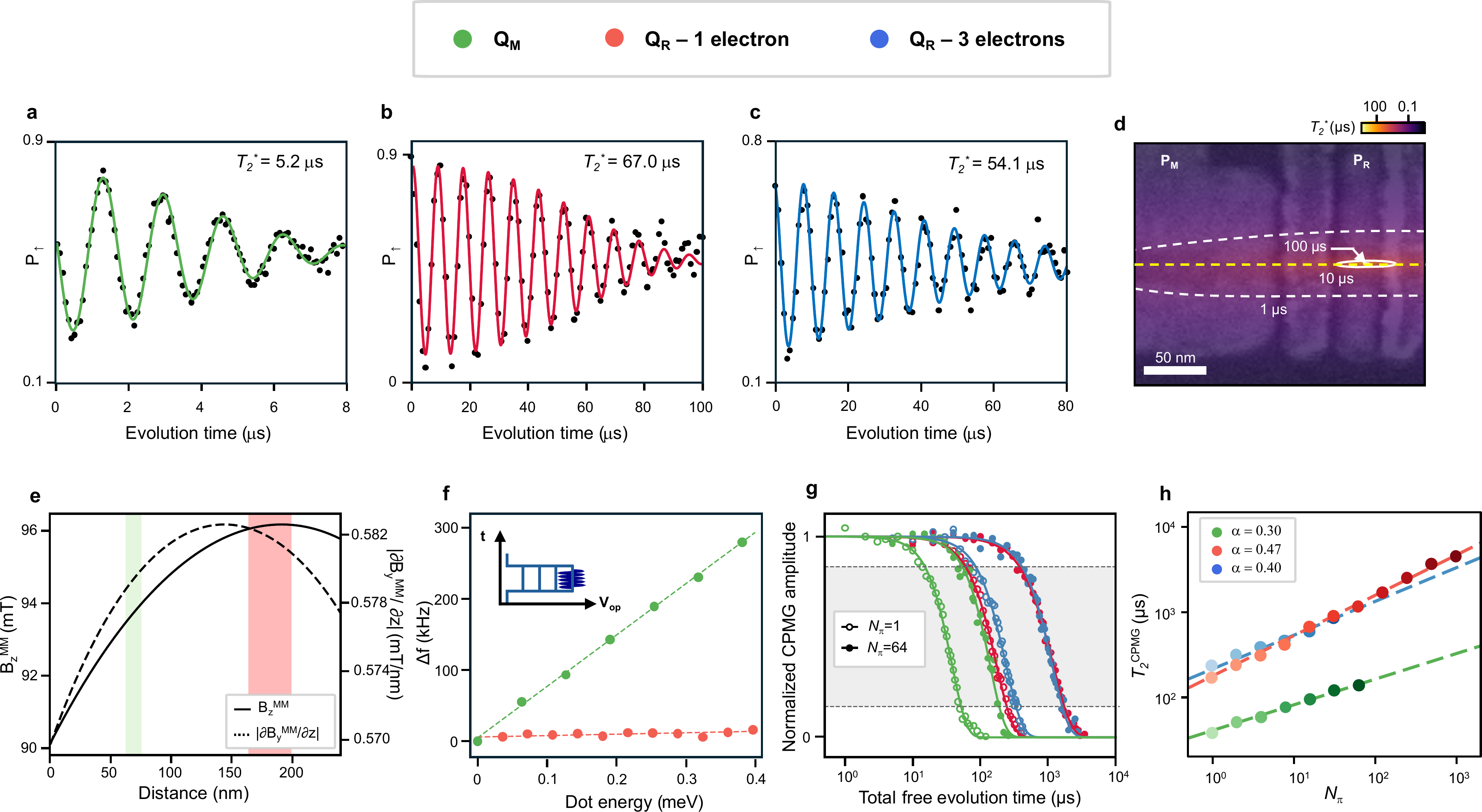}
    \caption{
        Coherence of the two qubits. (a-c) Representative Ramsey measurements obtained under three operating conditions. The qubit $Q_M$ in (a), operated away from the sweet spot, exhibits a shorter $T_2^*$ than $Q_R$ at the sweet spot in (b) and (c). Each point represents an average over 500 repetitions. (d) Simulated $T_2^*$ map calculated from the micromagnet stray field, showing a sweet spot beneath gate $P_R$. Dashed contours indicate calculated coherence times, and the dashed yellow horizontal line indicates the line cut used in (e). (e) Calculated z-component of the micromagnet field, $B_z^{\mathrm{MM}}$ (left axis, solid line), and the transverse driving gradient $|\partial B_y^{\mathrm{MM}}/\partial z|$ (right axis, dashed line), along the line cut in (d). Green and red shaded regions mark the estimated positions of $Q_M$ and $Q_R$, obtained from the gate geometry and measured resonance frequency difference in Fig.~\ref{fig:device}(c). (f) Susceptibility of the qubit resonance frequency to dot energy for $Q_R$ and $Q_M$. (g) Normalized spin-echo and CPMG amplitude as a function of total free evolution time for different numbers of refocusing pulses: $N_\pi = 1$ (spin echo, open circles) and $N_\pi = 64$ (CPMG, filled circles). The shaded region between 0.15 and 0.85 indicates the window used for noise extraction. (h) $T_2^{\mathrm{CPMG}}$ as a function of $N_\pi$. The data follow a power-law scaling $T_2^{\mathrm{CPMG}}\propto N_\pi^\alpha$.
        }
    \label{fig:coherence}
\end{figure*}

Ramsey measurements show a clear coherence contrast between $Q_R$ and $Q_M$ [Figs.~\ref{fig:coherence}(a)--(c)]. $Q_M$, which is exposed to a large decoherence gradient, shows $T_2^*$(30 s) = 5.2~$\mu$s [Fig.~\ref{fig:coherence}(a)] and an ergodic value $T_2^*(\infty)$ = 4.6~$\mu$s (Appendix B), consistent with values reported for micromagnet-based $^{28}$Si/SiGe spin qubits \cite{philips_universal_2022,fernandez_de_fuentes_running_2026,rojas-arias_scaling_2026}. In contrast, under optimized tuning, $Q_R$ reaches $T_2^*(31~\mathrm{s}) = 67~\mu$s [Fig.~\ref{fig:coherence}(b)] and $T_2^*(\infty) = 35.6~\mu$s—both longer than the corresponding values for $Q_M$ by approximately an order of magnitude, and, to our knowledge, among the longest values reported for micromagnet-based $^{28}$Si/SiGe spin qubits \cite{philips_universal_2022,yoneda_quantum-dot_2018,struck_low-frequency_2020,xue_quantum_2022,neyens_probing_2024}. Because both qubits share the same heterostructure and fabrication process, this contrast points to distinct local environments of the two qubits rather than any global device property.

To interpret this contrast, we perform micromagnet simulations using MuMax${^\mathrm3}$ \cite{vansteenkiste_design_2014} and quantify the susceptibility of each qubit frequency to spatial fluctuations driven by charge noise. From the simulated stray field profile [Fig.~\ref{fig:coherence}(d)], the local field gradient at each dot position is extracted, and $T_2^*$ is estimated as \cite{kha_micromagnets_2015,dumoulin_stuyck_low_2021}

\begin{equation}
T_2^*
=
\frac{\hbar\sqrt{2}}{g\mu_B}
\left[
\left(\frac{\partial B_z^{\mathrm{MM}}}{\partial x}\right)^2 \Delta_x^2
+
\left(\frac{\partial B_z^{\mathrm{MM}}}{\partial z}\right)^2 \Delta_z^2
\right]^{-1/2}
\label{eq:t2star_gradient}
\end{equation}

where $\Delta_x$ and $\Delta_z$ denote the root-mean-square in-plane displacements of the electron due to charge noise, and we assume an isotropic displacement $\Delta$ = $\Delta_x$ = $\Delta_z$ for simplicity. The map in Fig.~\ref{fig:coherence}(d) uses $\Delta$ = 0.1 nm, which sets the overall scale of $T_2^*$ without changing the spatial structure. Under this isotropic assumption, the map shows that $T_2^*$ can increase by up to two orders of magnitude within a narrow spatial window beneath $P_R$ where the decoherence gradient is minimized. The same simulation gives nearly identical transverse driving gradients of 0.582~mT/nm at $Q_R$ and 0.579~mT/nm at $Q_M$, with a relative difference below 1\% [Fig.~\ref{fig:coherence}(e)]—indicating that operation near the decoherence gradient minimum does not require a loss of EDSR drive strength. The comparable Rabi frequencies measured at the two positions [Fig.~\ref{fig:device}(c)] agree with these nearly equal driving gradients. Because this minimum is spatially narrow, we verify that $Q_R$ is positioned at this minimum by measuring the resonance frequency of each qubit as a function of the plunger gate voltage \cite{yoneda_quantum-dot_2018} and converting the gate voltage to dot energy using the lever arm extracted from the stability diagram [Fig.~\ref{fig:coherence}(f)]. Since charge noise perturbs the electrostatic potential of the dot, this susceptibility measures the coupling between electric-field fluctuations and the qubit frequency. The resonance frequency of $Q_M$ shifts with a slope of 723~kHz/meV, whereas $Q_R$ exhibits a slope of only 21~kHz/meV. This strong reduction of the qubit frequency susceptibility at the sweet spot is expected from the reduced decoherence gradient and accounts for the coherence contrast between the two qubits.

The coherence data presented above are obtained using energy-selective tunneling (EST) readout in the single-electron regime \cite{hanson_spins_2007,jang_robust_2020}, for which $Q_M$ is measured using shuttling. To confirm that the longer coherence at the sweet spot is not confined to a particular readout method or electron occupation, we also measure $T_2^*$ of $Q_R$ in the three-electron regime using both EST and Pauli spin blockade (PSB) readout \cite{takeda_rapid_2024}. EST readout gives $T_2^*$(30 s) = 54.1~$\mu$s [Fig.~\ref{fig:coherence}(c)] and $T_2^*(\infty)$ = 30.5~$\mu$s, and PSB readout gives $T_2^*$(28 s) = 35~$\mu$s and $T_2^*(\infty)$ = 25.9~$\mu$s, both nearly an order of magnitude longer than the $T_2^*$ of $Q_M$. These values are moderately lower than the single-electron $T_2^*$(31 s) of 67~$\mu$s, and two effects can contribute to this reduction. Three-electron operation requires gate voltage adjustments that modify the confinement potential, which may shift the quantum dot slightly within the micromagnet field profile and away from the optimal sweet spot position. In addition, changes in the multi-electron wave function, such as a modified charge distribution or orbital extent, could alter the coupling to the local noise environment. Even with this reduction, the $T_2^*$ of $Q_R$ remains roughly an order of magnitude longer than that of $Q_M$ across all configurations measured, including $Q_M$ under PSB readout where $T_2^*$ is further reduced (Appendix B), indicating that the primary factor behind the longer coherence is the position of the quantum dot within the micromagnet stray field, not the readout method or electron number.

Because the sweet spot already reduces the transduced charge noise, we examine how the coherence of $Q_R$ behaves once a refocusing sequence removes the remaining slow fluctuations. We probe this with spin echo and Carr--Purcell--Meiboom--Gill (CPMG) measurements for all three configurations. At $N_{\pi}$ = 64, the coherence time $T_2^\mathrm{CPMG}$ is approximately 1.1 ms for $Q_R$ and 138 $\mu$s for $Q_M$ [Fig.~\ref{fig:coherence}(g)]—an order-of-magnitude contrast consistent with the Ramsey data. For single-electron $Q_R$, $T_2^\mathrm{CPMG}$ reaches 4.6 ms at $N_{\pi}$ = 1024 as the refocusing pulses progressively filter out low-frequency noise. The power-law scaling $T_2^\mathrm{CPMG} \propto N_{\pi}^\alpha$ yields $\alpha$ = 0.47 and 0.40 for $Q_R$ in the single- and three-electron configurations [Fig.~\ref{fig:coherence}(h)]. From the relation $\alpha = \beta/(1+\beta)$ \cite{medford_scaling_2012}, where $\beta$ is the exponent of the $1/f^\beta$ noise spectrum, these correspond to $\beta$ $\approx$ 0.9 and 0.7; thus, the high-frequency noise sampled by the CPMG filter follows approximately $1/f$ scaling regardless of electron number at the sweet spot \cite{yoneda_quantum-dot_2018,cywinski_dynamical-decoupling_2014,connors_charge-noise_2022}. The exponent for $Q_M$ is lower, $\alpha \approx$ 0.30, which we attribute to a time dependent detuning of the qubit frequency introduced by bias tee distortion of the shuttling pulses used in the $Q_M$ measurement cycle (see Sec. S1). This detuning changes the effective rotation axis and angle during the refocusing pulses, introducing rotation errors that are not generally corrected by the refocusing of phase accumulated during free evolution. These errors provide a mechanism for the weaker CPMG scaling observed at $Q_M$.

\subsection{Qubit noise spectroscopy}
Ramsey and CPMG measurements quantify the overall dephasing time but do not isolate the individual noise contributions that limit coherence. Noise spectroscopy reconstructs the frequency-resolved noise power spectral density (PSD) \cite{yoneda_quantum-dot_2018,struck_low-frequency_2020,connors_charge-noise_2022} and, when combined with cross-correlation analysis \cite{yoneda_noise-correlation_2023,rojas-arias_spatial_2023,rojas-arias_origins_2025}, can distinguish locally generated noise from shared environmental fluctuations. In previous studies on micromagnet-based $^{28}$Si/SiGe qubits, the reconstructed spectra were dominated by charge noise transduced through a large decoherence gradient, and resolving the residual nuclear spin noise contributions remains an open challenge. In our device, the qubit at the sweet spot and the qubit in the large-gradient region share a single micromagnet, and we use this contrast to probe the dominant noise sources in each regime.

\begin{figure}[t]
    \centering
    \includegraphics[width=\linewidth]{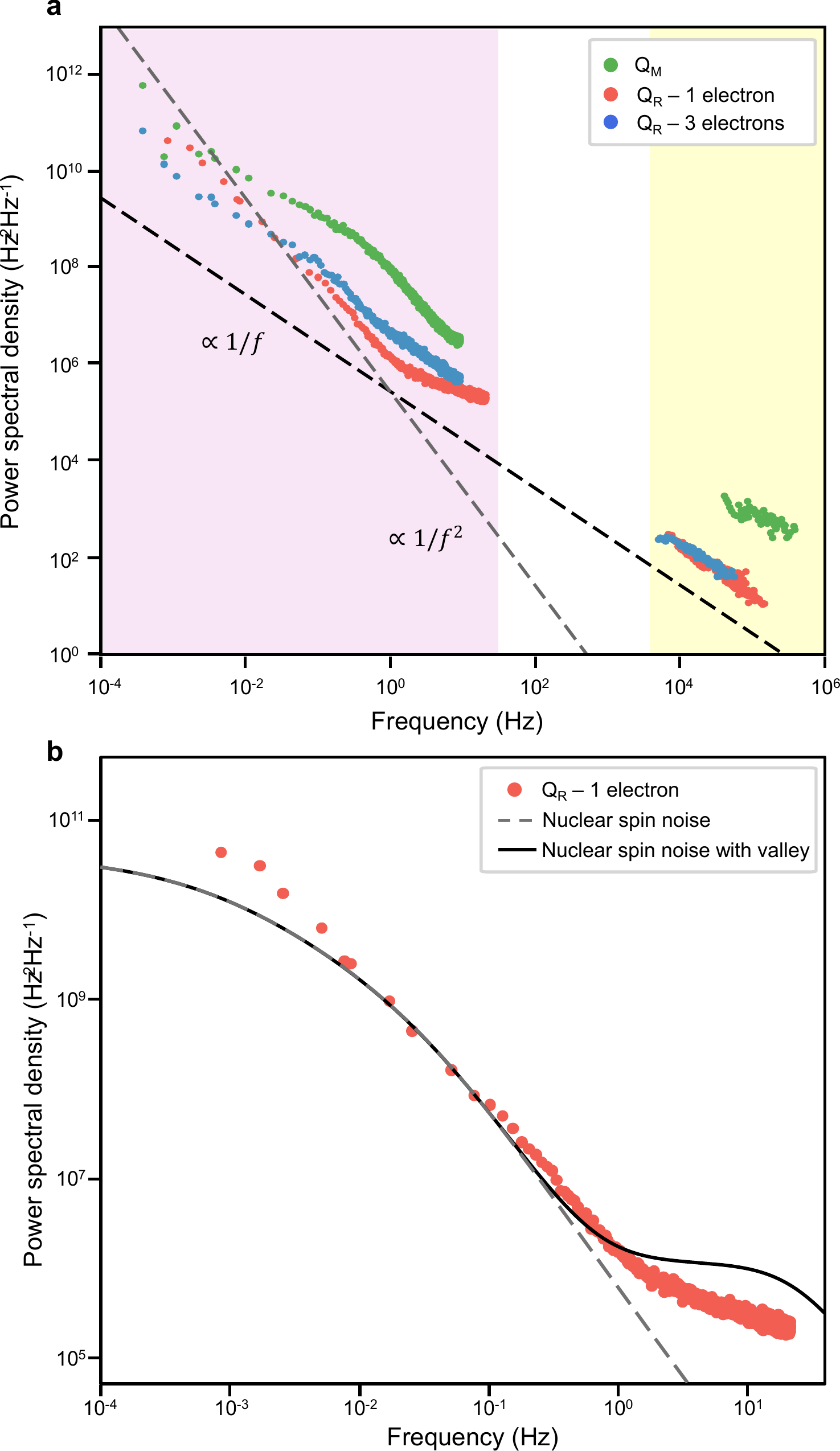}
    \caption{
        PSD of the two qubits. (a) PSD extracted from repeated Ramsey measurements based on Bayesian estimation (purple region) and CPMG measurements (yellow region). Narrow peaks from the measurement setup and their aliases are excluded. Black and gray dashed lines indicate reference slopes proportional to $1/f$ and $1/f^2$. (b) PSD of $Q_R$ in the single-electron regime (red). The gray dashed line shows the $^{29}$Si nuclear spin noise prediction without valley modulation, and the black solid line shows the same prediction with valley modulation.
        }
    \label{fig:auto-psd}
\end{figure}

We reconstruct the auto-power spectral densities (auto-PSDs) of the three operating configurations over nearly 10 decades in frequency, from approximately $10^{-4}$ to $10^6$ Hz. All spectral densities $S(f)$ in this work are single-sided and defined for $f\geq 0$. We combine Ramsey measurements based on Bayesian estimation \cite{yoneda_noise-correlation_2023,park_passive_2025} at low frequencies with CPMG-based spectroscopy at higher frequencies \cite{yoneda_quantum-dot_2018,cywinski_dynamical-decoupling_2014,connors_charge-noise_2022} [Fig.~\ref{fig:auto-psd}(a); experimental details are provided in Secs. S2 and S3]. Over most of the measured bandwidth, the noise power of $Q_R$ remains below that of $Q_M$, and the separation reaches nearly two orders of magnitude near 1 Hz. Above approximately $10^3$ Hz, all three spectra follow a common $1/f^\beta$ scaling—a charge noise feature widely reported at these frequencies \cite{yoneda_quantum-dot_2018,kepa_simulation_2023,paladino_1_2014}. In contrast, the clearest spectral difference appears at low frequencies. Single-electron $Q_R$ follows a $1/f^2$ trend that flattens near 1~Hz, while the three-electron $Q_R$ and $Q_M$ follow a combination of $1/f$ and $1/f^2$ behaviors. The low-frequency structure in the $Q_M$ and three-electron $Q_R$ spectra reflects their stronger coupling to charge fluctuations, possibly including contributions from nearby two-level fluctuators (TLFs). This coupling arises from the large decoherence gradient at $Q_M$ and from the displacement of the dot from the sweet spot during three-electron operation. Among the three configurations, the single-electron $Q_R$ shows both the lowest noise power and a clear $1/f^2$ spectral shape with a plateau. Hence, we focus on this configuration to identify the dominant noise sources in the long-coherence regime.

The $1/f^2$ spectrum of single-electron $Q_R$ can originate from a small number of TLFs \cite{struck_low-frequency_2020,cheng_modeling_2025} or from $^{29}$Si nuclear spin fluctuations coupled through the hyperfine interaction \cite{yoneda_noise-correlation_2023,rojas-arias_spatial_2023}. A small number of charge fluctuators produces a Lorentzian spectrum, a plateau followed by a $1/f^2$ roll-off, and nuclear spin diffusion produces the same form. In silicon, fast valley oscillations in the electron wave function modulate the hyperfine coupling and add a second plateau to the nuclear spectrum, so the two origins are distinguished by the presence of a plateau between two $1/f^2$ regions \cite{rojas-arias_origins_2025}. We evaluate this nuclear spin noise model with and without valley modulation using the nuclear spin diffusion constant $D = 0.4~\mathrm{nm}^2/\mathrm{s}$, which is appropriate for the 800~ppm $^{29}$Si concentration of our heterostructure \cite{rojas-arias_spatial_2023,rojas-arias_origins_2025,hayashi_nuclear_2008}, and in-plane and out-of-plane confinement lengths $l_p = 14~\mathrm{nm}$ and $l_y = 2.5~\mathrm{nm}$, respectively, typical of gate-defined Si/SiGe quantum dots [Fig.~\ref{fig:auto-psd}(b); details of the model are given in Appendix~C]. The measured spectrum follows the predicted magnitude and $1/f^2$ decay across the low-frequency range. Near 1~Hz the two predictions separate, decaying as $1/f^2$ without the valley term and flattening into a plateau with it. In this range the measured spectrum falls between the two predictions. The residual low-frequency noise of single-electron $Q_R$ approaches the level set by $^{29}$Si nuclear spin fluctuations, which the strongly reduced transduction of charge noise at the sweet spot makes observable. The second $1/f^2$ region above the plateau, however, is predicted near the Nyquist frequency of the Ramsey sampling rate and is not resolved, so a charge origin cannot be excluded from the spectral shape alone. We therefore examine the remaining charge noise contribution through cross-correlation analysis.

\subsection{Noise correlation}
\begin{figure}[t]
    \centering
    \includegraphics[width=\linewidth]{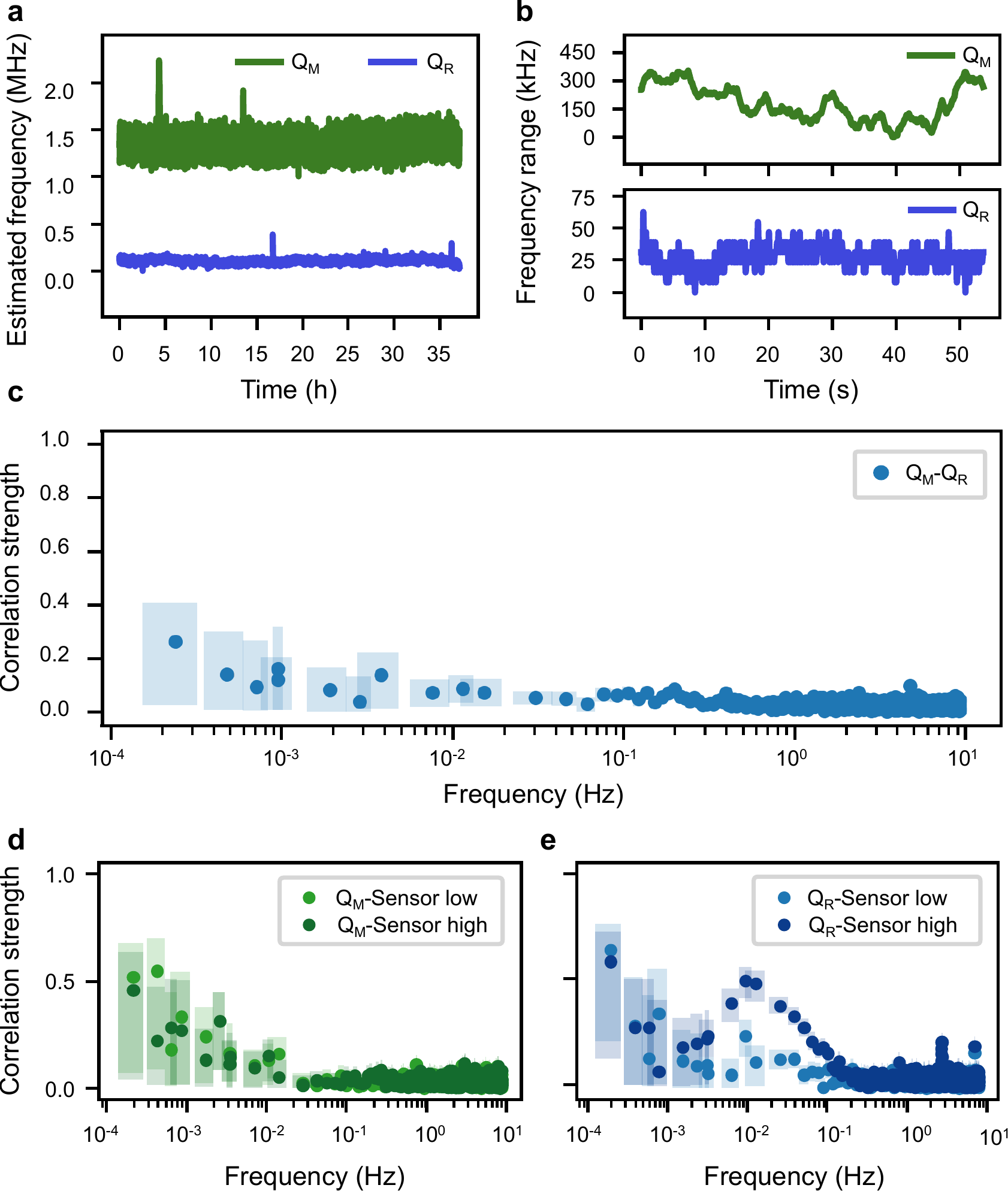}
    \caption{
        Qubit frequencies and cross-correlation strength. (a) Time traces of estimated qubit frequencies, simultaneously measured for qubits $Q_M$ and $Q_R$. $Q_R$, operating at the decoherence gradient sweet spot, exhibits markedly narrower frequency fluctuations than $Q_M$. (b) Frequency-fluctuation ranges for the two qubits, extracted over a representative time window in (a). The fluctuation range of $Q_M$ is approximately 6 times wider than that of $Q_R$. (c) Correlation strength between qubits $Q_M$ and $Q_R$. Despite being nearest-neighbor qubits, the data show overall low correlation across the frequency range. (d, e) Correlation strengths between the sensor signal and qubits $Q_M$ and $Q_R$. The correlations are calculated separately for the low (even parity) and high (odd parity) levels, shown in light and dark colors, respectively. The $Q_R$–sensor correlation in (e) exhibits a broad finite correlation between $10^{-3}$~Hz and $10^{-1}$~Hz, reaching 0.5. The shaded regions represent 90~\% confidence intervals.
    }
    \label{fig:correlation}
\end{figure}

Cross-correlation analysis between qubits reveals noise sources shared across the device and is an established tool for identifying spatially correlated noise contributions in semiconductor spin qubit platforms \cite{yoneda_noise-correlation_2023,rojas-arias_spatial_2023}. In these studies, charge noise or strongly coupled TLFs shared by neighboring qubits produce high interqubit correlation strengths. At the sweet spot, where transduced charge noise is strongly reduced, the auto-PSD is consistent with a nuclear origin but cannot rule out a residual charge contribution on its own. The interqubit correlation provides an additional test because charge noise that reaches both qubits appears as a shared signature, while nuclear spin noise, generated independently at each qubit site through the local hyperfine interaction, does not. Using PSB readout with $Q_R$ in the three-electron regime, we measure the frequencies of $Q_M$ and $Q_R$ simultaneously over 37 h through interleaved Bayesian Ramsey sequences and compute the normalized cross-PSD between their frequency fluctuations [Figs.~\ref{fig:correlation}(a) and \ref{fig:correlation}(b)]. We remove a slow monotonic drift originating from the superconducting magnet before computing the cross-PSD, so that the correlations reflect the qubit frequency fluctuations rather than this instrumental drift \cite{rojas-arias_scaling_2026}. The time traces show that $Q_R$ fluctuates within a 75-kHz range on the timescale of seconds, while $Q_M$ spans approximately 450 kHz, reflecting the large difference in decoherence gradient between the two positions. 

We denote the cross-PSD between two recorded traces $i$ and $j$ as

\begin{equation}
C_{ij}(f)
=
2\int_{-\infty}^{\infty}
d\tau\,
e^{2\pi i f\tau}
\left\langle
\delta x_i(t)\,
\delta x_j(t+\tau)
\right\rangle,
\qquad
f\geq 0.
\label{eq:cross_psd}
\end{equation}

Here, $\tau$ denotes the time delay between the two recorded traces $i$ and $j$, and $\delta x_i(t)$ is the fluctuation of the trace $i$. In this work these traces are qubit frequency fluctuations and charge sensor signals. The operator $\langle \cdots \rangle$ denotes a statistical average over time $t$, and the factor of two makes $C_{ij}$ a single-sided density defined for $f\geq0$. In general, the complex values of the cross-PSD provide information about the spatial correlations of the noise. We characterize the cross-PSD through the normalized correlation coefficient 
$r(f)=C_{MR}(f)/\sqrt{S_M(f)S_R(f)}$, with labels 
$i,j\in\{M,R\}$ indicating the two qubits, where 
$S_M\equiv C_{MM}$ and $S_R\equiv C_{RR}$. 
The normalized cross-PSD is a complex function 
$r(f)=|r(f)|e^{i\phi(f)}$, whose phase $\phi$ indicates whether the noise enters the two qubits in phase or out of phase, while its magnitude $|r|$, which is referred to as the correlation strength, measures the proportion of correlated noise relative to the total noise. At a given frequency, a correlation $|r| = 1$ corresponds to perfect spatial correlation, while $|r| = 0$ corresponds to independent (uncorrelated) noise. A phase of $\phi = 0$ corresponds to positive correlations, $\phi = \pi$ corresponds to negative correlations, and other values of $\phi$ indicate a time lag between the two signals.

Fig.~\ref{fig:correlation}(c) shows the correlation strength $|r(f)|$ between $Q_M$ and $Q_R$. In previous studies on comparable devices, nearest-neighbor qubits separated by approximately 100 to 200~nm exhibited correlation strengths as high as 0.7 \cite{yoneda_noise-correlation_2023,rojas-arias_spatial_2023}, reflecting charge noise shared across the qubit array. In our device, $|r(f)|$ is consistent with zero across most of the measured frequency range, and even its largest values, near 0.3, remain well below this level. These largest values occur below approximately $10^{-3}$ Hz, where the calculation uses few segments and the confidence intervals are wide. This weak interqubit correlation is consistent with $Q_R$ operating at the sweet spot, where the reduced decoherence gradient weakens the coupling to the electric-field fluctuations that generate shared noise between neighboring qubits, and with residual noise generated locally at each qubit site, as expected for hyperfine coupling to the $^{29}$Si bath.

Figs.~\ref{fig:correlation}(d) and \ref{fig:correlation}(e) show the correlation strength between each qubit frequency and the sensor signal. In PSB readout the sensor signal takes two levels that correspond to the even- and odd-parity outcomes, and we compute the correlation for the low and high levels separately to separate the shared fluctuation from the readout background \cite{rojas-arias_spatial_2023}. The charge sensor is sensitive to electrostatic fluctuations, so a fluctuation shared with a qubit frequency originates from an electrostatic source. The $Q_R$--sensor correlation exhibits a broad peak between $10^{-3}$ and $10^{-1}$~Hz, reaching 0.5. The high level, where the local slope of the Coulomb peak is approximately four times steeper, carries the stronger correlation, as the steeper slope converts the shared fluctuation into a larger part of the recorded signal (see Sec.~S4). Across the measured frequency range, the $Q_M$--sensor and $Q_M$--$Q_R$ correlations show no distinct signature of noise shared with $Q_M$. The magnitude of the $Q_R$--sensor cross-PSD follows a Lorentzian form that rolls off where the correlation is largest. A Lorentzian of this kind usually arises from a shared TLF with a constant switching rate \cite{connors_low-frequency_2019}, although the charge sensor stays on throughout the measurement and the correlated signal may include a contribution originating in the sensor dot itself \cite{park_passive_2025,hell_qubit_2016}.

The near-zero interqubit correlation indicates that the dephasing due to charge noise shared between $Q_R$ and $Q_M$ is strongly reduced at the sweet spot, while the $Q_R$--sensor correlation points to a local source for the charge noise that remains at $Q_R$. A finite correlation reflects the relative weight of the shared fluctuation in each channel and does not by itself set the absolute noise power. Although the spectra vary between operating configurations, the noise power at $Q_R$ remains below that at $Q_M$ over most of the measured frequency range [Fig.~\ref{fig:auto-psd}(a)], with the low-frequency spectrum in Fig.~\ref{fig:auto-psd}(b) approaching the level predicted for $^{29}$Si nuclear spin noise. These results support the conclusion that the coherence of $Q_R$ approaches the nuclear spin limit of the device, with charge-noise-induced dephasing reduced to the scale set by nuclear spin fluctuations.

\section{Discussion}
Our results show that operating at a minimum of the decoherence gradient substantially extends $T_2^*$ in a $^{28}$Si/SiGe device with a micromagnet. The two qubits, $Q_R$ and $Q_M$, are separated by approximately 100~nm on the same heterostructure; thus, the large coherence contrast between them cannot arise from a global device property. The measured susceptibility of the qubit frequency to the dot energy is more than an order of magnitude smaller at $Q_R$ than at $Q_M$, indicating that reduced transduction of charge noise through the decoherence gradient is a key factor in the enhanced $T_2^*$ at $Q_R$. Despite this large difference in decoherence gradient, the measured Rabi frequencies are comparable at the two positions; hence, the coherence extension does not come at the cost of slower electrical control.

At the sweet spot, noise spectroscopy and cross-correlation analysis characterize the residual noise environment. The auto-PSD of single-electron $Q_R$ approaches the level predicted for $^{29}$Si nuclear spin noise in this material. The interqubit correlation is weak, as expected for nuclear noise generated independently at each qubit site, while the qubit–sensor correlation points to a local source for the charge noise that remains at $Q_R$. The transduced charge noise is therefore strongly reduced but not fully removed, and the residual charge and nuclear contributions coexist near the predicted nuclear spin noise floor of the material.

Theoretical estimates of nuclear hyperfine dephasing predict a nuclear-noise-limited $T_2^*$ of a few to a few tens of microseconds at this level of isotopic purification \cite{witzel_nuclear_2012,kerckhoff_magnetic_2021,cvitkovich_coherence_2024}. Whether a device reaches this limit has been difficult to test, as the transduced charge noise in micromagnet-based devices remains above the nuclear level \cite{yoneda_quantum-dot_2018,struck_low-frequency_2020}. At the sweet spot, $Q_R$ reaches $T_2^*(\infty) = 35.6~\mu \mathrm{s}$ and the two contributions become comparable. Because residual charge noise still contributes at this operating point, the intrinsic nuclear limit of the device may lie above this value.

These results point to several directions for further work. Minimizing the residual charge noise, through improved oxide quality \cite{kepa_simulation_2023,connors_low-frequency_2019}, optimized gate geometry, or screening of the surrounding charge noise \cite{barnes_screening_2011,higginbotham_coherent_2014,choi_ballast_2025}, would resolve the nuclear limit more clearly and allow the nuclear noise models to be tested directly. Pulsing the charge sensor into Coulomb blockade during qubit manipulation \cite{park_passive_2025,hell_qubit_2016} would distinguish between charge sensor backaction and a fluctuator in the device, and would indicate where further reduction of the residual charge noise should be directed. Repeating the auto-PSD and cross-PSD analyses on devices with different $^{29}$Si concentrations would map how the nuclear coherence limit scales with isotopic purity and would quantify the $^{29}$Si and $^{73}$Ge contributions in an environment where transduced charge noise is strongly reduced \cite{cvitkovich_coherence_2024}. More broadly, the coherence contrast we observe arises from local factors that vary over 100 nm within a single device. Building a scalable quantum computer calls for an understanding of these factors at each qubit site, which include not only the field gradient produced by the micromagnet but also heterostructure inhomogeneity and strain variation. Toward that end, our results show that sweet spot operation is effective at a single dot position, and this is a starting point for extending it across an array.

\begin{acknowledgments}
This work was supported by a National Research Foundation of Korea (NRF) grant funded by the Korean Government (Ministry of Science and ICT (MSIT)) (RS-2023-00283291, RS-2024-00413957, SRC Center for Quantum Coherence in Condensed Matter RS-2023-00207732, RS-2023-NR077112 and Quantum Technology R\&D Leading Program (Quantum Computing) RS-2024-00442994) and a core center program grant funded by the Ministry of Education (No. 2021R1A6C101B418).
J. Yoneda acknowledges financial support from Kakenhi (JP26K01331) and JST FOREST Grant No. JPMJFR244D.
\end{acknowledgments}

\section*{AUTHOR CONTRIBUTIONS}
S.L. and H.S. performed the measurements, analyzed the data, and wrote the manuscript. J.P. fabricated the device. H.J. developed the Bayesian estimation software. J. Yun performed the micromagnet simulations. J. Yoneda contributed to the discussion of the noise correlation analysis. L.E.A.S. and D.D.E. developed and characterized the heterostructure under the supervision of G.S. D.K. supervised the project.

\section*{DATA AVAILABILITY}
The data that support the findings of this study are available from the corresponding author upon request.

\appendix
\section{EXPERIMENTAL SETUP}
The device is cooled in a dilution refrigerator (Oxford Instruments Triton-500) with the mixing chamber plate stabilized at approximately 45~mK and an electron temperature of 126 mK. DC voltages are applied to all metal gates and ohmic contacts via voltage sources (SIM928, Stanford Research Systems), with selected lines further connected to an arbitrary waveform generator (AWG, Quantum Machines OPX+ and SDT QCU) or a signal generator (Quantum Machines Octave) through bias tees. I/Q signals generated by the AWG are routed to the Octave, which serves as a local oscillator and up-converter to produce I/Q modulated microwave pulses for spin qubit manipulation. Additional baseband pulses are used to control the quantum dot energy levels and tunnel coupling. RF signals at the resonant frequencies of the LC tank circuits are supplied by a lock-in amplifier (Zurich Instruments UHFLI) and directed to two separate ohmic contacts. The reflected signals, modulated by the conductance of the sensor dots, are amplified by 45 dB through two series-connected cryogenic amplifiers (Cosmic Microwave Technology CITLF2) and by a further 25 dB via a room-temperature amplifier (Lotus Communication Systems LNA100M2P0G), before being demodulated by the lock-in amplifier and digitized by the OPX+.

\section{ACQUISITION-TIME DEPENDENCE OF $T_2^*$ AND THE ERGODIC LIMIT}

The inhomogeneous dephasing time extracted from a Ramsey experiment depends on the total data acquisition time $t_{\mathrm{acq}}$ because a finite acquisition samples only noise components at frequencies above the effective low-frequency cutoff $1/t_{\mathrm{acq}}$ \cite{struck_low-frequency_2020,neyens_probing_2024,delbecq_quantum_2016}. In the quasi-static Gaussian approximation, the frequency variance that determines $T_2^*$ is the integral of the noise PSD from $1/t_{\mathrm{acq}}$ upward; hence, a longer acquisition admits more low-frequency noise power, and $T_2^*(t_{\mathrm{acq}})$ decreases monotonically. Once $t_{\mathrm{acq}}$ exceeds the correlation time of the dominant slow components, the additional noise power gained by extending the acquisition time becomes negligible, and $T_2^*(t_{\mathrm{acq}})$ approaches an ergodic value. We denote this value $T_2^*(\infty)$ and define it operationally as $T_2^*$ measured at the longest acquisition time of each data set.

\begin{figure}[t]
    \centering
    \includegraphics[width=\linewidth]{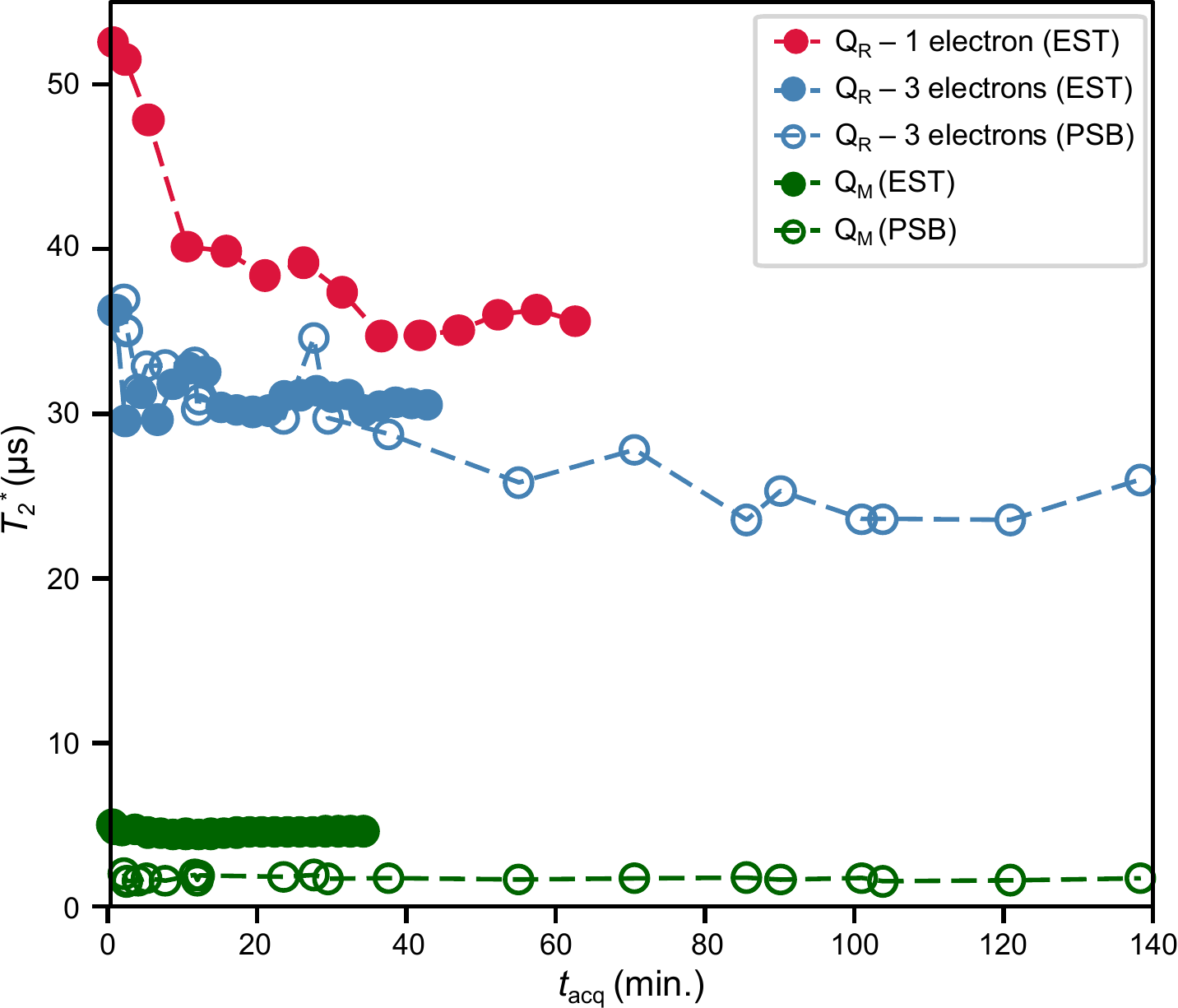}
    \caption{
        Ramsey dephasing time $T_2^*$ as a function of acquisition time $t_{\mathrm{acq}}$ for $Q_R$ and $Q_M$. Red markers show $Q_R$ in the single-electron regime with EST readout. Blue markers show $Q_R$ in the three-electron regime (filled for EST and open for PSB). Green markers show $Q_M$ (filled for EST and open for PSB). For every data set, $T_2^*$ decreases as $t_{\mathrm{acq}}$ increases because a longer acquisition lowers the effective low-frequency cutoff $1/t_{\mathrm{acq}}$ and admits more low-frequency noise into the measured frequency variance.
    }
    \label{fig:acq}
\end{figure}

Fig.~\ref{fig:acq} shows $T_2^*(t_{\mathrm{acq}})$ for the five operating configurations: $Q_R$ in the single-electron regime with EST readout, $Q_R$ in the three-electron regime with EST and PSB readout, and $Q_M$ with EST and PSB readout. For every data set, $T_2^*$ decreases with $t_{\mathrm{acq}}$ and approaches a configuration-dependent plateau. The resulting ergodic values are $T_2^*(\infty) = 35.6~\mu\mathrm{s}$ for single-electron $Q_R$ ($t_{\mathrm{acq}} = 63~\mathrm{min}$), 30.5 and 25.9 $~\mu\mathrm{s}$ for three-electron $Q_R$ with EST and PSB readout ($t_{\mathrm{acq}}$ = 43 min and 138 min), and 4.6 and 1.7 $~\mu\mathrm{s}$ for $Q_M$ with EST and PSB readout ($t_{\mathrm{acq}}$ = 34 min and 138 min). The contrast between $Q_R$ and $Q_M$ observed at short acquisition times therefore persists in the ergodic limit; hence, the coherence extension at the sweet spot is not limited to short acquisition times.

\section{$^{29}$Si NUCLEAR SPIN NOISE MODEL}
\begin{figure}[t]
    \centering
    \includegraphics[width=\linewidth]{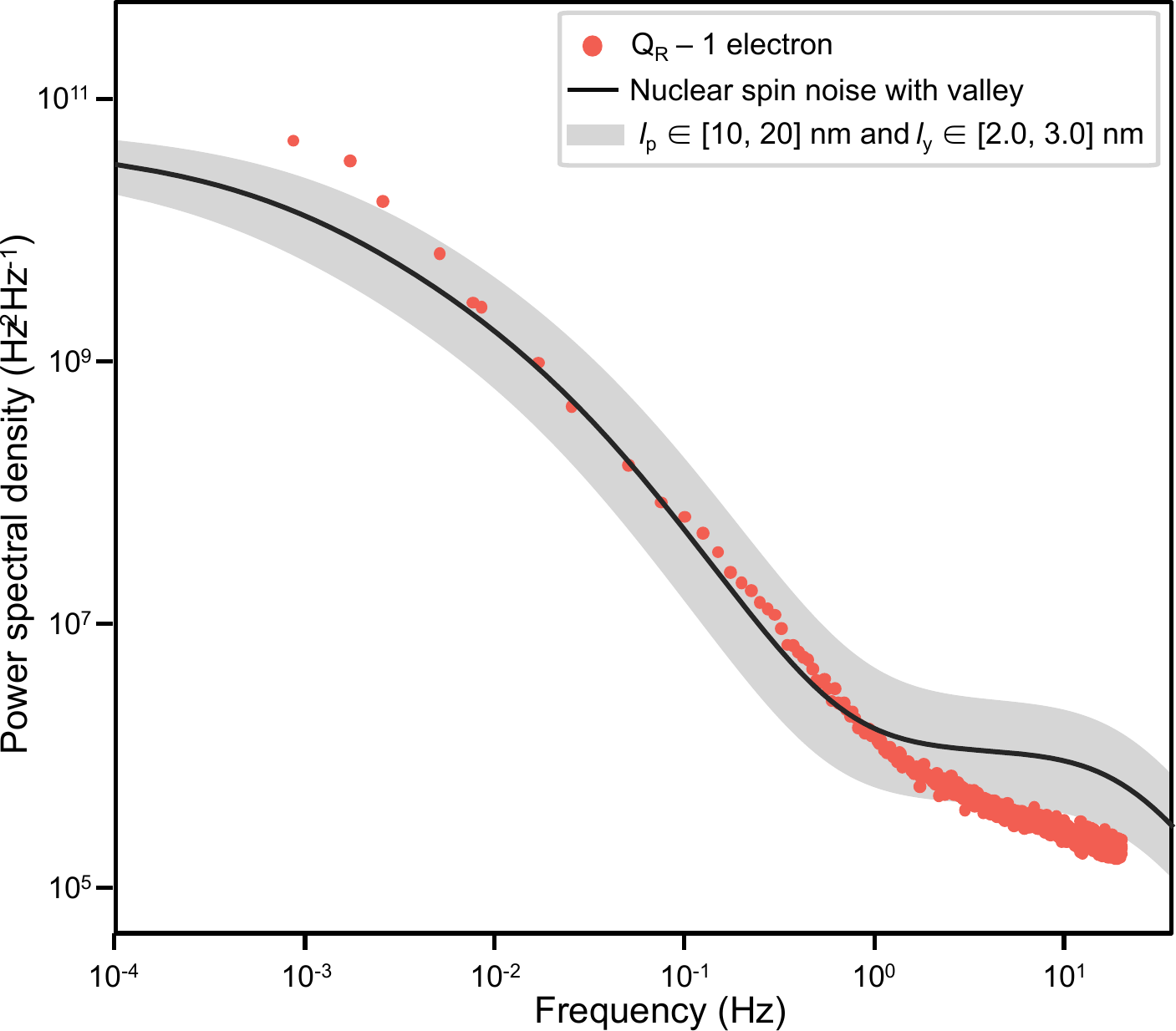}
    \caption{
        Ramsey noise spectrum of $Q_R$ in the single-electron regime with the
        valley-modulated nuclear spin prediction. The solid line uses
        $l_p = 14~\mathrm{nm}$ and $l_y = 2.5~\mathrm{nm}$, and the band spans
        $l_p \in [10,20]~\mathrm{nm}$ and
        $l_y \in [2.0,3.0]~\mathrm{nm}$.
        }
    \label{fig:nuclear_noise}
\end{figure}

We model the $^{29}$Si nuclear spin noise following the nuclear spin
diffusion approach of Refs.~\cite{rojas-arias_spatial_2023,rojas-arias_origins_2025,reilly_measurement_2008,malinowski_spectrum_2017}. The hyperfine shift of the
qubit frequency is
$\delta f(t) = (pA/h)\int d^3x\,P(x,t)|\psi(x)|^2$,
where $p$ is the residual $^{29}$Si fraction,
$A = 2.4~\mu\mathrm{eV}$ is the contact hyperfine coupling strength
in silicon \cite{philippopoulos_first-principles_2020,assali_hyperfine_2011,schliemann_electron_2003}, and $P(x,t)$ is the local nuclear polarization,
whose dynamics obey a diffusion equation with diffusion constant $D$
and are driven by a spatially and temporally uncorrelated stochastic
force. In silicon, the electron density combines a Gaussian envelope
with fast valley oscillations:

\begin{equation}
|\psi(x)|^2
\propto
\exp\left[
-\frac{x^2+z^2}{l_p^2}
-\frac{y^2}{l_y^2}
\right]
\cos^2(k_v y).
\label{eq:nuclear_wavefunction}
\end{equation}

Here, $l_p$ and $l_y$ denote the in-plane and out-of-plane confinement lengths, and the valley wave vector $k_v = 0.85 \times 2\pi/a_0$, with $a_0 = 0.543~\mathrm{nm}$, denoting the silicon lattice constant, is fixed by the band structure \cite{rojas-arias_spatial_2023}. With this wave function, the frequency correlator $\langle \delta f(t)\,\delta f(t+\tau)\rangle$ takes a closed form, and its Fourier transform gives the auto-PSD plotted in Figs.~\ref{fig:auto-psd}(b) and \ref{fig:nuclear_noise}. Evaluating this correlator with and without the valley term gives the two curves in Fig.~\ref{fig:auto-psd}. With the valley term, the spectrum develops a plateau between two $1/f^2$ regions, while setting $k_v \rightarrow 0$ leaves a single $1/f^2$ decay without this feature.

For our heterostructure, $p = 800~\mathrm{ppm}$, and we use $D = 0.4~\mathrm{nm}^2/\mathrm{s}$, the value adopted at this concentration in Refs.~\cite{rojas-arias_spatial_2023,rojas-arias_origins_2025} based on the concentration dependence of the nuclear spin diffusion constants measured in isotopically controlled silicon \cite{hayashi_nuclear_2008}. The confinement lengths are typical values for gate-defined Si/SiGe quantum dots, $l_p = 14~\mathrm{nm}$ and $l_y = 2.5~\mathrm{nm}$, and are not independently measured for our device. Fig.~\ref{fig:nuclear_noise} shows the prediction for $l_p \in [10,20]~\mathrm{nm}$ and $l_y \in [2.0,3.0]~\mathrm{nm}$. Varying the confinement lengths within this range shifts the predicted magnitude, while the plateau remains in the same frequency range.

With these parameters, the valley-modulated prediction captures the overall magnitude of the single-electron $Q_R$ spectrum and the flattening of its decay near 1~Hz, which the prediction without the valley term does not capture [Fig.~\ref{fig:auto-psd}(b)]. In this range, the measured PSD lies below the valley-modulated prediction and above the prediction without the valley term. The residual low-frequency noise of $Q_R$ at the sweet spot therefore approaches the level set by the $^{29}$Si nuclear spin bath.

\bibliography{references}

\begin{thebibliography}{11}%
\makeatletter
\providecommand \@ifxundefined [1]{%
 \@ifx{#1\undefined}
}%
\providecommand \@ifnum [1]{%
 \ifnum #1\expandafter \@firstoftwo
 \else \expandafter \@secondoftwo
 \fi
}%
\providecommand \@ifx [1]{%
 \ifx #1\expandafter \@firstoftwo
 \else \expandafter \@secondoftwo
 \fi
}%
\providecommand \natexlab [1]{#1}%
\providecommand \enquote  [1]{``#1''}%
\providecommand \bibnamefont  [1]{#1}%
\providecommand \bibfnamefont [1]{#1}%
\providecommand \citenamefont [1]{#1}%
\providecommand \href@noop [0]{\@secondoftwo}%
\providecommand \href [0]{\begingroup \@sanitize@url \@href}%
\providecommand \@href[1]{\@@startlink{#1}\@@href}%
\providecommand \@@href[1]{\endgroup#1\@@endlink}%
\providecommand \@sanitize@url [0]{\catcode `\\12\catcode `\$12\catcode `\&12\catcode `\#12\catcode `\^12\catcode `\_12\catcode `\%12\relax}%
\providecommand \@@startlink[1]{}%
\providecommand \@@endlink[0]{}%
\providecommand \url  [0]{\begingroup\@sanitize@url \@url }%
\providecommand \@url [1]{\endgroup\@href {#1}{\urlprefix }}%
\providecommand \urlprefix  [0]{URL }%
\providecommand \Eprint [0]{\href }%
\providecommand \doibase [0]{https://doi.org/}%
\providecommand \selectlanguage [0]{\@gobble}%
\providecommand \bibinfo  [0]{\@secondoftwo}%
\providecommand \bibfield  [0]{\@secondoftwo}%
\providecommand \translation [1]{[#1]}%
\providecommand \BibitemOpen [0]{}%
\providecommand \bibitemStop [0]{}%
\providecommand \bibitemNoStop [0]{.\EOS\space}%
\providecommand \EOS [0]{\spacefactor3000\relax}%
\providecommand \BibitemShut  [1]{\csname bibitem#1\endcsname}%
\let\auto@bib@innerbib\@empty
\bibitem [{\citenamefont {Cywi{\'n}ski}(2014)}]{cywinski_dynamical-decoupling_2014}%
  \BibitemOpen
  \bibfield  {author} {\bibinfo {author} {\bibfnamefont {{\L}.}~\bibnamefont {Cywi{\'n}ski}},\ }\href {https://doi.org/10.1103/PhysRevA.90.042307} {\bibfield  {journal} {\bibinfo  {journal} {Physical Review A}\ }\textbf {\bibinfo {volume} {90}},\ \bibinfo {pages} {042307} (\bibinfo {year} {2014})}\BibitemShut {NoStop}%
\bibitem [{\citenamefont {Yoneda}\ \emph {et~al.}(2023)\citenamefont {Yoneda}, \citenamefont {Rojas-Arias}, \citenamefont {Stano}, \citenamefont {Takeda}, \citenamefont {Noiri}, \citenamefont {Nakajima}, \citenamefont {Loss},\ and\ \citenamefont {Tarucha}}]{yoneda_noise-correlation_2023}%
  \BibitemOpen
  \bibfield  {author} {\bibinfo {author} {\bibfnamefont {J.}~\bibnamefont {Yoneda}}, \bibinfo {author} {\bibfnamefont {J.~S.}\ \bibnamefont {Rojas-Arias}}, \bibinfo {author} {\bibfnamefont {P.}~\bibnamefont {Stano}}, \bibinfo {author} {\bibfnamefont {K.}~\bibnamefont {Takeda}}, \bibinfo {author} {\bibfnamefont {A.}~\bibnamefont {Noiri}}, \bibinfo {author} {\bibfnamefont {T.}~\bibnamefont {Nakajima}}, \bibinfo {author} {\bibfnamefont {D.}~\bibnamefont {Loss}},\ and\ \bibinfo {author} {\bibfnamefont {S.}~\bibnamefont {Tarucha}},\ }\href {https://doi.org/10.1038/s41567-023-02238-6} {\bibfield  {journal} {\bibinfo  {journal} {Nature Physics}\ }\textbf {\bibinfo {volume} {19}},\ \bibinfo {pages} {1793} (\bibinfo {year} {2023})}\BibitemShut {NoStop}%
\bibitem [{\citenamefont {Park}\ \emph {et~al.}(2025)\citenamefont {Park}, \citenamefont {Jang}, \citenamefont {Sohn}, \citenamefont {Yun}, \citenamefont {Song}, \citenamefont {Kang}, \citenamefont {Stehouwer}, \citenamefont {Esposti}, \citenamefont {Scappucci},\ and\ \citenamefont {Kim}}]{park_passive_2025}%
  \BibitemOpen
  \bibfield  {author} {\bibinfo {author} {\bibfnamefont {J.}~\bibnamefont {Park}}, \bibinfo {author} {\bibfnamefont {H.}~\bibnamefont {Jang}}, \bibinfo {author} {\bibfnamefont {H.}~\bibnamefont {Sohn}}, \bibinfo {author} {\bibfnamefont {J.}~\bibnamefont {Yun}}, \bibinfo {author} {\bibfnamefont {Y.}~\bibnamefont {Song}}, \bibinfo {author} {\bibfnamefont {B.}~\bibnamefont {Kang}}, \bibinfo {author} {\bibfnamefont {L.~E.~A.}\ \bibnamefont {Stehouwer}}, \bibinfo {author} {\bibfnamefont {D.~D.}\ \bibnamefont {Esposti}}, \bibinfo {author} {\bibfnamefont {G.}~\bibnamefont {Scappucci}},\ and\ \bibinfo {author} {\bibfnamefont {D.}~\bibnamefont {Kim}},\ }\href {https://doi.org/10.1038/s41467-024-55338-z} {\bibfield  {journal} {\bibinfo  {journal} {Nature Communications}\ }\textbf {\bibinfo {volume} {16}},\ \bibinfo {pages} {78} (\bibinfo {year} {2025})}\BibitemShut {NoStop}%
\bibitem [{\citenamefont {Bartlett}(1948)}]{bartlett_smoothing_1948}%
  \BibitemOpen
  \bibfield  {author} {\bibinfo {author} {\bibfnamefont {M.~S.}\ \bibnamefont {Bartlett}},\ }\href {https://doi.org/10.1038/161686a0} {\bibfield  {journal} {\bibinfo  {journal} {Nature}\ }\textbf {\bibinfo {volume} {161}},\ \bibinfo {pages} {686} (\bibinfo {year} {1948})}\BibitemShut {NoStop}%
\bibitem [{\citenamefont {Percival}\ and\ \citenamefont {Walden}(1993)}]{percival_spectral_1993}%
  \BibitemOpen
  \bibfield  {author} {\bibinfo {author} {\bibfnamefont {D.~B.}\ \bibnamefont {Percival}}\ and\ \bibinfo {author} {\bibfnamefont {A.~T.}\ \bibnamefont {Walden}},\ }\href@noop {} {\emph {\bibinfo {title} {Spectral {Analysis} for {Physical} {Applications}}}}\ (\bibinfo  {publisher} {Cambridge University Press},\ \bibinfo {address} {Cambridge, UK},\ \bibinfo {year} {1993})\BibitemShut {NoStop}%
\bibitem [{\citenamefont {Wang}(2016)}]{wang_4_2016}%
  \BibitemOpen
  \bibfield  {author} {\bibinfo {author} {\bibfnamefont {C.}~\bibnamefont {Wang}},\ }in\ \href@noop {} {\emph {\bibinfo {booktitle} {Cryocoolers 19}}},\ \bibinfo {editor} {edited by\ \bibinfo {editor} {\bibfnamefont {S.~D.}\ \bibnamefont {Miller}}\ and\ \bibinfo {editor} {\bibfnamefont {R.~G.}\ \bibnamefont {Ross}}}\ (\bibinfo  {publisher} {International Cryocooler Conference, Inc.},\ \bibinfo {address} {Boulder, CO},\ \bibinfo {year} {2016})\ p.\ \bibinfo {pages} {299}\BibitemShut {NoStop}%
\bibitem [{\citenamefont {Connors}\ \emph {et~al.}(2022)\citenamefont {Connors}, \citenamefont {Nelson}, \citenamefont {Edge},\ and\ \citenamefont {Nichol}}]{connors_charge-noise_2022}%
  \BibitemOpen
  \bibfield  {author} {\bibinfo {author} {\bibfnamefont {E.~J.}\ \bibnamefont {Connors}}, \bibinfo {author} {\bibfnamefont {J.}~\bibnamefont {Nelson}}, \bibinfo {author} {\bibfnamefont {L.~F.}\ \bibnamefont {Edge}},\ and\ \bibinfo {author} {\bibfnamefont {J.~M.}\ \bibnamefont {Nichol}},\ }\href {https://doi.org/10.1038/s41467-022-28519-x} {\bibfield  {journal} {\bibinfo  {journal} {Nature Communications}\ }\textbf {\bibinfo {volume} {13}},\ \bibinfo {pages} {940} (\bibinfo {year} {2022})}\BibitemShut {NoStop}%
\bibitem [{\citenamefont {Paladino}\ \emph {et~al.}(2014)\citenamefont {Paladino}, \citenamefont {Galperin}, \citenamefont {Falci},\ and\ \citenamefont {Altshuler}}]{paladino_1_2014}%
  \BibitemOpen
  \bibfield  {author} {\bibinfo {author} {\bibfnamefont {E.}~\bibnamefont {Paladino}}, \bibinfo {author} {\bibfnamefont {Y.~M.}\ \bibnamefont {Galperin}}, \bibinfo {author} {\bibfnamefont {G.}~\bibnamefont {Falci}},\ and\ \bibinfo {author} {\bibfnamefont {B.~L.}\ \bibnamefont {Altshuler}},\ }\href {https://doi.org/10.1103/RevModPhys.86.361} {\bibfield  {journal} {\bibinfo  {journal} {Reviews of Modern Physics}\ }\textbf {\bibinfo {volume} {86}},\ \bibinfo {pages} {361} (\bibinfo {year} {2014})}\BibitemShut {NoStop}%
\bibitem [{\citenamefont {Rojas-Arias}\ \emph {et~al.}(2023)\citenamefont {Rojas-Arias}, \citenamefont {Noiri}, \citenamefont {Stano}, \citenamefont {Nakajima}, \citenamefont {Yoneda}, \citenamefont {Takeda}, \citenamefont {Kobayashi}, \citenamefont {Sammak}, \citenamefont {Scappucci}, \citenamefont {Loss},\ and\ \citenamefont {Tarucha}}]{rojas-arias_spatial_2023}%
  \BibitemOpen
  \bibfield  {author} {\bibinfo {author} {\bibfnamefont {J.}~\bibnamefont {Rojas-Arias}}, \bibinfo {author} {\bibfnamefont {A.}~\bibnamefont {Noiri}}, \bibinfo {author} {\bibfnamefont {P.}~\bibnamefont {Stano}}, \bibinfo {author} {\bibfnamefont {T.}~\bibnamefont {Nakajima}}, \bibinfo {author} {\bibfnamefont {J.}~\bibnamefont {Yoneda}}, \bibinfo {author} {\bibfnamefont {K.}~\bibnamefont {Takeda}}, \bibinfo {author} {\bibfnamefont {T.}~\bibnamefont {Kobayashi}}, \bibinfo {author} {\bibfnamefont {A.}~\bibnamefont {Sammak}}, \bibinfo {author} {\bibfnamefont {G.}~\bibnamefont {Scappucci}}, \bibinfo {author} {\bibfnamefont {D.}~\bibnamefont {Loss}},\ and\ \bibinfo {author} {\bibfnamefont {S.}~\bibnamefont {Tarucha}},\ }\href {https://doi.org/10.1103/PhysRevApplied.20.054024} {\bibfield  {journal} {\bibinfo  {journal} {Physical Review Applied}\ }\textbf {\bibinfo {volume} {20}},\ \bibinfo {pages} {054024} (\bibinfo {year} {2023})}\BibitemShut {NoStop}%
\bibitem [{\citenamefont {Connors}\ \emph {et~al.}(2019)\citenamefont {Connors}, \citenamefont {Nelson}, \citenamefont {Qiao}, \citenamefont {Edge},\ and\ \citenamefont {Nichol}}]{connors_low-frequency_2019}%
  \BibitemOpen
  \bibfield  {author} {\bibinfo {author} {\bibfnamefont {E.~J.}\ \bibnamefont {Connors}}, \bibinfo {author} {\bibfnamefont {J.}~\bibnamefont {Nelson}}, \bibinfo {author} {\bibfnamefont {H.}~\bibnamefont {Qiao}}, \bibinfo {author} {\bibfnamefont {L.~F.}\ \bibnamefont {Edge}},\ and\ \bibinfo {author} {\bibfnamefont {J.~M.}\ \bibnamefont {Nichol}},\ }\href {https://doi.org/10.1103/PhysRevB.100.165305} {\bibfield  {journal} {\bibinfo  {journal} {Physical Review B}\ }\textbf {\bibinfo {volume} {100}},\ \bibinfo {pages} {165305} (\bibinfo {year} {2019})}\BibitemShut {NoStop}%
\bibitem [{\citenamefont {Hell}\ \emph {et~al.}(2016)\citenamefont {Hell}, \citenamefont {Wegewijs},\ and\ \citenamefont {DiVincenzo}}]{hell_qubit_2016}%
  \BibitemOpen
  \bibfield  {author} {\bibinfo {author} {\bibfnamefont {M.}~\bibnamefont {Hell}}, \bibinfo {author} {\bibfnamefont {M.~R.}\ \bibnamefont {Wegewijs}},\ and\ \bibinfo {author} {\bibfnamefont {D.~P.}\ \bibnamefont {DiVincenzo}},\ }\href {https://doi.org/10.1103/PhysRevB.93.045418} {\bibfield  {journal} {\bibinfo  {journal} {Physical Review B}\ }\textbf {\bibinfo {volume} {93}},\ \bibinfo {pages} {045418} (\bibinfo {year} {2016})}\BibitemShut {NoStop}%
\end{thebibliography}%


\begin{thebibliography}{53}%
\makeatletter
\providecommand \@ifxundefined [1]{%
 \@ifx{#1\undefined}
}%
\providecommand \@ifnum [1]{%
 \ifnum #1\expandafter \@firstoftwo
 \else \expandafter \@secondoftwo
 \fi
}%
\providecommand \@ifx [1]{%
 \ifx #1\expandafter \@firstoftwo
 \else \expandafter \@secondoftwo
 \fi
}%
\providecommand \natexlab [1]{#1}%
\providecommand \enquote  [1]{``#1''}%
\providecommand \bibnamefont  [1]{#1}%
\providecommand \bibfnamefont [1]{#1}%
\providecommand \citenamefont [1]{#1}%
\providecommand \href@noop [0]{\@secondoftwo}%
\providecommand \href [0]{\begingroup \@sanitize@url \@href}%
\providecommand \@href[1]{\@@startlink{#1}\@@href}%
\providecommand \@@href[1]{\endgroup#1\@@endlink}%
\providecommand \@sanitize@url [0]{\catcode `\\12\catcode `\$12\catcode `\&12\catcode `\#12\catcode `\^12\catcode `\_12\catcode `\%12\relax}%
\providecommand \@@startlink[1]{}%
\providecommand \@@endlink[0]{}%
\providecommand \url  [0]{\begingroup\@sanitize@url \@url }%
\providecommand \@url [1]{\endgroup\@href {#1}{\urlprefix }}%
\providecommand \urlprefix  [0]{URL }%
\providecommand \Eprint [0]{\href }%
\providecommand \doibase [0]{https://doi.org/}%
\providecommand \selectlanguage [0]{\@gobble}%
\providecommand \bibinfo  [0]{\@secondoftwo}%
\providecommand \bibfield  [0]{\@secondoftwo}%
\providecommand \translation [1]{[#1]}%
\providecommand \BibitemOpen [0]{}%
\providecommand \bibitemStop [0]{}%
\providecommand \bibitemNoStop [0]{.\EOS\space}%
\providecommand \EOS [0]{\spacefactor3000\relax}%
\providecommand \BibitemShut  [1]{\csname bibitem#1\endcsname}%
\let\auto@bib@innerbib\@empty
\bibitem [{\citenamefont {Loss}\ and\ \citenamefont {DiVincenzo}(1998)}]{loss_quantum_1998}%
  \BibitemOpen
  \bibfield  {author} {\bibinfo {author} {\bibfnamefont {D.}~\bibnamefont {Loss}}\ and\ \bibinfo {author} {\bibfnamefont {D.~P.}\ \bibnamefont {DiVincenzo}},\ }\bibfield  {title} {\bibinfo {title} {Quantum computation with quantum dots},\ }\href {https://doi.org/10.1103/PhysRevA.57.120} {\bibfield  {journal} {\bibinfo  {journal} {Physical Review A}\ }\textbf {\bibinfo {volume} {57}},\ \bibinfo {pages} {120} (\bibinfo {year} {1998})}\BibitemShut {NoStop}%
\bibitem [{\citenamefont {Vandersypen}\ and\ \citenamefont {Eriksson}(2019)}]{vandersypen_quantum_2019}%
  \BibitemOpen
  \bibfield  {author} {\bibinfo {author} {\bibfnamefont {L.~M.~K.}\ \bibnamefont {Vandersypen}}\ and\ \bibinfo {author} {\bibfnamefont {M.~A.}\ \bibnamefont {Eriksson}},\ }\bibfield  {title} {\bibinfo {title} {Quantum computing with semiconductor spins},\ }\href {https://doi.org/10.1063/PT.3.4270} {\bibfield  {journal} {\bibinfo  {journal} {Physics Today}\ }\textbf {\bibinfo {volume} {72}},\ \bibinfo {pages} {38} (\bibinfo {year} {2019})}\BibitemShut {NoStop}%
\bibitem [{\citenamefont {Maurand}\ \emph {et~al.}(2016)\citenamefont {Maurand}, \citenamefont {Jehl}, \citenamefont {Kotekar-Patil}, \citenamefont {Corna}, \citenamefont {Bohuslavskyi}, \citenamefont {Lavi{\'e}ville}, \citenamefont {Hutin}, \citenamefont {Barraud}, \citenamefont {Vinet}, \citenamefont {Sanquer},\ and\ \citenamefont {De~Franceschi}}]{maurand_cmos_2016}%
  \BibitemOpen
  \bibfield  {author} {\bibinfo {author} {\bibfnamefont {R.}~\bibnamefont {Maurand}}, \bibinfo {author} {\bibfnamefont {X.}~\bibnamefont {Jehl}}, \bibinfo {author} {\bibfnamefont {D.}~\bibnamefont {Kotekar-Patil}}, \bibinfo {author} {\bibfnamefont {A.}~\bibnamefont {Corna}}, \bibinfo {author} {\bibfnamefont {H.}~\bibnamefont {Bohuslavskyi}}, \bibinfo {author} {\bibfnamefont {R.}~\bibnamefont {Lavi{\'e}ville}}, \bibinfo {author} {\bibfnamefont {L.}~\bibnamefont {Hutin}}, \bibinfo {author} {\bibfnamefont {S.}~\bibnamefont {Barraud}}, \bibinfo {author} {\bibfnamefont {M.}~\bibnamefont {Vinet}}, \bibinfo {author} {\bibfnamefont {M.}~\bibnamefont {Sanquer}},\ and\ \bibinfo {author} {\bibfnamefont {S.}~\bibnamefont {De~Franceschi}},\ }\bibfield  {title} {\bibinfo {title} {A {CMOS} silicon spin qubit},\ }\href {https://doi.org/10.1038/ncomms13575} {\bibfield  {journal} {\bibinfo  {journal} {Nature Communications}\ }\textbf {\bibinfo {volume} {7}},\ \bibinfo {pages} {13575} (\bibinfo {year} {2016})}\BibitemShut
  {NoStop}%
\bibitem [{\citenamefont {Steinacker}\ \emph {et~al.}(2025)\citenamefont {Steinacker}, \citenamefont {Dumoulin~Stuyck}, \citenamefont {Lim}, \citenamefont {Tanttu}, \citenamefont {Feng}, \citenamefont {Serrano}, \citenamefont {Nickl}, \citenamefont {Candido}, \citenamefont {Cifuentes}, \citenamefont {Vahapoglu}, \citenamefont {Bartee}, \citenamefont {Hudson}, \citenamefont {Chan}, \citenamefont {Kubicek}, \citenamefont {Jussot}, \citenamefont {Canvel}, \citenamefont {Beyne}, \citenamefont {Shimura}, \citenamefont {Loo}, \citenamefont {Godfrin}, \citenamefont {Raes}, \citenamefont {Baudot}, \citenamefont {Wan}, \citenamefont {Laucht}, \citenamefont {Yang}, \citenamefont {Saraiva}, \citenamefont {Escott}, \citenamefont {De~Greve},\ and\ \citenamefont {Dzurak}}]{steinacker_industry-compatible_2025}%
  \BibitemOpen
  \bibfield  {author} {\bibinfo {author} {\bibfnamefont {P.}~\bibnamefont {Steinacker}}, \bibinfo {author} {\bibfnamefont {N.}~\bibnamefont {Dumoulin~Stuyck}}, \bibinfo {author} {\bibfnamefont {W.~H.}\ \bibnamefont {Lim}}, \bibinfo {author} {\bibfnamefont {T.}~\bibnamefont {Tanttu}}, \bibinfo {author} {\bibfnamefont {M.}~\bibnamefont {Feng}}, \bibinfo {author} {\bibfnamefont {S.}~\bibnamefont {Serrano}}, \bibinfo {author} {\bibfnamefont {A.}~\bibnamefont {Nickl}}, \bibinfo {author} {\bibfnamefont {M.}~\bibnamefont {Candido}}, \bibinfo {author} {\bibfnamefont {J.~D.}\ \bibnamefont {Cifuentes}}, \bibinfo {author} {\bibfnamefont {E.}~\bibnamefont {Vahapoglu}}, \bibinfo {author} {\bibfnamefont {S.~K.}\ \bibnamefont {Bartee}}, \bibinfo {author} {\bibfnamefont {F.~E.}\ \bibnamefont {Hudson}}, \bibinfo {author} {\bibfnamefont {K.~W.}\ \bibnamefont {Chan}}, \bibinfo {author} {\bibfnamefont {S.}~\bibnamefont {Kubicek}}, \bibinfo {author} {\bibfnamefont {J.}~\bibnamefont {Jussot}}, \bibinfo {author} {\bibfnamefont
  {Y.}~\bibnamefont {Canvel}}, \bibinfo {author} {\bibfnamefont {S.}~\bibnamefont {Beyne}}, \bibinfo {author} {\bibfnamefont {Y.}~\bibnamefont {Shimura}}, \bibinfo {author} {\bibfnamefont {R.}~\bibnamefont {Loo}}, \bibinfo {author} {\bibfnamefont {C.}~\bibnamefont {Godfrin}}, \bibinfo {author} {\bibfnamefont {B.}~\bibnamefont {Raes}}, \bibinfo {author} {\bibfnamefont {S.}~\bibnamefont {Baudot}}, \bibinfo {author} {\bibfnamefont {D.}~\bibnamefont {Wan}}, \bibinfo {author} {\bibfnamefont {A.}~\bibnamefont {Laucht}}, \bibinfo {author} {\bibfnamefont {C.~H.}\ \bibnamefont {Yang}}, \bibinfo {author} {\bibfnamefont {A.}~\bibnamefont {Saraiva}}, \bibinfo {author} {\bibfnamefont {C.~C.}\ \bibnamefont {Escott}}, \bibinfo {author} {\bibfnamefont {K.}~\bibnamefont {De~Greve}},\ and\ \bibinfo {author} {\bibfnamefont {A.~S.}\ \bibnamefont {Dzurak}},\ }\bibfield  {title} {\bibinfo {title} {Industry-compatible silicon spin-qubit unit cells exceeding 99\% fidelity},\ }\href {https://doi.org/10.1038/s41586-025-09531-9}
  {\bibfield  {journal} {\bibinfo  {journal} {Nature}\ }\textbf {\bibinfo {volume} {646}},\ \bibinfo {pages} {81} (\bibinfo {year} {2025})}\BibitemShut {NoStop}%
\bibitem [{\citenamefont {George}\ \emph {et~al.}(2025)\citenamefont {George}, \citenamefont {M{\k a}dzik}, \citenamefont {Henry}, \citenamefont {Wagner}, \citenamefont {Islam}, \citenamefont {Borjans}, \citenamefont {Connors}, \citenamefont {Corrigan}, \citenamefont {Curry}, \citenamefont {Harper}, \citenamefont {Keith}, \citenamefont {Lampert}, \citenamefont {Luthi}, \citenamefont {Mohiyaddin}, \citenamefont {Murcia}, \citenamefont {Nair}, \citenamefont {Nahm}, \citenamefont {Nethwewala}, \citenamefont {Neyens}, \citenamefont {Patra}, \citenamefont {Raharjo}, \citenamefont {Rogan}, \citenamefont {Savytskyy}, \citenamefont {Watson}, \citenamefont {Ziegler}, \citenamefont {Zietz}, \citenamefont {Pellerano}, \citenamefont {Pillarisetty}, \citenamefont {Bishop}, \citenamefont {Bojarski}, \citenamefont {Roberts},\ and\ \citenamefont {Clarke}}]{george_12-spin-qubit_2025}%
  \BibitemOpen
  \bibfield  {author} {\bibinfo {author} {\bibfnamefont {H.~C.}\ \bibnamefont {George}}, \bibinfo {author} {\bibfnamefont {M.~T.}\ \bibnamefont {M{\k a}dzik}}, \bibinfo {author} {\bibfnamefont {E.~M.}\ \bibnamefont {Henry}}, \bibinfo {author} {\bibfnamefont {A.~J.}\ \bibnamefont {Wagner}}, \bibinfo {author} {\bibfnamefont {M.~M.}\ \bibnamefont {Islam}}, \bibinfo {author} {\bibfnamefont {F.}~\bibnamefont {Borjans}}, \bibinfo {author} {\bibfnamefont {E.~J.}\ \bibnamefont {Connors}}, \bibinfo {author} {\bibfnamefont {J.}~\bibnamefont {Corrigan}}, \bibinfo {author} {\bibfnamefont {M.}~\bibnamefont {Curry}}, \bibinfo {author} {\bibfnamefont {M.~K.}\ \bibnamefont {Harper}}, \bibinfo {author} {\bibfnamefont {D.}~\bibnamefont {Keith}}, \bibinfo {author} {\bibfnamefont {L.}~\bibnamefont {Lampert}}, \bibinfo {author} {\bibfnamefont {F.}~\bibnamefont {Luthi}}, \bibinfo {author} {\bibfnamefont {F.~A.}\ \bibnamefont {Mohiyaddin}}, \bibinfo {author} {\bibfnamefont {S.}~\bibnamefont {Murcia}}, \bibinfo {author}
  {\bibfnamefont {R.}~\bibnamefont {Nair}}, \bibinfo {author} {\bibfnamefont {R.}~\bibnamefont {Nahm}}, \bibinfo {author} {\bibfnamefont {A.}~\bibnamefont {Nethwewala}}, \bibinfo {author} {\bibfnamefont {S.}~\bibnamefont {Neyens}}, \bibinfo {author} {\bibfnamefont {B.}~\bibnamefont {Patra}}, \bibinfo {author} {\bibfnamefont {R.~D.}\ \bibnamefont {Raharjo}}, \bibinfo {author} {\bibfnamefont {C.}~\bibnamefont {Rogan}}, \bibinfo {author} {\bibfnamefont {R.}~\bibnamefont {Savytskyy}}, \bibinfo {author} {\bibfnamefont {T.~F.}\ \bibnamefont {Watson}}, \bibinfo {author} {\bibfnamefont {J.}~\bibnamefont {Ziegler}}, \bibinfo {author} {\bibfnamefont {O.~K.}\ \bibnamefont {Zietz}}, \bibinfo {author} {\bibfnamefont {S.}~\bibnamefont {Pellerano}}, \bibinfo {author} {\bibfnamefont {R.}~\bibnamefont {Pillarisetty}}, \bibinfo {author} {\bibfnamefont {N.~C.}\ \bibnamefont {Bishop}}, \bibinfo {author} {\bibfnamefont {S.~A.}\ \bibnamefont {Bojarski}}, \bibinfo {author} {\bibfnamefont {J.}~\bibnamefont {Roberts}},\ and\ \bibinfo
  {author} {\bibfnamefont {J.~S.}\ \bibnamefont {Clarke}},\ }\bibfield  {title} {\bibinfo {title} {12-{Spin}-{Qubit} {Arrays} {Fabricated} on a 300 mm {Semiconductor} {Manufacturing} {Line}},\ }\href {https://doi.org/10.1021/acs.nanolett.4c05205} {\bibfield  {journal} {\bibinfo  {journal} {Nano Letters}\ }\textbf {\bibinfo {volume} {25}},\ \bibinfo {pages} {793} (\bibinfo {year} {2025})}\BibitemShut {NoStop}%
\bibitem [{\citenamefont {Zwanenburg}\ \emph {et~al.}(2013)\citenamefont {Zwanenburg}, \citenamefont {Dzurak}, \citenamefont {Morello}, \citenamefont {Simmons}, \citenamefont {Hollenberg}, \citenamefont {Klimeck}, \citenamefont {Rogge}, \citenamefont {Coppersmith},\ and\ \citenamefont {Eriksson}}]{zwanenburg_silicon_2013}%
  \BibitemOpen
  \bibfield  {author} {\bibinfo {author} {\bibfnamefont {F.~A.}\ \bibnamefont {Zwanenburg}}, \bibinfo {author} {\bibfnamefont {A.~S.}\ \bibnamefont {Dzurak}}, \bibinfo {author} {\bibfnamefont {A.}~\bibnamefont {Morello}}, \bibinfo {author} {\bibfnamefont {M.~Y.}\ \bibnamefont {Simmons}}, \bibinfo {author} {\bibfnamefont {L.~C.~L.}\ \bibnamefont {Hollenberg}}, \bibinfo {author} {\bibfnamefont {G.}~\bibnamefont {Klimeck}}, \bibinfo {author} {\bibfnamefont {S.}~\bibnamefont {Rogge}}, \bibinfo {author} {\bibfnamefont {S.~N.}\ \bibnamefont {Coppersmith}},\ and\ \bibinfo {author} {\bibfnamefont {M.~A.}\ \bibnamefont {Eriksson}},\ }\bibfield  {title} {\bibinfo {title} {Silicon quantum electronics},\ }\href {https://doi.org/10.1103/RevModPhys.85.961} {\bibfield  {journal} {\bibinfo  {journal} {Reviews of Modern Physics}\ }\textbf {\bibinfo {volume} {85}},\ \bibinfo {pages} {961} (\bibinfo {year} {2013})}\BibitemShut {NoStop}%
\bibitem [{\citenamefont {Veldhorst}\ \emph {et~al.}(2014)\citenamefont {Veldhorst}, \citenamefont {Hwang}, \citenamefont {Yang}, \citenamefont {Leenstra}, \citenamefont {Ronde}, \citenamefont {Dehollain}, \citenamefont {Muhonen}, \citenamefont {Hudson}, \citenamefont {Itoh}, \citenamefont {Morello},\ and\ \citenamefont {Dzurak}}]{veldhorst_addressable_2014}%
  \BibitemOpen
  \bibfield  {author} {\bibinfo {author} {\bibfnamefont {M.}~\bibnamefont {Veldhorst}}, \bibinfo {author} {\bibfnamefont {J.~C.~C.}\ \bibnamefont {Hwang}}, \bibinfo {author} {\bibfnamefont {C.~H.}\ \bibnamefont {Yang}}, \bibinfo {author} {\bibfnamefont {A.~W.}\ \bibnamefont {Leenstra}}, \bibinfo {author} {\bibfnamefont {B.}~\bibnamefont {Ronde}}, \bibinfo {author} {\bibfnamefont {J.~P.}\ \bibnamefont {Dehollain}}, \bibinfo {author} {\bibfnamefont {J.~T.}\ \bibnamefont {Muhonen}}, \bibinfo {author} {\bibfnamefont {F.~E.}\ \bibnamefont {Hudson}}, \bibinfo {author} {\bibfnamefont {K.~M.}\ \bibnamefont {Itoh}}, \bibinfo {author} {\bibfnamefont {A.}~\bibnamefont {Morello}},\ and\ \bibinfo {author} {\bibfnamefont {A.~S.}\ \bibnamefont {Dzurak}},\ }\bibfield  {title} {\bibinfo {title} {An addressable quantum dot qubit with fault-tolerant control-fidelity},\ }\href {https://doi.org/10.1038/nnano.2014.216} {\bibfield  {journal} {\bibinfo  {journal} {Nature Nanotechnology}\ }\textbf {\bibinfo {volume} {9}},\ \bibinfo
  {pages} {981} (\bibinfo {year} {2014})}\BibitemShut {NoStop}%
\bibitem [{\citenamefont {Obata}\ \emph {et~al.}(2010)\citenamefont {Obata}, \citenamefont {Pioro-Ladri{\`e}re}, \citenamefont {Tokura}, \citenamefont {Shin}, \citenamefont {Kubo}, \citenamefont {Yoshida}, \citenamefont {Taniyama},\ and\ \citenamefont {Tarucha}}]{obata_coherent_2010}%
  \BibitemOpen
  \bibfield  {author} {\bibinfo {author} {\bibfnamefont {T.}~\bibnamefont {Obata}}, \bibinfo {author} {\bibfnamefont {M.}~\bibnamefont {Pioro-Ladri{\`e}re}}, \bibinfo {author} {\bibfnamefont {Y.}~\bibnamefont {Tokura}}, \bibinfo {author} {\bibfnamefont {Y.-S.}\ \bibnamefont {Shin}}, \bibinfo {author} {\bibfnamefont {T.}~\bibnamefont {Kubo}}, \bibinfo {author} {\bibfnamefont {K.}~\bibnamefont {Yoshida}}, \bibinfo {author} {\bibfnamefont {T.}~\bibnamefont {Taniyama}},\ and\ \bibinfo {author} {\bibfnamefont {S.}~\bibnamefont {Tarucha}},\ }\bibfield  {title} {\bibinfo {title} {Coherent manipulation of individual electron spin in a double quantum dot integrated with a micromagnet},\ }\href {https://doi.org/10.1103/PhysRevB.81.085317} {\bibfield  {journal} {\bibinfo  {journal} {Physical Review B}\ }\textbf {\bibinfo {volume} {81}},\ \bibinfo {pages} {085317} (\bibinfo {year} {2010})}\BibitemShut {NoStop}%
\bibitem [{\citenamefont {Yoneda}\ \emph {et~al.}(2015)\citenamefont {Yoneda}, \citenamefont {Otsuka}, \citenamefont {Takakura}, \citenamefont {Pioro-Ladri{\`e}re}, \citenamefont {Brunner}, \citenamefont {Lu}, \citenamefont {Nakajima}, \citenamefont {Obata}, \citenamefont {Noiri}, \citenamefont {Palmstr{\o}m}, \citenamefont {Gossard},\ and\ \citenamefont {Tarucha}}]{yoneda_robust_2015}%
  \BibitemOpen
  \bibfield  {author} {\bibinfo {author} {\bibfnamefont {J.}~\bibnamefont {Yoneda}}, \bibinfo {author} {\bibfnamefont {T.}~\bibnamefont {Otsuka}}, \bibinfo {author} {\bibfnamefont {T.}~\bibnamefont {Takakura}}, \bibinfo {author} {\bibfnamefont {M.}~\bibnamefont {Pioro-Ladri{\`e}re}}, \bibinfo {author} {\bibfnamefont {R.}~\bibnamefont {Brunner}}, \bibinfo {author} {\bibfnamefont {H.}~\bibnamefont {Lu}}, \bibinfo {author} {\bibfnamefont {T.}~\bibnamefont {Nakajima}}, \bibinfo {author} {\bibfnamefont {T.}~\bibnamefont {Obata}}, \bibinfo {author} {\bibfnamefont {A.}~\bibnamefont {Noiri}}, \bibinfo {author} {\bibfnamefont {C.~J.}\ \bibnamefont {Palmstr{\o}m}}, \bibinfo {author} {\bibfnamefont {A.~C.}\ \bibnamefont {Gossard}},\ and\ \bibinfo {author} {\bibfnamefont {S.}~\bibnamefont {Tarucha}},\ }\bibfield  {title} {\bibinfo {title} {Robust micromagnet design for fast electrical manipulations of single spins in quantum dots},\ }\href {https://doi.org/10.7567/APEX.8.084401} {\bibfield  {journal} {\bibinfo  {journal}
  {Applied Physics Express}\ }\textbf {\bibinfo {volume} {8}},\ \bibinfo {pages} {084401} (\bibinfo {year} {2015})}\BibitemShut {NoStop}%
\bibitem [{\citenamefont {Dumoulin~Stuyck}\ \emph {et~al.}(2021)\citenamefont {Dumoulin~Stuyck}, \citenamefont {Mohiyaddin}, \citenamefont {Li}, \citenamefont {Heyns}, \citenamefont {Govoreanu},\ and\ \citenamefont {Radu}}]{dumoulin_stuyck_low_2021}%
  \BibitemOpen
  \bibfield  {author} {\bibinfo {author} {\bibfnamefont {N.~I.}\ \bibnamefont {Dumoulin~Stuyck}}, \bibinfo {author} {\bibfnamefont {F.~A.}\ \bibnamefont {Mohiyaddin}}, \bibinfo {author} {\bibfnamefont {R.}~\bibnamefont {Li}}, \bibinfo {author} {\bibfnamefont {M.}~\bibnamefont {Heyns}}, \bibinfo {author} {\bibfnamefont {B.}~\bibnamefont {Govoreanu}},\ and\ \bibinfo {author} {\bibfnamefont {I.~P.}\ \bibnamefont {Radu}},\ }\bibfield  {title} {\bibinfo {title} {Low dephasing and robust micromagnet designs for silicon spin qubits},\ }\href {https://doi.org/10.1063/5.0059939} {\bibfield  {journal} {\bibinfo  {journal} {Applied Physics Letters}\ }\textbf {\bibinfo {volume} {119}},\ \bibinfo {pages} {094001} (\bibinfo {year} {2021})}\BibitemShut {NoStop}%
\bibitem [{\citenamefont {Philips}\ \emph {et~al.}(2022)\citenamefont {Philips}, \citenamefont {M{\k a}dzik}, \citenamefont {Amitonov}, \citenamefont {Snoo}, \citenamefont {Russ}, \citenamefont {Kalhor}, \citenamefont {Volk}, \citenamefont {Lawrie}, \citenamefont {Brousse}, \citenamefont {Tryputen}, \citenamefont {Wuetz}, \citenamefont {Sammak}, \citenamefont {Veldhorst}, \citenamefont {Scappucci},\ and\ \citenamefont {Vandersypen}}]{philips_universal_2022}%
  \BibitemOpen
  \bibfield  {author} {\bibinfo {author} {\bibfnamefont {S.~G.~J.}\ \bibnamefont {Philips}}, \bibinfo {author} {\bibfnamefont {M.~T.}\ \bibnamefont {M{\k a}dzik}}, \bibinfo {author} {\bibfnamefont {S.~V.}\ \bibnamefont {Amitonov}}, \bibinfo {author} {\bibfnamefont {S.~L.}\ \bibnamefont {Snoo}}, \bibinfo {author} {\bibfnamefont {M.}~\bibnamefont {Russ}}, \bibinfo {author} {\bibfnamefont {N.}~\bibnamefont {Kalhor}}, \bibinfo {author} {\bibfnamefont {C.}~\bibnamefont {Volk}}, \bibinfo {author} {\bibfnamefont {W.~I.~L.}\ \bibnamefont {Lawrie}}, \bibinfo {author} {\bibfnamefont {D.}~\bibnamefont {Brousse}}, \bibinfo {author} {\bibfnamefont {L.}~\bibnamefont {Tryputen}}, \bibinfo {author} {\bibfnamefont {B.~P.}\ \bibnamefont {Wuetz}}, \bibinfo {author} {\bibfnamefont {A.}~\bibnamefont {Sammak}}, \bibinfo {author} {\bibfnamefont {M.}~\bibnamefont {Veldhorst}}, \bibinfo {author} {\bibfnamefont {G.}~\bibnamefont {Scappucci}},\ and\ \bibinfo {author} {\bibfnamefont {L.~M.~K.}\ \bibnamefont {Vandersypen}},\ }\bibfield
  {title} {\bibinfo {title} {Universal control of a six-qubit quantum processor in silicon},\ }\href {https://doi.org/10.1038/s41586-022-05117-x} {\bibfield  {journal} {\bibinfo  {journal} {Nature}\ }\textbf {\bibinfo {volume} {609}},\ \bibinfo {pages} {919} (\bibinfo {year} {2022})}\BibitemShut {NoStop}%
\bibitem [{\citenamefont {Pioro-Ladri{\`e}re}\ \emph {et~al.}(2008)\citenamefont {Pioro-Ladri{\`e}re}, \citenamefont {Obata}, \citenamefont {Tokura}, \citenamefont {Shin}, \citenamefont {Kubo}, \citenamefont {Yoshida}, \citenamefont {Taniyama},\ and\ \citenamefont {Tarucha}}]{pioro-ladriere_electrically_2008}%
  \BibitemOpen
  \bibfield  {author} {\bibinfo {author} {\bibfnamefont {M.}~\bibnamefont {Pioro-Ladri{\`e}re}}, \bibinfo {author} {\bibfnamefont {T.}~\bibnamefont {Obata}}, \bibinfo {author} {\bibfnamefont {Y.}~\bibnamefont {Tokura}}, \bibinfo {author} {\bibfnamefont {Y.-S.}\ \bibnamefont {Shin}}, \bibinfo {author} {\bibfnamefont {T.}~\bibnamefont {Kubo}}, \bibinfo {author} {\bibfnamefont {K.}~\bibnamefont {Yoshida}}, \bibinfo {author} {\bibfnamefont {T.}~\bibnamefont {Taniyama}},\ and\ \bibinfo {author} {\bibfnamefont {S.}~\bibnamefont {Tarucha}},\ }\bibfield  {title} {\bibinfo {title} {Electrically driven single-electron spin resonance in a slanting {Zeeman} field},\ }\href {https://doi.org/10.1038/nphys1053} {\bibfield  {journal} {\bibinfo  {journal} {Nature Physics}\ }\textbf {\bibinfo {volume} {4}},\ \bibinfo {pages} {776} (\bibinfo {year} {2008})}\BibitemShut {NoStop}%
\bibitem [{\citenamefont {Yoneda}\ \emph {et~al.}(2018)\citenamefont {Yoneda}, \citenamefont {Takeda}, \citenamefont {Otsuka}, \citenamefont {Nakajima}, \citenamefont {Delbecq}, \citenamefont {Allison}, \citenamefont {Honda}, \citenamefont {Kodera}, \citenamefont {Oda}, \citenamefont {Hoshi}, \citenamefont {Usami}, \citenamefont {Itoh},\ and\ \citenamefont {Tarucha}}]{yoneda_quantum-dot_2018}%
  \BibitemOpen
  \bibfield  {author} {\bibinfo {author} {\bibfnamefont {J.}~\bibnamefont {Yoneda}}, \bibinfo {author} {\bibfnamefont {K.}~\bibnamefont {Takeda}}, \bibinfo {author} {\bibfnamefont {T.}~\bibnamefont {Otsuka}}, \bibinfo {author} {\bibfnamefont {T.}~\bibnamefont {Nakajima}}, \bibinfo {author} {\bibfnamefont {M.~R.}\ \bibnamefont {Delbecq}}, \bibinfo {author} {\bibfnamefont {G.}~\bibnamefont {Allison}}, \bibinfo {author} {\bibfnamefont {T.}~\bibnamefont {Honda}}, \bibinfo {author} {\bibfnamefont {T.}~\bibnamefont {Kodera}}, \bibinfo {author} {\bibfnamefont {S.}~\bibnamefont {Oda}}, \bibinfo {author} {\bibfnamefont {Y.}~\bibnamefont {Hoshi}}, \bibinfo {author} {\bibfnamefont {N.}~\bibnamefont {Usami}}, \bibinfo {author} {\bibfnamefont {K.~M.}\ \bibnamefont {Itoh}},\ and\ \bibinfo {author} {\bibfnamefont {S.}~\bibnamefont {Tarucha}},\ }\bibfield  {title} {\bibinfo {title} {A quantum-dot spin qubit with coherence limited by charge noise and fidelity higher than 99.9\%},\ }\href
  {https://doi.org/10.1038/s41565-017-0014-x} {\bibfield  {journal} {\bibinfo  {journal} {Nature Nanotechnology}\ }\textbf {\bibinfo {volume} {13}},\ \bibinfo {pages} {102} (\bibinfo {year} {2018})}\BibitemShut {NoStop}%
\bibitem [{\citenamefont {Kha}\ \emph {et~al.}(2015)\citenamefont {Kha}, \citenamefont {Joynt},\ and\ \citenamefont {Culcer}}]{kha_micromagnets_2015}%
  \BibitemOpen
  \bibfield  {author} {\bibinfo {author} {\bibfnamefont {A.}~\bibnamefont {Kha}}, \bibinfo {author} {\bibfnamefont {R.}~\bibnamefont {Joynt}},\ and\ \bibinfo {author} {\bibfnamefont {D.}~\bibnamefont {Culcer}},\ }\bibfield  {title} {\bibinfo {title} {Do micromagnets expose spin qubits to charge and {Johnson} noise?},\ }\href {https://doi.org/10.1063/1.4934693} {\bibfield  {journal} {\bibinfo  {journal} {Applied Physics Letters}\ }\textbf {\bibinfo {volume} {107}},\ \bibinfo {pages} {172101} (\bibinfo {year} {2015})}\BibitemShut {NoStop}%
\bibitem [{\citenamefont {Struck}\ \emph {et~al.}(2020)\citenamefont {Struck}, \citenamefont {Hollmann}, \citenamefont {Schauer}, \citenamefont {Fedorets}, \citenamefont {Schmidbauer}, \citenamefont {Sawano}, \citenamefont {Riemann}, \citenamefont {Abrosimov}, \citenamefont {Cywi{\'n}ski}, \citenamefont {Bougeard},\ and\ \citenamefont {Schreiber}}]{struck_low-frequency_2020}%
  \BibitemOpen
  \bibfield  {author} {\bibinfo {author} {\bibfnamefont {T.}~\bibnamefont {Struck}}, \bibinfo {author} {\bibfnamefont {A.}~\bibnamefont {Hollmann}}, \bibinfo {author} {\bibfnamefont {F.}~\bibnamefont {Schauer}}, \bibinfo {author} {\bibfnamefont {O.}~\bibnamefont {Fedorets}}, \bibinfo {author} {\bibfnamefont {A.}~\bibnamefont {Schmidbauer}}, \bibinfo {author} {\bibfnamefont {K.}~\bibnamefont {Sawano}}, \bibinfo {author} {\bibfnamefont {H.}~\bibnamefont {Riemann}}, \bibinfo {author} {\bibfnamefont {N.~V.}\ \bibnamefont {Abrosimov}}, \bibinfo {author} {\bibfnamefont {{\L}.}~\bibnamefont {Cywi{\'n}ski}}, \bibinfo {author} {\bibfnamefont {D.}~\bibnamefont {Bougeard}},\ and\ \bibinfo {author} {\bibfnamefont {L.~R.}\ \bibnamefont {Schreiber}},\ }\bibfield  {title} {\bibinfo {title} {Low-frequency spin qubit energy splitting noise in highly purified {28Si}/{SiGe}},\ }\href {https://doi.org/10.1038/s41534-020-0276-2} {\bibfield  {journal} {\bibinfo  {journal} {npj Quantum Information}\ }\textbf {\bibinfo {volume}
  {6}},\ \bibinfo {pages} {40} (\bibinfo {year} {2020})}\BibitemShut {NoStop}%
\bibitem [{\citenamefont {Paquelet~Wuetz}\ \emph {et~al.}(2023)\citenamefont {Paquelet~Wuetz}, \citenamefont {Degli~Esposti}, \citenamefont {Zwerver}, \citenamefont {Amitonov}, \citenamefont {Botifoll}, \citenamefont {Arbiol}, \citenamefont {Sammak}, \citenamefont {Vandersypen}, \citenamefont {Russ},\ and\ \citenamefont {Scappucci}}]{paquelet_wuetz_reducing_2023}%
  \BibitemOpen
  \bibfield  {author} {\bibinfo {author} {\bibfnamefont {B.}~\bibnamefont {Paquelet~Wuetz}}, \bibinfo {author} {\bibfnamefont {D.}~\bibnamefont {Degli~Esposti}}, \bibinfo {author} {\bibfnamefont {A.-M.~J.}\ \bibnamefont {Zwerver}}, \bibinfo {author} {\bibfnamefont {S.~V.}\ \bibnamefont {Amitonov}}, \bibinfo {author} {\bibfnamefont {M.}~\bibnamefont {Botifoll}}, \bibinfo {author} {\bibfnamefont {J.}~\bibnamefont {Arbiol}}, \bibinfo {author} {\bibfnamefont {A.}~\bibnamefont {Sammak}}, \bibinfo {author} {\bibfnamefont {L.~M.~K.}\ \bibnamefont {Vandersypen}}, \bibinfo {author} {\bibfnamefont {M.}~\bibnamefont {Russ}},\ and\ \bibinfo {author} {\bibfnamefont {G.}~\bibnamefont {Scappucci}},\ }\bibfield  {title} {\bibinfo {title} {Reducing charge noise in quantum dots by using thin silicon quantum wells},\ }\href {https://doi.org/10.1038/s41467-023-36951-w} {\bibfield  {journal} {\bibinfo  {journal} {Nature Communications}\ }\textbf {\bibinfo {volume} {14}},\ \bibinfo {pages} {1385} (\bibinfo {year}
  {2023})}\BibitemShut {NoStop}%
\bibitem [{\citenamefont {Hanson}\ \emph {et~al.}(2007)\citenamefont {Hanson}, \citenamefont {Kouwenhoven}, \citenamefont {Petta}, \citenamefont {Tarucha},\ and\ \citenamefont {Vandersypen}}]{hanson_spins_2007}%
  \BibitemOpen
  \bibfield  {author} {\bibinfo {author} {\bibfnamefont {R.}~\bibnamefont {Hanson}}, \bibinfo {author} {\bibfnamefont {L.~P.}\ \bibnamefont {Kouwenhoven}}, \bibinfo {author} {\bibfnamefont {J.~R.}\ \bibnamefont {Petta}}, \bibinfo {author} {\bibfnamefont {S.}~\bibnamefont {Tarucha}},\ and\ \bibinfo {author} {\bibfnamefont {L.~M.~K.}\ \bibnamefont {Vandersypen}},\ }\bibfield  {title} {\bibinfo {title} {Spins in few-electron quantum dots},\ }\href {https://doi.org/10.1103/RevModPhys.79.1217} {\bibfield  {journal} {\bibinfo  {journal} {Reviews of Modern Physics}\ }\textbf {\bibinfo {volume} {79}},\ \bibinfo {pages} {1217} (\bibinfo {year} {2007})}\BibitemShut {NoStop}%
\bibitem [{\citenamefont {Cao}\ \emph {et~al.}(2016)\citenamefont {Cao}, \citenamefont {Li}, \citenamefont {Yu}, \citenamefont {Wang}, \citenamefont {Chen}, \citenamefont {Song}, \citenamefont {Xiao}, \citenamefont {Guo}, \citenamefont {Jiang}, \citenamefont {Hu},\ and\ \citenamefont {Guo}}]{cao_tunable_2016}%
  \BibitemOpen
  \bibfield  {author} {\bibinfo {author} {\bibfnamefont {G.}~\bibnamefont {Cao}}, \bibinfo {author} {\bibfnamefont {H.-O.}\ \bibnamefont {Li}}, \bibinfo {author} {\bibfnamefont {G.-D.}\ \bibnamefont {Yu}}, \bibinfo {author} {\bibfnamefont {B.-C.}\ \bibnamefont {Wang}}, \bibinfo {author} {\bibfnamefont {B.-B.}\ \bibnamefont {Chen}}, \bibinfo {author} {\bibfnamefont {X.-X.}\ \bibnamefont {Song}}, \bibinfo {author} {\bibfnamefont {M.}~\bibnamefont {Xiao}}, \bibinfo {author} {\bibfnamefont {G.-C.}\ \bibnamefont {Guo}}, \bibinfo {author} {\bibfnamefont {H.-W.}\ \bibnamefont {Jiang}}, \bibinfo {author} {\bibfnamefont {X.}~\bibnamefont {Hu}},\ and\ \bibinfo {author} {\bibfnamefont {G.-P.}\ \bibnamefont {Guo}},\ }\bibfield  {title} {\bibinfo {title} {Tunable {Hybrid} {Qubit} in a {GaAs} {Double} {Quantum} {Dot}},\ }\href {https://doi.org/10.1103/PhysRevLett.116.086801} {\bibfield  {journal} {\bibinfo  {journal} {Physical Review Letters}\ }\textbf {\bibinfo {volume} {116}},\ \bibinfo {pages} {086801} (\bibinfo {year}
  {2016})}\BibitemShut {NoStop}%
\bibitem [{\citenamefont {Kawakami}\ \emph {et~al.}(2014)\citenamefont {Kawakami}, \citenamefont {Scarlino}, \citenamefont {Ward}, \citenamefont {Braakman}, \citenamefont {Savage}, \citenamefont {Lagally}, \citenamefont {Friesen}, \citenamefont {Coppersmith}, \citenamefont {Eriksson},\ and\ \citenamefont {Vandersypen}}]{kawakami_electrical_2014}%
  \BibitemOpen
  \bibfield  {author} {\bibinfo {author} {\bibfnamefont {E.}~\bibnamefont {Kawakami}}, \bibinfo {author} {\bibfnamefont {P.}~\bibnamefont {Scarlino}}, \bibinfo {author} {\bibfnamefont {D.~R.}\ \bibnamefont {Ward}}, \bibinfo {author} {\bibfnamefont {F.~R.}\ \bibnamefont {Braakman}}, \bibinfo {author} {\bibfnamefont {D.~E.}\ \bibnamefont {Savage}}, \bibinfo {author} {\bibfnamefont {M.~G.}\ \bibnamefont {Lagally}}, \bibinfo {author} {\bibfnamefont {M.}~\bibnamefont {Friesen}}, \bibinfo {author} {\bibfnamefont {S.~N.}\ \bibnamefont {Coppersmith}}, \bibinfo {author} {\bibfnamefont {M.~A.}\ \bibnamefont {Eriksson}},\ and\ \bibinfo {author} {\bibfnamefont {L.~M.~K.}\ \bibnamefont {Vandersypen}},\ }\bibfield  {title} {\bibinfo {title} {Electrical control of a long-lived spin qubit in a {Si}/{SiGe} quantum dot},\ }\href {https://doi.org/10.1038/nnano.2014.153} {\bibfield  {journal} {\bibinfo  {journal} {Nature Nanotechnology}\ }\textbf {\bibinfo {volume} {9}},\ \bibinfo {pages} {666} (\bibinfo {year}
  {2014})}\BibitemShut {NoStop}%
\bibitem [{\citenamefont {Witzel}\ \emph {et~al.}(2012)\citenamefont {Witzel}, \citenamefont {Rahman},\ and\ \citenamefont {Carroll}}]{witzel_nuclear_2012}%
  \BibitemOpen
  \bibfield  {author} {\bibinfo {author} {\bibfnamefont {W.~M.}\ \bibnamefont {Witzel}}, \bibinfo {author} {\bibfnamefont {R.}~\bibnamefont {Rahman}},\ and\ \bibinfo {author} {\bibfnamefont {M.~S.}\ \bibnamefont {Carroll}},\ }\bibfield  {title} {\bibinfo {title} {Nuclear spin induced decoherence of a quantum dot in {Si} confined at a {SiGe} interface: {Decoherence} dependence on {\ensuremath{{}^{73}}}{Ge}},\ }\href {https://doi.org/10.1103/PhysRevB.85.205312} {\bibfield  {journal} {\bibinfo  {journal} {Physical Review B}\ }\textbf {\bibinfo {volume} {85}},\ \bibinfo {pages} {205312} (\bibinfo {year} {2012})}\BibitemShut {NoStop}%
\bibitem [{\citenamefont {Kerckhoff}\ \emph {et~al.}(2021)\citenamefont {Kerckhoff}, \citenamefont {Sun}, \citenamefont {Fong}, \citenamefont {Jones}, \citenamefont {Kiselev}, \citenamefont {Barnes}, \citenamefont {Noah}, \citenamefont {Acuna}, \citenamefont {Akmal}, \citenamefont {Ha}, \citenamefont {Wright}, \citenamefont {Thomas}, \citenamefont {Jackson}, \citenamefont {Edge}, \citenamefont {Eng}, \citenamefont {Ross},\ and\ \citenamefont {Ladd}}]{kerckhoff_magnetic_2021}%
  \BibitemOpen
  \bibfield  {author} {\bibinfo {author} {\bibfnamefont {J.}~\bibnamefont {Kerckhoff}}, \bibinfo {author} {\bibfnamefont {B.}~\bibnamefont {Sun}}, \bibinfo {author} {\bibfnamefont {B.}~\bibnamefont {Fong}}, \bibinfo {author} {\bibfnamefont {C.}~\bibnamefont {Jones}}, \bibinfo {author} {\bibfnamefont {A.}~\bibnamefont {Kiselev}}, \bibinfo {author} {\bibfnamefont {D.}~\bibnamefont {Barnes}}, \bibinfo {author} {\bibfnamefont {R.}~\bibnamefont {Noah}}, \bibinfo {author} {\bibfnamefont {E.}~\bibnamefont {Acuna}}, \bibinfo {author} {\bibfnamefont {M.}~\bibnamefont {Akmal}}, \bibinfo {author} {\bibfnamefont {S.}~\bibnamefont {Ha}}, \bibinfo {author} {\bibfnamefont {J.}~\bibnamefont {Wright}}, \bibinfo {author} {\bibfnamefont {B.}~\bibnamefont {Thomas}}, \bibinfo {author} {\bibfnamefont {C.}~\bibnamefont {Jackson}}, \bibinfo {author} {\bibfnamefont {L.}~\bibnamefont {Edge}}, \bibinfo {author} {\bibfnamefont {K.}~\bibnamefont {Eng}}, \bibinfo {author} {\bibfnamefont {R.}~\bibnamefont {Ross}},\ and\ \bibinfo {author}
  {\bibfnamefont {T.}~\bibnamefont {Ladd}},\ }\bibfield  {title} {\bibinfo {title} {Magnetic {Gradient} {Fluctuations} from {Quadrupolar} {\ensuremath{{}^{73}}}{Ge} in {Si}/{SiGe} {Exchange}-{Only} {Qubits}},\ }\href {https://doi.org/10.1103/PRXQuantum.2.010347} {\bibfield  {journal} {\bibinfo  {journal} {PRX Quantum}\ }\textbf {\bibinfo {volume} {2}},\ \bibinfo {pages} {010347} (\bibinfo {year} {2021})}\BibitemShut {NoStop}%
\bibitem [{\citenamefont {Cvitkovich}\ \emph {et~al.}(2024)\citenamefont {Cvitkovich}, \citenamefont {Stano}, \citenamefont {Wilhelmer}, \citenamefont {Waldh{\"o}r}, \citenamefont {Loss}, \citenamefont {Niquet},\ and\ \citenamefont {Grasser}}]{cvitkovich_coherence_2024}%
  \BibitemOpen
  \bibfield  {author} {\bibinfo {author} {\bibfnamefont {L.}~\bibnamefont {Cvitkovich}}, \bibinfo {author} {\bibfnamefont {P.}~\bibnamefont {Stano}}, \bibinfo {author} {\bibfnamefont {C.}~\bibnamefont {Wilhelmer}}, \bibinfo {author} {\bibfnamefont {D.}~\bibnamefont {Waldh{\"o}r}}, \bibinfo {author} {\bibfnamefont {D.}~\bibnamefont {Loss}}, \bibinfo {author} {\bibfnamefont {Y.-M.}\ \bibnamefont {Niquet}},\ and\ \bibinfo {author} {\bibfnamefont {T.}~\bibnamefont {Grasser}},\ }\bibfield  {title} {\bibinfo {title} {Coherence limit due to hyperfine interaction with nuclei in the barrier material of {Si} spin qubits},\ }\href {https://doi.org/10.1103/PhysRevApplied.22.064089} {\bibfield  {journal} {\bibinfo  {journal} {Physical Review Applied}\ }\textbf {\bibinfo {volume} {22}},\ \bibinfo {pages} {064089} (\bibinfo {year} {2024})}\BibitemShut {NoStop}%
\bibitem [{\citenamefont {Degli~Esposti}\ \emph {et~al.}(2022)\citenamefont {Degli~Esposti}, \citenamefont {Paquelet~Wuetz}, \citenamefont {Fezzi}, \citenamefont {Lodari}, \citenamefont {Sammak},\ and\ \citenamefont {Scappucci}}]{degli_esposti_wafer-scale_2022}%
  \BibitemOpen
  \bibfield  {author} {\bibinfo {author} {\bibfnamefont {D.}~\bibnamefont {Degli~Esposti}}, \bibinfo {author} {\bibfnamefont {B.}~\bibnamefont {Paquelet~Wuetz}}, \bibinfo {author} {\bibfnamefont {V.}~\bibnamefont {Fezzi}}, \bibinfo {author} {\bibfnamefont {M.}~\bibnamefont {Lodari}}, \bibinfo {author} {\bibfnamefont {A.}~\bibnamefont {Sammak}},\ and\ \bibinfo {author} {\bibfnamefont {G.}~\bibnamefont {Scappucci}},\ }\bibfield  {title} {\bibinfo {title} {Wafer-scale low-disorder {2DEG} in {28Si}/{SiGe} without an epitaxial {Si} cap},\ }\href {https://doi.org/10.1063/5.0088576} {\bibfield  {journal} {\bibinfo  {journal} {Applied Physics Letters}\ }\textbf {\bibinfo {volume} {120}},\ \bibinfo {pages} {184003} (\bibinfo {year} {2022})}\BibitemShut {NoStop}%
\bibitem [{\citenamefont {Park}\ \emph {et~al.}(2026)\citenamefont {Park}, \citenamefont {Jang}, \citenamefont {Sohn}, \citenamefont {Song}, \citenamefont {Stehouwer}, \citenamefont {Degli~Esposti}, \citenamefont {Scappucci},\ and\ \citenamefont {Kim}}]{park_highly_2026}%
  \BibitemOpen
  \bibfield  {author} {\bibinfo {author} {\bibfnamefont {J.}~\bibnamefont {Park}}, \bibinfo {author} {\bibfnamefont {H.}~\bibnamefont {Jang}}, \bibinfo {author} {\bibfnamefont {H.}~\bibnamefont {Sohn}}, \bibinfo {author} {\bibfnamefont {Y.}~\bibnamefont {Song}}, \bibinfo {author} {\bibfnamefont {L.~E.~A.}\ \bibnamefont {Stehouwer}}, \bibinfo {author} {\bibfnamefont {D.}~\bibnamefont {Degli~Esposti}}, \bibinfo {author} {\bibfnamefont {G.}~\bibnamefont {Scappucci}},\ and\ \bibinfo {author} {\bibfnamefont {D.}~\bibnamefont {Kim}},\ }\bibfield  {title} {\bibinfo {title} {Highly {Tunable} {Two}-{Qubit} {Interactions} in {Si}/{SiGe} {Quantum} {Dots} by {Interchanging} the {Roles} of {Qubit}-{Defining} {Gates}},\ }\href {https://doi.org/10.1021/acs.nanolett.6c00044} {\bibfield  {journal} {\bibinfo  {journal} {Nano Letters}\ }\textbf {\bibinfo {volume} {26}},\ \bibinfo {pages} {7493} (\bibinfo {year} {2026})}\BibitemShut {NoStop}%
\bibitem [{\citenamefont {Fern{\'a}ndez~de Fuentes}\ \emph {et~al.}(2026)\citenamefont {Fern{\'a}ndez~de Fuentes}, \citenamefont {Raymenants}, \citenamefont {Undseth}, \citenamefont {Pietx-Casas}, \citenamefont {Philips}, \citenamefont {M{\k a}dzik}, \citenamefont {Snoo}, \citenamefont {Amitonov}, \citenamefont {Tryputen}, \citenamefont {Schmitz}, \citenamefont {Matsuura}, \citenamefont {Scappucci},\ and\ \citenamefont {Vandersypen}}]{fernandez_de_fuentes_running_2026}%
  \BibitemOpen
  \bibfield  {author} {\bibinfo {author} {\bibfnamefont {I.}~\bibnamefont {Fern{\'a}ndez~de Fuentes}}, \bibinfo {author} {\bibfnamefont {E.}~\bibnamefont {Raymenants}}, \bibinfo {author} {\bibfnamefont {B.}~\bibnamefont {Undseth}}, \bibinfo {author} {\bibfnamefont {O.}~\bibnamefont {Pietx-Casas}}, \bibinfo {author} {\bibfnamefont {S.}~\bibnamefont {Philips}}, \bibinfo {author} {\bibfnamefont {M.}~\bibnamefont {M{\k a}dzik}}, \bibinfo {author} {\bibfnamefont {S.}~\bibnamefont {Snoo}}, \bibinfo {author} {\bibfnamefont {S.}~\bibnamefont {Amitonov}}, \bibinfo {author} {\bibfnamefont {L.}~\bibnamefont {Tryputen}}, \bibinfo {author} {\bibfnamefont {A.}~\bibnamefont {Schmitz}}, \bibinfo {author} {\bibfnamefont {A.}~\bibnamefont {Matsuura}}, \bibinfo {author} {\bibfnamefont {G.}~\bibnamefont {Scappucci}},\ and\ \bibinfo {author} {\bibfnamefont {L.}~\bibnamefont {Vandersypen}},\ }\bibfield  {title} {\bibinfo {title} {Running a {Six}-{Qubit} {Quantum} {Circuit} on a {Silicon} {Spin}-{Qubit} {Array}},\ }\href
  {https://doi.org/10.1103/f285-l2v5} {\bibfield  {journal} {\bibinfo  {journal} {PRX Quantum}\ }\textbf {\bibinfo {volume} {7}},\ \bibinfo {pages} {010308} (\bibinfo {year} {2026})}\BibitemShut {NoStop}%
\bibitem [{\citenamefont {Rojas-Arias}\ \emph {et~al.}(2026{\natexlab{a}})\citenamefont {Rojas-Arias}, \citenamefont {Camenzind}, \citenamefont {Wu}, \citenamefont {Stano}, \citenamefont {Noiri}, \citenamefont {Takeda}, \citenamefont {Nakajima}, \citenamefont {Kobayashi}, \citenamefont {Scappucci}, \citenamefont {Loss},\ and\ \citenamefont {Tarucha}}]{rojas-arias_scaling_2026}%
  \BibitemOpen
  \bibfield  {author} {\bibinfo {author} {\bibfnamefont {J.~S.}\ \bibnamefont {Rojas-Arias}}, \bibinfo {author} {\bibfnamefont {L.~C.}\ \bibnamefont {Camenzind}}, \bibinfo {author} {\bibfnamefont {Y.-H.}\ \bibnamefont {Wu}}, \bibinfo {author} {\bibfnamefont {P.}~\bibnamefont {Stano}}, \bibinfo {author} {\bibfnamefont {A.}~\bibnamefont {Noiri}}, \bibinfo {author} {\bibfnamefont {K.}~\bibnamefont {Takeda}}, \bibinfo {author} {\bibfnamefont {T.}~\bibnamefont {Nakajima}}, \bibinfo {author} {\bibfnamefont {T.}~\bibnamefont {Kobayashi}}, \bibinfo {author} {\bibfnamefont {G.}~\bibnamefont {Scappucci}}, \bibinfo {author} {\bibfnamefont {D.}~\bibnamefont {Loss}},\ and\ \bibinfo {author} {\bibfnamefont {S.}~\bibnamefont {Tarucha}},\ }\href {https://doi.org/10.48550/arXiv.2603.03051} {\bibinfo {title} {Scaling of silicon spin qubits under correlated noise}} (\bibinfo {year} {2026}{\natexlab{a}}),\ \bibinfo {note} {arXiv:2603.03051 [cond-mat.mes-hall]}\BibitemShut {NoStop}%
\bibitem [{\citenamefont {Xue}\ \emph {et~al.}(2022)\citenamefont {Xue}, \citenamefont {Russ}, \citenamefont {Samkharadze}, \citenamefont {Undseth}, \citenamefont {Sammak}, \citenamefont {Scappucci},\ and\ \citenamefont {Vandersypen}}]{xue_quantum_2022}%
  \BibitemOpen
  \bibfield  {author} {\bibinfo {author} {\bibfnamefont {X.}~\bibnamefont {Xue}}, \bibinfo {author} {\bibfnamefont {M.}~\bibnamefont {Russ}}, \bibinfo {author} {\bibfnamefont {N.}~\bibnamefont {Samkharadze}}, \bibinfo {author} {\bibfnamefont {B.}~\bibnamefont {Undseth}}, \bibinfo {author} {\bibfnamefont {A.}~\bibnamefont {Sammak}}, \bibinfo {author} {\bibfnamefont {G.}~\bibnamefont {Scappucci}},\ and\ \bibinfo {author} {\bibfnamefont {L.~M.~K.}\ \bibnamefont {Vandersypen}},\ }\bibfield  {title} {\bibinfo {title} {Quantum logic with spin qubits crossing the surface code threshold},\ }\href {https://doi.org/10.1038/s41586-021-04273-w} {\bibfield  {journal} {\bibinfo  {journal} {Nature}\ }\textbf {\bibinfo {volume} {601}},\ \bibinfo {pages} {343} (\bibinfo {year} {2022})}\BibitemShut {NoStop}%
\bibitem [{\citenamefont {Neyens}\ \emph {et~al.}(2024)\citenamefont {Neyens}, \citenamefont {Zietz}, \citenamefont {Watson}, \citenamefont {Luthi}, \citenamefont {Nethwewala}, \citenamefont {George}, \citenamefont {Henry}, \citenamefont {Islam}, \citenamefont {Wagner}, \citenamefont {Borjans}, \citenamefont {Connors}, \citenamefont {Corrigan}, \citenamefont {Curry}, \citenamefont {Keith}, \citenamefont {Kotlyar}, \citenamefont {Lampert}, \citenamefont {M{\k a}dzik}, \citenamefont {Millard}, \citenamefont {Mohiyaddin}, \citenamefont {Pellerano}, \citenamefont {Pillarisetty}, \citenamefont {Ramsey}, \citenamefont {Savytskyy}, \citenamefont {Schaal}, \citenamefont {Zheng}, \citenamefont {Ziegler}, \citenamefont {Bishop}, \citenamefont {Bojarski}, \citenamefont {Roberts},\ and\ \citenamefont {Clarke}}]{neyens_probing_2024}%
  \BibitemOpen
  \bibfield  {author} {\bibinfo {author} {\bibfnamefont {S.}~\bibnamefont {Neyens}}, \bibinfo {author} {\bibfnamefont {O.~K.}\ \bibnamefont {Zietz}}, \bibinfo {author} {\bibfnamefont {T.~F.}\ \bibnamefont {Watson}}, \bibinfo {author} {\bibfnamefont {F.}~\bibnamefont {Luthi}}, \bibinfo {author} {\bibfnamefont {A.}~\bibnamefont {Nethwewala}}, \bibinfo {author} {\bibfnamefont {H.~C.}\ \bibnamefont {George}}, \bibinfo {author} {\bibfnamefont {E.}~\bibnamefont {Henry}}, \bibinfo {author} {\bibfnamefont {M.}~\bibnamefont {Islam}}, \bibinfo {author} {\bibfnamefont {A.~J.}\ \bibnamefont {Wagner}}, \bibinfo {author} {\bibfnamefont {F.}~\bibnamefont {Borjans}}, \bibinfo {author} {\bibfnamefont {E.~J.}\ \bibnamefont {Connors}}, \bibinfo {author} {\bibfnamefont {J.}~\bibnamefont {Corrigan}}, \bibinfo {author} {\bibfnamefont {M.~J.}\ \bibnamefont {Curry}}, \bibinfo {author} {\bibfnamefont {D.}~\bibnamefont {Keith}}, \bibinfo {author} {\bibfnamefont {R.}~\bibnamefont {Kotlyar}}, \bibinfo {author} {\bibfnamefont {L.~F.}\
  \bibnamefont {Lampert}}, \bibinfo {author} {\bibfnamefont {M.~T.}\ \bibnamefont {M{\k a}dzik}}, \bibinfo {author} {\bibfnamefont {K.}~\bibnamefont {Millard}}, \bibinfo {author} {\bibfnamefont {F.~A.}\ \bibnamefont {Mohiyaddin}}, \bibinfo {author} {\bibfnamefont {S.}~\bibnamefont {Pellerano}}, \bibinfo {author} {\bibfnamefont {R.}~\bibnamefont {Pillarisetty}}, \bibinfo {author} {\bibfnamefont {M.}~\bibnamefont {Ramsey}}, \bibinfo {author} {\bibfnamefont {R.}~\bibnamefont {Savytskyy}}, \bibinfo {author} {\bibfnamefont {S.}~\bibnamefont {Schaal}}, \bibinfo {author} {\bibfnamefont {G.}~\bibnamefont {Zheng}}, \bibinfo {author} {\bibfnamefont {J.}~\bibnamefont {Ziegler}}, \bibinfo {author} {\bibfnamefont {N.~C.}\ \bibnamefont {Bishop}}, \bibinfo {author} {\bibfnamefont {S.}~\bibnamefont {Bojarski}}, \bibinfo {author} {\bibfnamefont {J.}~\bibnamefont {Roberts}},\ and\ \bibinfo {author} {\bibfnamefont {J.~S.}\ \bibnamefont {Clarke}},\ }\bibfield  {title} {\bibinfo {title} {Probing single electrons across 300-mm
  spin qubit wafers},\ }\href {https://doi.org/10.1038/s41586-024-07275-6} {\bibfield  {journal} {\bibinfo  {journal} {Nature}\ }\textbf {\bibinfo {volume} {629}},\ \bibinfo {pages} {80} (\bibinfo {year} {2024})}\BibitemShut {NoStop}%
\bibitem [{\citenamefont {Vansteenkiste}\ \emph {et~al.}(2014)\citenamefont {Vansteenkiste}, \citenamefont {Leliaert}, \citenamefont {Dvornik}, \citenamefont {Helsen}, \citenamefont {Garcia-Sanchez},\ and\ \citenamefont {Van~Waeyenberge}}]{vansteenkiste_design_2014}%
  \BibitemOpen
  \bibfield  {author} {\bibinfo {author} {\bibfnamefont {A.}~\bibnamefont {Vansteenkiste}}, \bibinfo {author} {\bibfnamefont {J.}~\bibnamefont {Leliaert}}, \bibinfo {author} {\bibfnamefont {M.}~\bibnamefont {Dvornik}}, \bibinfo {author} {\bibfnamefont {M.}~\bibnamefont {Helsen}}, \bibinfo {author} {\bibfnamefont {F.}~\bibnamefont {Garcia-Sanchez}},\ and\ \bibinfo {author} {\bibfnamefont {B.}~\bibnamefont {Van~Waeyenberge}},\ }\bibfield  {title} {\bibinfo {title} {The design and verification of {MuMax3}},\ }\href {https://doi.org/10.1063/1.4899186} {\bibfield  {journal} {\bibinfo  {journal} {AIP Advances}\ }\textbf {\bibinfo {volume} {4}},\ \bibinfo {pages} {107133} (\bibinfo {year} {2014})}\BibitemShut {NoStop}%
\bibitem [{\citenamefont {Jang}\ \emph {et~al.}(2020)\citenamefont {Jang}, \citenamefont {Kim}, \citenamefont {Cho}, \citenamefont {Chung}, \citenamefont {Park}, \citenamefont {Eom}, \citenamefont {Umansky}, \citenamefont {Chung},\ and\ \citenamefont {Kim}}]{jang_robust_2020}%
  \BibitemOpen
  \bibfield  {author} {\bibinfo {author} {\bibfnamefont {W.}~\bibnamefont {Jang}}, \bibinfo {author} {\bibfnamefont {J.}~\bibnamefont {Kim}}, \bibinfo {author} {\bibfnamefont {M.-K.}\ \bibnamefont {Cho}}, \bibinfo {author} {\bibfnamefont {H.}~\bibnamefont {Chung}}, \bibinfo {author} {\bibfnamefont {S.}~\bibnamefont {Park}}, \bibinfo {author} {\bibfnamefont {J.}~\bibnamefont {Eom}}, \bibinfo {author} {\bibfnamefont {V.}~\bibnamefont {Umansky}}, \bibinfo {author} {\bibfnamefont {Y.}~\bibnamefont {Chung}},\ and\ \bibinfo {author} {\bibfnamefont {D.}~\bibnamefont {Kim}},\ }\bibfield  {title} {\bibinfo {title} {Robust energy-selective tunneling readout of singlet-triplet qubits under large magnetic field gradient},\ }\href {https://doi.org/10.1038/s41534-020-00295-w} {\bibfield  {journal} {\bibinfo  {journal} {npj Quantum Information}\ }\textbf {\bibinfo {volume} {6}},\ \bibinfo {pages} {64} (\bibinfo {year} {2020})}\BibitemShut {NoStop}%
\bibitem [{\citenamefont {Takeda}\ \emph {et~al.}(2024)\citenamefont {Takeda}, \citenamefont {Noiri}, \citenamefont {Nakajima}, \citenamefont {Camenzind}, \citenamefont {Kobayashi}, \citenamefont {Sammak}, \citenamefont {Scappucci},\ and\ \citenamefont {Tarucha}}]{takeda_rapid_2024}%
  \BibitemOpen
  \bibfield  {author} {\bibinfo {author} {\bibfnamefont {K.}~\bibnamefont {Takeda}}, \bibinfo {author} {\bibfnamefont {A.}~\bibnamefont {Noiri}}, \bibinfo {author} {\bibfnamefont {T.}~\bibnamefont {Nakajima}}, \bibinfo {author} {\bibfnamefont {L.~C.}\ \bibnamefont {Camenzind}}, \bibinfo {author} {\bibfnamefont {T.}~\bibnamefont {Kobayashi}}, \bibinfo {author} {\bibfnamefont {A.}~\bibnamefont {Sammak}}, \bibinfo {author} {\bibfnamefont {G.}~\bibnamefont {Scappucci}},\ and\ \bibinfo {author} {\bibfnamefont {S.}~\bibnamefont {Tarucha}},\ }\bibfield  {title} {\bibinfo {title} {Rapid single-shot parity spin readout in a silicon double quantum dot with fidelity exceeding 99\%},\ }\href {https://doi.org/10.1038/s41534-024-00813-0} {\bibfield  {journal} {\bibinfo  {journal} {npj Quantum Information}\ }\textbf {\bibinfo {volume} {10}},\ \bibinfo {pages} {22} (\bibinfo {year} {2024})}\BibitemShut {NoStop}%
\bibitem [{\citenamefont {Medford}\ \emph {et~al.}(2012)\citenamefont {Medford}, \citenamefont {Cywi{\'n}ski}, \citenamefont {Barthel}, \citenamefont {Marcus}, \citenamefont {Hanson},\ and\ \citenamefont {Gossard}}]{medford_scaling_2012}%
  \BibitemOpen
  \bibfield  {author} {\bibinfo {author} {\bibfnamefont {J.}~\bibnamefont {Medford}}, \bibinfo {author} {\bibfnamefont {{\L}.}~\bibnamefont {Cywi{\'n}ski}}, \bibinfo {author} {\bibfnamefont {C.}~\bibnamefont {Barthel}}, \bibinfo {author} {\bibfnamefont {C.~M.}\ \bibnamefont {Marcus}}, \bibinfo {author} {\bibfnamefont {M.~P.}\ \bibnamefont {Hanson}},\ and\ \bibinfo {author} {\bibfnamefont {A.~C.}\ \bibnamefont {Gossard}},\ }\bibfield  {title} {\bibinfo {title} {Scaling of {Dynamical} {Decoupling} for {Spin} {Qubits}},\ }\href {https://doi.org/10.1103/PhysRevLett.108.086802} {\bibfield  {journal} {\bibinfo  {journal} {Physical Review Letters}\ }\textbf {\bibinfo {volume} {108}},\ \bibinfo {pages} {086802} (\bibinfo {year} {2012})}\BibitemShut {NoStop}%
\bibitem [{\citenamefont {Cywi{\'n}ski}(2014)}]{cywinski_dynamical-decoupling_2014}%
  \BibitemOpen
  \bibfield  {author} {\bibinfo {author} {\bibfnamefont {{\L}.}~\bibnamefont {Cywi{\'n}ski}},\ }\bibfield  {title} {\bibinfo {title} {Dynamical-decoupling noise spectroscopy at an optimal working point of a qubit},\ }\href {https://doi.org/10.1103/PhysRevA.90.042307} {\bibfield  {journal} {\bibinfo  {journal} {Physical Review A}\ }\textbf {\bibinfo {volume} {90}},\ \bibinfo {pages} {042307} (\bibinfo {year} {2014})}\BibitemShut {NoStop}%
\bibitem [{\citenamefont {Connors}\ \emph {et~al.}(2022)\citenamefont {Connors}, \citenamefont {Nelson}, \citenamefont {Edge},\ and\ \citenamefont {Nichol}}]{connors_charge-noise_2022}%
  \BibitemOpen
  \bibfield  {author} {\bibinfo {author} {\bibfnamefont {E.~J.}\ \bibnamefont {Connors}}, \bibinfo {author} {\bibfnamefont {J.}~\bibnamefont {Nelson}}, \bibinfo {author} {\bibfnamefont {L.~F.}\ \bibnamefont {Edge}},\ and\ \bibinfo {author} {\bibfnamefont {J.~M.}\ \bibnamefont {Nichol}},\ }\bibfield  {title} {\bibinfo {title} {Charge-noise spectroscopy of {Si}/{SiGe} quantum dots via dynamically-decoupled exchange oscillations},\ }\href {https://doi.org/10.1038/s41467-022-28519-x} {\bibfield  {journal} {\bibinfo  {journal} {Nature Communications}\ }\textbf {\bibinfo {volume} {13}},\ \bibinfo {pages} {940} (\bibinfo {year} {2022})}\BibitemShut {NoStop}%
\bibitem [{\citenamefont {Yoneda}\ \emph {et~al.}(2023)\citenamefont {Yoneda}, \citenamefont {Rojas-Arias}, \citenamefont {Stano}, \citenamefont {Takeda}, \citenamefont {Noiri}, \citenamefont {Nakajima}, \citenamefont {Loss},\ and\ \citenamefont {Tarucha}}]{yoneda_noise-correlation_2023}%
  \BibitemOpen
  \bibfield  {author} {\bibinfo {author} {\bibfnamefont {J.}~\bibnamefont {Yoneda}}, \bibinfo {author} {\bibfnamefont {J.~S.}\ \bibnamefont {Rojas-Arias}}, \bibinfo {author} {\bibfnamefont {P.}~\bibnamefont {Stano}}, \bibinfo {author} {\bibfnamefont {K.}~\bibnamefont {Takeda}}, \bibinfo {author} {\bibfnamefont {A.}~\bibnamefont {Noiri}}, \bibinfo {author} {\bibfnamefont {T.}~\bibnamefont {Nakajima}}, \bibinfo {author} {\bibfnamefont {D.}~\bibnamefont {Loss}},\ and\ \bibinfo {author} {\bibfnamefont {S.}~\bibnamefont {Tarucha}},\ }\bibfield  {title} {\bibinfo {title} {Noise-correlation spectrum for a pair of spin qubits in silicon},\ }\href {https://doi.org/10.1038/s41567-023-02238-6} {\bibfield  {journal} {\bibinfo  {journal} {Nature Physics}\ }\textbf {\bibinfo {volume} {19}},\ \bibinfo {pages} {1793} (\bibinfo {year} {2023})}\BibitemShut {NoStop}%
\bibitem [{\citenamefont {Rojas-Arias}\ \emph {et~al.}(2023)\citenamefont {Rojas-Arias}, \citenamefont {Noiri}, \citenamefont {Stano}, \citenamefont {Nakajima}, \citenamefont {Yoneda}, \citenamefont {Takeda}, \citenamefont {Kobayashi}, \citenamefont {Sammak}, \citenamefont {Scappucci}, \citenamefont {Loss},\ and\ \citenamefont {Tarucha}}]{rojas-arias_spatial_2023}%
  \BibitemOpen
  \bibfield  {author} {\bibinfo {author} {\bibfnamefont {J.}~\bibnamefont {Rojas-Arias}}, \bibinfo {author} {\bibfnamefont {A.}~\bibnamefont {Noiri}}, \bibinfo {author} {\bibfnamefont {P.}~\bibnamefont {Stano}}, \bibinfo {author} {\bibfnamefont {T.}~\bibnamefont {Nakajima}}, \bibinfo {author} {\bibfnamefont {J.}~\bibnamefont {Yoneda}}, \bibinfo {author} {\bibfnamefont {K.}~\bibnamefont {Takeda}}, \bibinfo {author} {\bibfnamefont {T.}~\bibnamefont {Kobayashi}}, \bibinfo {author} {\bibfnamefont {A.}~\bibnamefont {Sammak}}, \bibinfo {author} {\bibfnamefont {G.}~\bibnamefont {Scappucci}}, \bibinfo {author} {\bibfnamefont {D.}~\bibnamefont {Loss}},\ and\ \bibinfo {author} {\bibfnamefont {S.}~\bibnamefont {Tarucha}},\ }\bibfield  {title} {\bibinfo {title} {Spatial noise correlations beyond nearest neighbors in {\ensuremath{{}^{28}}}{Si}/{SiGe} spin qubits},\ }\href {https://doi.org/10.1103/PhysRevApplied.20.054024} {\bibfield  {journal} {\bibinfo  {journal} {Physical Review Applied}\ }\textbf {\bibinfo {volume}
  {20}},\ \bibinfo {pages} {054024} (\bibinfo {year} {2023})}\BibitemShut {NoStop}%
\bibitem [{\citenamefont {Rojas-Arias}\ \emph {et~al.}(2026{\natexlab{b}})\citenamefont {Rojas-Arias}, \citenamefont {Kojima}, \citenamefont {Takeda}, \citenamefont {Stano}, \citenamefont {Nakajima}, \citenamefont {Yoneda}, \citenamefont {Noiri}, \citenamefont {Kobayashi}, \citenamefont {Loss},\ and\ \citenamefont {Tarucha}}]{rojas-arias_origins_2025}%
  \BibitemOpen
  \bibfield  {author} {\bibinfo {author} {\bibfnamefont {J.~S.}\ \bibnamefont {Rojas-Arias}}, \bibinfo {author} {\bibfnamefont {Y.}~\bibnamefont {Kojima}}, \bibinfo {author} {\bibfnamefont {K.}~\bibnamefont {Takeda}}, \bibinfo {author} {\bibfnamefont {P.}~\bibnamefont {Stano}}, \bibinfo {author} {\bibfnamefont {T.}~\bibnamefont {Nakajima}}, \bibinfo {author} {\bibfnamefont {J.}~\bibnamefont {Yoneda}}, \bibinfo {author} {\bibfnamefont {A.}~\bibnamefont {Noiri}}, \bibinfo {author} {\bibfnamefont {T.}~\bibnamefont {Kobayashi}}, \bibinfo {author} {\bibfnamefont {D.}~\bibnamefont {Loss}},\ and\ \bibinfo {author} {\bibfnamefont {S.}~\bibnamefont {Tarucha}},\ }\bibfield  {title} {\bibinfo {title} {The origins of noise in the {Zeeman} splitting of spin qubits in natural-silicon devices},\ }\href {https://doi.org/10.1038/s41534-025-01150-6} {\bibfield  {journal} {\bibinfo  {journal} {npj Quantum Information}\ }\textbf {\bibinfo {volume} {12}},\ \bibinfo {pages} {9} (\bibinfo {year} {2026}{\natexlab{b}})}\BibitemShut
  {NoStop}%
\bibitem [{\citenamefont {Park}\ \emph {et~al.}(2025)\citenamefont {Park}, \citenamefont {Jang}, \citenamefont {Sohn}, \citenamefont {Yun}, \citenamefont {Song}, \citenamefont {Kang}, \citenamefont {Stehouwer}, \citenamefont {Esposti}, \citenamefont {Scappucci},\ and\ \citenamefont {Kim}}]{park_passive_2025}%
  \BibitemOpen
  \bibfield  {author} {\bibinfo {author} {\bibfnamefont {J.}~\bibnamefont {Park}}, \bibinfo {author} {\bibfnamefont {H.}~\bibnamefont {Jang}}, \bibinfo {author} {\bibfnamefont {H.}~\bibnamefont {Sohn}}, \bibinfo {author} {\bibfnamefont {J.}~\bibnamefont {Yun}}, \bibinfo {author} {\bibfnamefont {Y.}~\bibnamefont {Song}}, \bibinfo {author} {\bibfnamefont {B.}~\bibnamefont {Kang}}, \bibinfo {author} {\bibfnamefont {L.~E.~A.}\ \bibnamefont {Stehouwer}}, \bibinfo {author} {\bibfnamefont {D.~D.}\ \bibnamefont {Esposti}}, \bibinfo {author} {\bibfnamefont {G.}~\bibnamefont {Scappucci}},\ and\ \bibinfo {author} {\bibfnamefont {D.}~\bibnamefont {Kim}},\ }\bibfield  {title} {\bibinfo {title} {Passive and active suppression of transduced noise in silicon spin qubits},\ }\href {https://doi.org/10.1038/s41467-024-55338-z} {\bibfield  {journal} {\bibinfo  {journal} {Nature Communications}\ }\textbf {\bibinfo {volume} {16}},\ \bibinfo {pages} {78} (\bibinfo {year} {2025})}\BibitemShut {NoStop}%
\bibitem [{\citenamefont {K{\k e}pa}\ \emph {et~al.}(2023)\citenamefont {K{\k e}pa}, \citenamefont {Focke}, \citenamefont {Cywi{\'n}ski},\ and\ \citenamefont {Krzywda}}]{kepa_simulation_2023}%
  \BibitemOpen
  \bibfield  {author} {\bibinfo {author} {\bibfnamefont {M.}~\bibnamefont {K{\k e}pa}}, \bibinfo {author} {\bibfnamefont {N.}~\bibnamefont {Focke}}, \bibinfo {author} {\bibfnamefont {{\L}.}~\bibnamefont {Cywi{\'n}ski}},\ and\ \bibinfo {author} {\bibfnamefont {J.~A.}\ \bibnamefont {Krzywda}},\ }\bibfield  {title} {\bibinfo {title} {Simulation of 1 / f charge noise affecting a quantum dot in a {Si}/{SiGe} structure},\ }\href {https://doi.org/10.1063/5.0151029} {\bibfield  {journal} {\bibinfo  {journal} {Applied Physics Letters}\ }\textbf {\bibinfo {volume} {123}},\ \bibinfo {pages} {034005} (\bibinfo {year} {2023})}\BibitemShut {NoStop}%
\bibitem [{\citenamefont {Paladino}\ \emph {et~al.}(2014)\citenamefont {Paladino}, \citenamefont {Galperin}, \citenamefont {Falci},\ and\ \citenamefont {Altshuler}}]{paladino_1_2014}%
  \BibitemOpen
  \bibfield  {author} {\bibinfo {author} {\bibfnamefont {E.}~\bibnamefont {Paladino}}, \bibinfo {author} {\bibfnamefont {Y.~M.}\ \bibnamefont {Galperin}}, \bibinfo {author} {\bibfnamefont {G.}~\bibnamefont {Falci}},\ and\ \bibinfo {author} {\bibfnamefont {B.~L.}\ \bibnamefont {Altshuler}},\ }\bibfield  {title} {\bibinfo {title} {1 / f noise: {Implications} for solid-state quantum information},\ }\href {https://doi.org/10.1103/RevModPhys.86.361} {\bibfield  {journal} {\bibinfo  {journal} {Reviews of Modern Physics}\ }\textbf {\bibinfo {volume} {86}},\ \bibinfo {pages} {361} (\bibinfo {year} {2014})}\BibitemShut {NoStop}%
\bibitem [{\citenamefont {Cheng}\ and\ \citenamefont {Guo}(2025)}]{cheng_modeling_2025}%
  \BibitemOpen
  \bibfield  {author} {\bibinfo {author} {\bibfnamefont {G.}~\bibnamefont {Cheng}}\ and\ \bibinfo {author} {\bibfnamefont {J.}~\bibnamefont {Guo}},\ }\bibfield  {title} {\bibinfo {title} {Modeling correlated-noise in silicon spin qubit device},\ }\href {https://doi.org/10.1063/5.0216833} {\bibfield  {journal} {\bibinfo  {journal} {APL Quantum}\ }\textbf {\bibinfo {volume} {2}},\ \bibinfo {pages} {016101} (\bibinfo {year} {2025})}\BibitemShut {NoStop}%
\bibitem [{\citenamefont {Hayashi}\ \emph {et~al.}(2008)\citenamefont {Hayashi}, \citenamefont {Itoh},\ and\ \citenamefont {Vlasenko}}]{hayashi_nuclear_2008}%
  \BibitemOpen
  \bibfield  {author} {\bibinfo {author} {\bibfnamefont {H.}~\bibnamefont {Hayashi}}, \bibinfo {author} {\bibfnamefont {K.~M.}\ \bibnamefont {Itoh}},\ and\ \bibinfo {author} {\bibfnamefont {L.~S.}\ \bibnamefont {Vlasenko}},\ }\bibfield  {title} {\bibinfo {title} {Nuclear magnetic resonance linewidth and spin diffusion in $^{\textrm{29}}${Si} isotopically controlled silicon},\ }\href {https://doi.org/10.1103/PhysRevB.78.153201} {\bibfield  {journal} {\bibinfo  {journal} {Physical Review B}\ }\textbf {\bibinfo {volume} {78}},\ \bibinfo {pages} {153201} (\bibinfo {year} {2008})}\BibitemShut {NoStop}%
\bibitem [{\citenamefont {Connors}\ \emph {et~al.}(2019)\citenamefont {Connors}, \citenamefont {Nelson}, \citenamefont {Qiao}, \citenamefont {Edge},\ and\ \citenamefont {Nichol}}]{connors_low-frequency_2019}%
  \BibitemOpen
  \bibfield  {author} {\bibinfo {author} {\bibfnamefont {E.~J.}\ \bibnamefont {Connors}}, \bibinfo {author} {\bibfnamefont {J.}~\bibnamefont {Nelson}}, \bibinfo {author} {\bibfnamefont {H.}~\bibnamefont {Qiao}}, \bibinfo {author} {\bibfnamefont {L.~F.}\ \bibnamefont {Edge}},\ and\ \bibinfo {author} {\bibfnamefont {J.~M.}\ \bibnamefont {Nichol}},\ }\bibfield  {title} {\bibinfo {title} {Low-frequency charge noise in {Si}/{SiGe} quantum dots},\ }\href {https://doi.org/10.1103/PhysRevB.100.165305} {\bibfield  {journal} {\bibinfo  {journal} {Physical Review B}\ }\textbf {\bibinfo {volume} {100}},\ \bibinfo {pages} {165305} (\bibinfo {year} {2019})}\BibitemShut {NoStop}%
\bibitem [{\citenamefont {Hell}\ \emph {et~al.}(2016)\citenamefont {Hell}, \citenamefont {Wegewijs},\ and\ \citenamefont {DiVincenzo}}]{hell_qubit_2016}%
  \BibitemOpen
  \bibfield  {author} {\bibinfo {author} {\bibfnamefont {M.}~\bibnamefont {Hell}}, \bibinfo {author} {\bibfnamefont {M.~R.}\ \bibnamefont {Wegewijs}},\ and\ \bibinfo {author} {\bibfnamefont {D.~P.}\ \bibnamefont {DiVincenzo}},\ }\bibfield  {title} {\bibinfo {title} {Qubit quantum-dot sensors: {Noise} cancellation by coherent backaction, initial slips, and elliptical precession},\ }\href {https://doi.org/10.1103/PhysRevB.93.045418} {\bibfield  {journal} {\bibinfo  {journal} {Physical Review B}\ }\textbf {\bibinfo {volume} {93}},\ \bibinfo {pages} {045418} (\bibinfo {year} {2016})}\BibitemShut {NoStop}%
\bibitem [{\citenamefont {Barnes}\ \emph {et~al.}(2011)\citenamefont {Barnes}, \citenamefont {Kestner}, \citenamefont {Nguyen},\ and\ \citenamefont {Das~Sarma}}]{barnes_screening_2011}%
  \BibitemOpen
  \bibfield  {author} {\bibinfo {author} {\bibfnamefont {E.}~\bibnamefont {Barnes}}, \bibinfo {author} {\bibfnamefont {J.~P.}\ \bibnamefont {Kestner}}, \bibinfo {author} {\bibfnamefont {N.~T.~T.}\ \bibnamefont {Nguyen}},\ and\ \bibinfo {author} {\bibfnamefont {S.}~\bibnamefont {Das~Sarma}},\ }\bibfield  {title} {\bibinfo {title} {Screening of charged impurities with multielectron singlet-triplet spin qubits in quantum dots},\ }\href {https://doi.org/10.1103/PhysRevB.84.235309} {\bibfield  {journal} {\bibinfo  {journal} {Physical Review B}\ }\textbf {\bibinfo {volume} {84}},\ \bibinfo {pages} {235309} (\bibinfo {year} {2011})}\BibitemShut {NoStop}%
\bibitem [{\citenamefont {Higginbotham}\ \emph {et~al.}(2014)\citenamefont {Higginbotham}, \citenamefont {Kuemmeth}, \citenamefont {Hanson}, \citenamefont {Gossard},\ and\ \citenamefont {Marcus}}]{higginbotham_coherent_2014}%
  \BibitemOpen
  \bibfield  {author} {\bibinfo {author} {\bibfnamefont {A.~P.}\ \bibnamefont {Higginbotham}}, \bibinfo {author} {\bibfnamefont {F.}~\bibnamefont {Kuemmeth}}, \bibinfo {author} {\bibfnamefont {M.~P.}\ \bibnamefont {Hanson}}, \bibinfo {author} {\bibfnamefont {A.~C.}\ \bibnamefont {Gossard}},\ and\ \bibinfo {author} {\bibfnamefont {C.~M.}\ \bibnamefont {Marcus}},\ }\bibfield  {title} {\bibinfo {title} {Coherent {Operations} and {Screening} in {Multielectron} {Spin} {Qubits}},\ }\href {https://doi.org/10.1103/PhysRevLett.112.026801} {\bibfield  {journal} {\bibinfo  {journal} {Physical Review Letters}\ }\textbf {\bibinfo {volume} {112}},\ \bibinfo {pages} {026801} (\bibinfo {year} {2014})}\BibitemShut {NoStop}%
\bibitem [{\citenamefont {Choi}\ \emph {et~al.}(2025)\citenamefont {Choi}, \citenamefont {Nichol},\ and\ \citenamefont {Barnes}}]{choi_ballast_2025}%
  \BibitemOpen
  \bibfield  {author} {\bibinfo {author} {\bibfnamefont {Y.}~\bibnamefont {Choi}}, \bibinfo {author} {\bibfnamefont {J.~M.}\ \bibnamefont {Nichol}},\ and\ \bibinfo {author} {\bibfnamefont {E.}~\bibnamefont {Barnes}},\ }\bibfield  {title} {\bibinfo {title} {Ballast {Charges} for {Semiconductor} {Spin} {Qubits}},\ }\href {https://doi.org/10.1103/81qt-48bl} {\bibfield  {journal} {\bibinfo  {journal} {Physical Review Letters}\ }\textbf {\bibinfo {volume} {134}},\ \bibinfo {pages} {237002} (\bibinfo {year} {2025})}\BibitemShut {NoStop}%
\bibitem [{\citenamefont {Delbecq}\ \emph {et~al.}(2016)\citenamefont {Delbecq}, \citenamefont {Nakajima}, \citenamefont {Stano}, \citenamefont {Otsuka}, \citenamefont {Amaha}, \citenamefont {Yoneda}, \citenamefont {Takeda}, \citenamefont {Allison}, \citenamefont {Ludwig}, \citenamefont {Wieck},\ and\ \citenamefont {Tarucha}}]{delbecq_quantum_2016}%
  \BibitemOpen
  \bibfield  {author} {\bibinfo {author} {\bibfnamefont {M.~R.}\ \bibnamefont {Delbecq}}, \bibinfo {author} {\bibfnamefont {T.}~\bibnamefont {Nakajima}}, \bibinfo {author} {\bibfnamefont {P.}~\bibnamefont {Stano}}, \bibinfo {author} {\bibfnamefont {T.}~\bibnamefont {Otsuka}}, \bibinfo {author} {\bibfnamefont {S.}~\bibnamefont {Amaha}}, \bibinfo {author} {\bibfnamefont {J.}~\bibnamefont {Yoneda}}, \bibinfo {author} {\bibfnamefont {K.}~\bibnamefont {Takeda}}, \bibinfo {author} {\bibfnamefont {G.}~\bibnamefont {Allison}}, \bibinfo {author} {\bibfnamefont {A.}~\bibnamefont {Ludwig}}, \bibinfo {author} {\bibfnamefont {A.~D.}\ \bibnamefont {Wieck}},\ and\ \bibinfo {author} {\bibfnamefont {S.}~\bibnamefont {Tarucha}},\ }\bibfield  {title} {\bibinfo {title} {Quantum {Dephasing} in a {Gated} {GaAs} {Triple} {Quantum} {Dot} due to {Nonergodic} {Noise}},\ }\href {https://doi.org/10.1103/PhysRevLett.116.046802} {\bibfield  {journal} {\bibinfo  {journal} {Physical Review Letters}\ }\textbf {\bibinfo {volume} {116}},\
  \bibinfo {pages} {046802} (\bibinfo {year} {2016})}\BibitemShut {NoStop}%
\bibitem [{\citenamefont {Reilly}\ \emph {et~al.}(2008)\citenamefont {Reilly}, \citenamefont {Taylor}, \citenamefont {Laird}, \citenamefont {Petta}, \citenamefont {Marcus}, \citenamefont {Hanson},\ and\ \citenamefont {Gossard}}]{reilly_measurement_2008}%
  \BibitemOpen
  \bibfield  {author} {\bibinfo {author} {\bibfnamefont {D.~J.}\ \bibnamefont {Reilly}}, \bibinfo {author} {\bibfnamefont {J.~M.}\ \bibnamefont {Taylor}}, \bibinfo {author} {\bibfnamefont {E.~A.}\ \bibnamefont {Laird}}, \bibinfo {author} {\bibfnamefont {J.~R.}\ \bibnamefont {Petta}}, \bibinfo {author} {\bibfnamefont {C.~M.}\ \bibnamefont {Marcus}}, \bibinfo {author} {\bibfnamefont {M.~P.}\ \bibnamefont {Hanson}},\ and\ \bibinfo {author} {\bibfnamefont {A.~C.}\ \bibnamefont {Gossard}},\ }\bibfield  {title} {\bibinfo {title} {Measurement of {Temporal} {Correlations} of the {Overhauser} {Field} in a {Double} {Quantum} {Dot}},\ }\href {https://doi.org/10.1103/PhysRevLett.101.236803} {\bibfield  {journal} {\bibinfo  {journal} {Physical Review Letters}\ }\textbf {\bibinfo {volume} {101}},\ \bibinfo {pages} {236803} (\bibinfo {year} {2008})}\BibitemShut {NoStop}%
\bibitem [{\citenamefont {Malinowski}\ \emph {et~al.}(2017)\citenamefont {Malinowski}, \citenamefont {Martins}, \citenamefont {Cywi{\'n}ski}, \citenamefont {Rudner}, \citenamefont {Nissen}, \citenamefont {Fallahi}, \citenamefont {Gardner}, \citenamefont {Manfra}, \citenamefont {Marcus},\ and\ \citenamefont {Kuemmeth}}]{malinowski_spectrum_2017}%
  \BibitemOpen
  \bibfield  {author} {\bibinfo {author} {\bibfnamefont {F.~K.}\ \bibnamefont {Malinowski}}, \bibinfo {author} {\bibfnamefont {F.}~\bibnamefont {Martins}}, \bibinfo {author} {\bibfnamefont {{\L}.}~\bibnamefont {Cywi{\'n}ski}}, \bibinfo {author} {\bibfnamefont {M.~S.}\ \bibnamefont {Rudner}}, \bibinfo {author} {\bibfnamefont {P.~D.}\ \bibnamefont {Nissen}}, \bibinfo {author} {\bibfnamefont {S.}~\bibnamefont {Fallahi}}, \bibinfo {author} {\bibfnamefont {G.~C.}\ \bibnamefont {Gardner}}, \bibinfo {author} {\bibfnamefont {M.~J.}\ \bibnamefont {Manfra}}, \bibinfo {author} {\bibfnamefont {C.~M.}\ \bibnamefont {Marcus}},\ and\ \bibinfo {author} {\bibfnamefont {F.}~\bibnamefont {Kuemmeth}},\ }\bibfield  {title} {\bibinfo {title} {Spectrum of the {Nuclear} {Environment} for {GaAs} {Spin} {Qubits}},\ }\href {https://doi.org/10.1103/PhysRevLett.118.177702} {\bibfield  {journal} {\bibinfo  {journal} {Physical Review Letters}\ }\textbf {\bibinfo {volume} {118}},\ \bibinfo {pages} {177702} (\bibinfo {year}
  {2017})}\BibitemShut {NoStop}%
\bibitem [{\citenamefont {Philippopoulos}\ \emph {et~al.}(2020)\citenamefont {Philippopoulos}, \citenamefont {Chesi},\ and\ \citenamefont {Coish}}]{philippopoulos_first-principles_2020}%
  \BibitemOpen
  \bibfield  {author} {\bibinfo {author} {\bibfnamefont {P.}~\bibnamefont {Philippopoulos}}, \bibinfo {author} {\bibfnamefont {S.}~\bibnamefont {Chesi}},\ and\ \bibinfo {author} {\bibfnamefont {W.~A.}\ \bibnamefont {Coish}},\ }\bibfield  {title} {\bibinfo {title} {First-principles hyperfine tensors for electrons and holes in {GaAs} and silicon},\ }\href {https://doi.org/10.1103/PhysRevB.101.115302} {\bibfield  {journal} {\bibinfo  {journal} {Physical Review B}\ }\textbf {\bibinfo {volume} {101}},\ \bibinfo {pages} {115302} (\bibinfo {year} {2020})}\BibitemShut {NoStop}%
\bibitem [{\citenamefont {Assali}\ \emph {et~al.}(2011)\citenamefont {Assali}, \citenamefont {Petrilli}, \citenamefont {Capaz}, \citenamefont {Koiller}, \citenamefont {Hu},\ and\ \citenamefont {Das~Sarma}}]{assali_hyperfine_2011}%
  \BibitemOpen
  \bibfield  {author} {\bibinfo {author} {\bibfnamefont {L.~V.~C.}\ \bibnamefont {Assali}}, \bibinfo {author} {\bibfnamefont {H.~M.}\ \bibnamefont {Petrilli}}, \bibinfo {author} {\bibfnamefont {R.~B.}\ \bibnamefont {Capaz}}, \bibinfo {author} {\bibfnamefont {B.}~\bibnamefont {Koiller}}, \bibinfo {author} {\bibfnamefont {X.}~\bibnamefont {Hu}},\ and\ \bibinfo {author} {\bibfnamefont {S.}~\bibnamefont {Das~Sarma}},\ }\bibfield  {title} {\bibinfo {title} {Hyperfine interactions in silicon quantum dots},\ }\href {https://doi.org/10.1103/PhysRevB.83.165301} {\bibfield  {journal} {\bibinfo  {journal} {Physical Review B}\ }\textbf {\bibinfo {volume} {83}},\ \bibinfo {pages} {165301} (\bibinfo {year} {2011})}\BibitemShut {NoStop}%
\bibitem [{\citenamefont {Schliemann}\ \emph {et~al.}(2003)\citenamefont {Schliemann}, \citenamefont {Khaetskii},\ and\ \citenamefont {Loss}}]{schliemann_electron_2003}%
  \BibitemOpen
  \bibfield  {author} {\bibinfo {author} {\bibfnamefont {J.}~\bibnamefont {Schliemann}}, \bibinfo {author} {\bibfnamefont {A.}~\bibnamefont {Khaetskii}},\ and\ \bibinfo {author} {\bibfnamefont {D.}~\bibnamefont {Loss}},\ }\bibfield  {title} {\bibinfo {title} {Electron spin dynamics in quantum dots and related nanostructures due to hyperfine interaction with nuclei},\ }\href {https://doi.org/10.1088/0953-8984/15/50/R01} {\bibfield  {journal} {\bibinfo  {journal} {Journal of Physics: Condensed Matter}\ }\textbf {\bibinfo {volume} {15}},\ \bibinfo {pages} {R1809} (\bibinfo {year} {2003})}\BibitemShut {NoStop}%
\end{thebibliography}%

\clearpage
\begin{bibunit}[apsrev4-2]
\onecolumngrid
\begin{center}
{\large\bfseries Supplemental Material}\\[0.6em]
{\bfseries Probing Residual Noise at a Decoherence Sweet Spot in a $^{28}$Si/SiGe Spin Qubit}
\end{center}

\setcounter{equation}{0}
\renewcommand{\theequation}{S\arabic{equation}}
\setcounter{figure}{0}
\renewcommand{\thefigure}{S\arabic{figure}}

\section*{S1. DISTORTION OF THE SHUTTLING PULSES BY THE BIAS TEE}
\begin{figure}[b]
    \centering
    \includegraphics[width=0.6\linewidth]{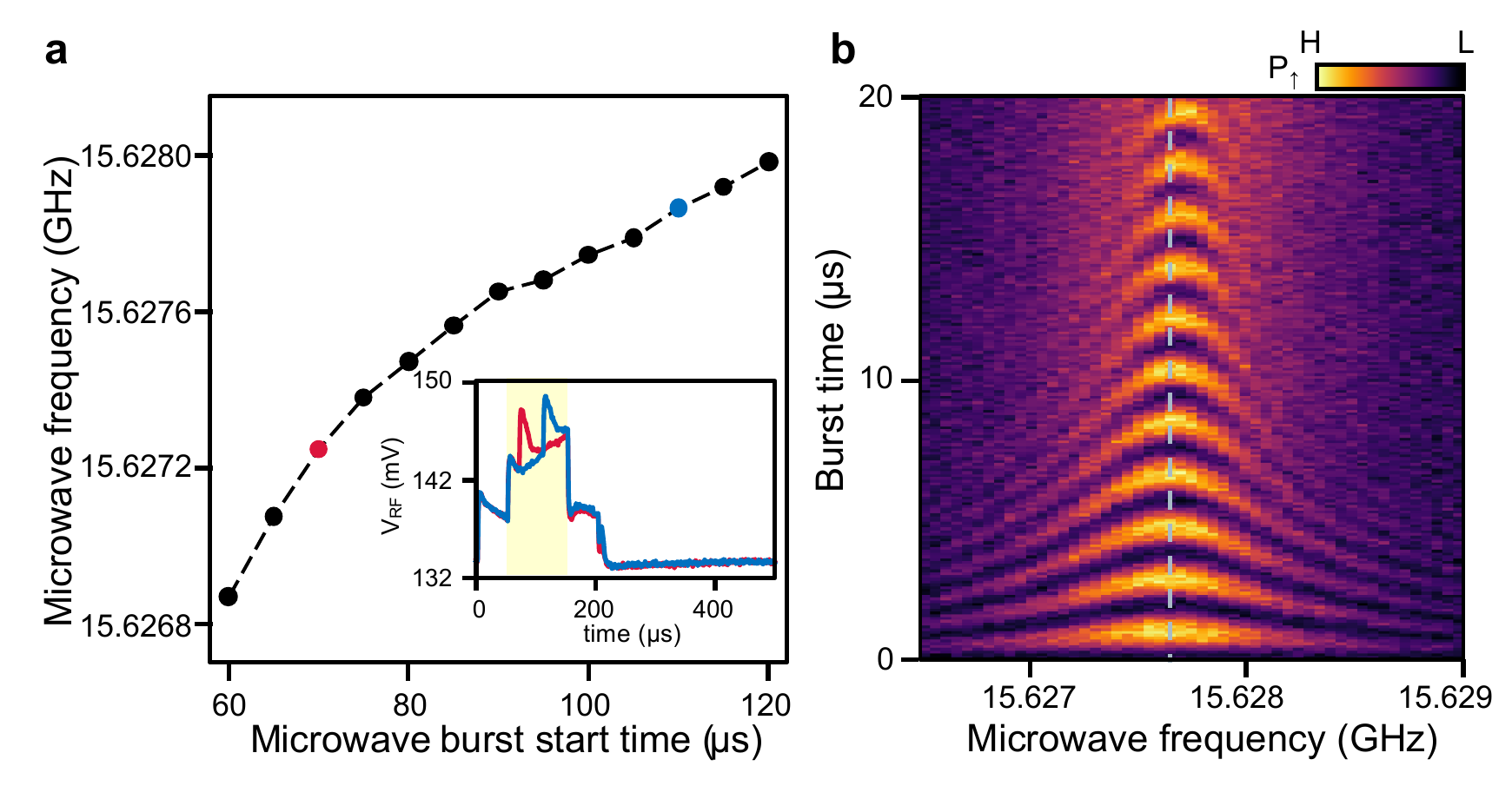}
    \caption{
        Distortion of the shuttling pulses by the bias tee. (a) Resonance frequency of $Q_M$ as a function of the microwave burst start time during the shuttled interval (yellow region in the inset). The inset shows the measured boxcar-averaged sensor dot signal versus time. The large peak is induced by the microwave burst applied to perform the X gate when the electron is at $Q_M$. The red and blue markers indicate the two representative microwave timings corresponding to boxcar traces in the inset. (b) Long-duration Rabi chevron measured at $Q_M$. The shift of the chevron center toward higher frequency reflects the time-dependent change in the qubit resonance frequency caused by pulse distortion. The light gray dashed line marks the initial resonance frequency determined from the first Rabi maximum.
    }
    \label{S1}
\end{figure}

The measurement cycle for $Q_M$ includes baseband shuttling pulses applied to $P_R$ and $P_M$, which move the electron from $Q_R$ to $Q_M$ for manipulation and return it to $Q_R$ for readout. These pulses pass through bias tees mounted inside the dilution refrigerator. Owing to the high pass response of the bias tees, the pulses delivered to the gates are distorted from an ideal square pulse shape. The inset of Fig.~\ref{S1}(a) shows the boxcar-averaged sensor response recorded during the measurement cycle. The signal level rises steadily across the shuttled interval (yellow region in the inset) instead of staying flat, which reflects the distorted pulse shape at the gates. 

Fig.~\ref{S1}(a) shows the resonance frequency of $Q_M$ as a function of the microwave burst start time during the shuttling pulse. The resonance frequency changes rapidly shortly after the shuttling pulse is applied, and the rate of change gradually decreases with time. In the measurement shown here, the resonance frequency shifts by approximately 1.2~MHz between 60 and 120~$\mu$s. We therefore place the microwave burst as late as allowed by the readout visibility to avoid the most rapidly varying part of the response. Nevertheless, a residual drift of the resonance frequency remains during qubit manipulation. This behavior is also visible in the long duration Rabi chevron in Fig.~\ref{S1}(b). Although the effect is small for short microwave bursts, the center of the chevron progressively bends toward higher frequency as the burst duration increases, reflecting the accumulated effect of the changing resonance frequency.

This drift can affect the refocusing pulses in a CPMG sequence. In the rotating frame of the drive frequency $f_{\mathrm{d}}$, the qubit Hamiltonian during a microwave pulse can be written as
\begin{equation}
\frac{H(t)}{h}
=
\frac{1}{2}
\left[
\delta(t)\sigma_z
+
\Omega\left(\cos\varphi\,\sigma_x+\sin\varphi\,\sigma_y\right)
\right],
\label{eq:supp-qm-pulse-hamiltonian}
\end{equation}
where $\delta(t)=f_q(t)-f_{\mathrm{d}}$ is the detuning between the qubit resonance frequency and the microwave drive, $\Omega$ is the Rabi frequency, and $\varphi$ specifies the rotation axis in the equatorial plane. When $\delta=0$, a refocusing pulse produces the intended $\pi$ rotation. A finite detuning changes both the effective rotation axis and the rotation angle.

CPMG suppresses dephasing from slowly varying frequency offsets by refocusing the phase accumulated between pulses~\cite{cywinski_dynamical-decoupling_2014}.
The resonance-frequency shift caused by distortion of the shuttling pulse is reproduced in each measurement and continues during the $\pi$ pulses, producing rotation errors.
Although CPMG is robust against rotation errors that are the same for every $\pi$ pulse within a sequence, the time-dependent detuning causes these errors to vary from pulse to pulse.
We attribute the reduction in CPMG signal amplitude to these rotation errors, which are not fully canceled by the sequence.

Because $T_2^*$ of $Q_M$ is approximately 5~$\mu$s, the shuttling pulse used for the Ramsey measurement is short and its level changes only slightly through the high pass response of the bias tee. The CPMG measurement instead requires a much longer shuttling pulse, over which the level changes progressively and a larger resonance frequency drift develops while many refocusing pulses are applied. This drift adds to the intrinsic frequency noise and can lower the CPMG scaling exponent measured for $Q_M$.

For $Q_R$, the baseband pulses used during the measurement have substantially smaller amplitudes than the shuttling pulses required for $Q_M$, so the absolute gate voltage drift associated with pulse distortion is expected to be smaller. In addition, the strongly reduced decoherence gradient at $Q_R$ converts residual gate voltage variations into much smaller qubit frequency shifts. Any remaining pulse distortion is therefore expected to have a much smaller influence on the resonance frequency and CPMG coherence of $Q_R$ than on those of $Q_M$.

\section*{S2. LOW-FREQUENCY NOISE SPECTROSCOPY}
We estimate the qubit resonance frequency from single shot Ramsey outcomes using a real time Bayesian estimation technique \cite{yoneda_noise-correlation_2023,park_passive_2025}, since extracting a frequency estimate from a full Ramsey fringe measurement would limit the sampling rate to the subhertz range.

The posterior distribution over the qubit frequency $f$ after $N$ single shot outcomes is
\begin{equation}
P(f\mid m_N,m_{N-1},\ldots,m_1)
\propto
P_0(f)
\prod_{k=1}^{N}
\frac{1}{2}
\left[1+r_k\left(\alpha_{\mathrm{B}}+\beta_{\mathrm{B}}\cos(2\pi f t_k+\theta)\right)\right],
\label{eq:supp-bayes}
\end{equation}
where $m_k$ is the outcome of the $k$th shot taken at free evolution time $t_k$, $r_k=1$ ($-1$) for $m_k=\downarrow$ ($\uparrow$), $\theta$ is the initial phase of the off resonant Ramsey oscillation, and $\alpha_{\mathrm{B}}$ and $\beta_{\mathrm{B}}$ account for the error in the axis of rotation and for the oscillation visibility. The prior $P_0(f)$ is a Gaussian centered at the preceding estimate, and a uniform distribution is used for the first estimate. We take the most probable value of the posterior distribution as the frequency estimate, and we obtain one estimate from $N=101$ single shot outcomes taken at 101 equally spaced free evolution times, with one outcome at each time. The range and the spacing of the free evolution times are set separately for each configuration.

The estimation runs in real time on the quantum controller (Quantum Machines OPX+) within the QUA framework. The manipulation pulse length is 100~$\mu$s for every configuration. The readout pulse length is 100~$\mu$s for $Q_R$ in the single-electron configuration and 400~$\mu$s for $Q_R$ in the three-electron configuration and for $Q_M$, so that one shot takes 200~$\mu$s and 500~$\mu$s, respectively. After each shot, the controller requires an additional 35~$\mu$s to update the posterior distribution. Including this update time, one frequency estimate takes approximately 24~ms and 54~ms, respectively. The resulting sampling rate is above 10~Hz in both cases.

Each frequency trace of total duration $T$ is converted into a single-sided noise power spectral density $S(f)$ using the Bartlett method \cite{bartlett_smoothing_1948,percival_spectral_1993}. The trace is divided into $N_W$ non overlapping segments of equal length $T_W=T/N_W$. For each segment, we subtract the segment mean and define the frequency fluctuation as $\delta f_q(t)=f_q(t)-\langle f_q\rangle$. For the $j$th segment, the periodogram at positive frequency is
\begin{equation}
S_j(f)=\frac{2}{T_W}\left|\int_{0}^{T_W}dt\,\delta f_q(t)e^{-2\pi i f t}\right|^2,
\label{eq:supp-bartlett-periodogram}
\end{equation}
and the spectral density is the average of the $N_W$ periodograms,
\begin{equation}
S(f)=\frac{1}{N_W}\sum_{j=1}^{N_W}S_j(f).
\label{eq:supp-bartlett-average}
\end{equation}
The averaging reduces the variance of the estimate by a factor of $N_W$, while the lowest accessible frequency rises from $1/T$ to $N_W/T$. The choice of $N_W$ therefore balances the variance of the estimate against the low-frequency reach.

We evaluate the spectrum with $N_W=1$, 3, 10, and 30 and combine the four results. Each spectrum in Fig.~\ref{fig:auto-psd}(a) is shown only below the lowest frequency accessible with the next larger $N_W$, so that adjacent segments do not overlap and every frequency is represented by the estimate with the largest number of averages available at that frequency. The combined spectrum extends up to the Nyquist frequency set by the sampling rate of the corresponding configuration.

To reduce the variance at the lowest frequencies, each low-frequency data set in Fig.~\ref{fig:auto-psd}(a) is the average of five measurements taken under identical conditions. Narrow peaks appear in the averaged spectrum and are removed. Their frequencies are consistent with harmonics of the approximately 1.4~Hz operating frequency of the pulse tube cooler and with aliases of those harmonics. The same fundamental frequency and its harmonics have been reported for a Cryomech PT415 pulse tube \cite{wang_4_2016}.

\section*{S3. NOISE EXTRACTION FROM THE CPMG FILTER FUNCTION}
\begin{figure}[b]
    \centering
    \includegraphics[width=\linewidth]{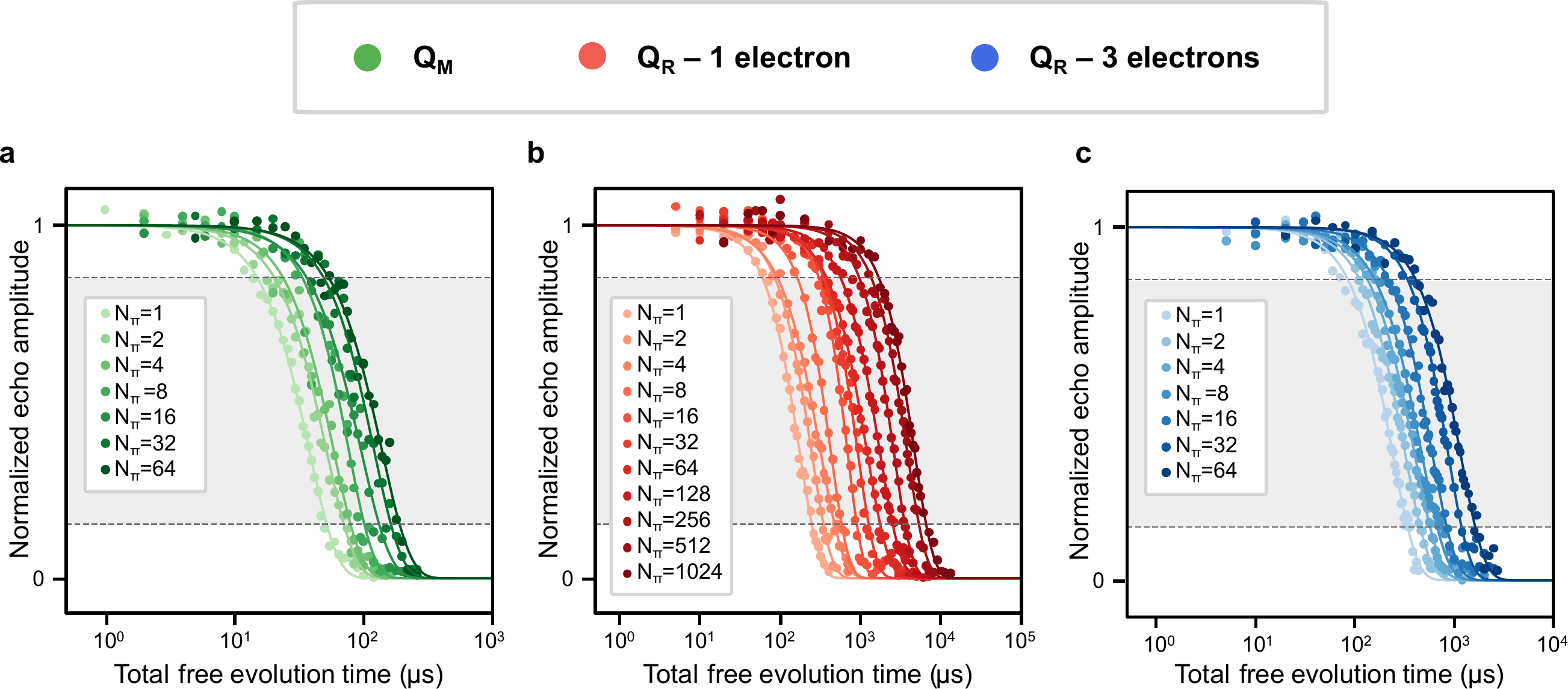}
    \caption{
        Additional data for CPMG experiments. (a) $Q_M$, (b) $Q_R$ in the single-electron configuration, and (c) $Q_R$ in the three-electron configuration. Each trace is fitted to $\exp[-(\tau_{\mathrm{tot}}/T_2^{\mathrm{CPMG}})^\gamma]$. The shaded region with $0.15<A_{\mathrm{CPMG}}<0.85$ indicates the window used for noise extraction.
    }
    \label{S3}
\end{figure}

Each refocusing pulse of a CPMG sequence reverses the sign with which the qubit accumulates phase, so that the accumulated phase is the frequency fluctuation weighted by the modulation function of the pulse sequence. We include the finite duration of the refocusing pulses following the filter-function treatment of Refs.~\cite{cywinski_dynamical-decoupling_2014,connors_charge-noise_2022}. For stationary Gaussian frequency noise and negligible pulse errors, the coherence after a total sequence time $t$ follows
\begin{equation}
W(t)=\exp\left[-2\pi^2\int_{0}^{\infty}df\,S(f)F_{N_\pi}(f,t)\right],
\label{eq:supp-cpmg-coherence}
\end{equation}
where
\begin{equation}
F_{N_\pi}(f,t)=\left|\int_{0}^{t}dt'\,y(t')e^{2\pi i f t'}\right|^2.
\label{eq:supp-cpmg-filter}
\end{equation}
The modulation function $y(t')$ alternates between $+1$ and $-1$ during free evolution and is taken to be zero during the finite duration of each refocusing pulse. For $N_\pi$ equally spaced pulses, the first maximum of $F_{N_\pi}$ is centered at
\begin{equation}
f_{\mathrm{CPMG}}=\frac{N_\pi D'}{2\tau_{\mathrm{tot}}}=\frac{N_\pi}{2t},
\qquad
t=\tau_{\mathrm{tot}}+N_\pi t_\pi,
\qquad
D'=\frac{\tau_{\mathrm{tot}}}{t},
\label{eq:supp-cpmg-frequency}
\end{equation}
where $\tau_{\mathrm{tot}}$ is the total free evolution time, $t_\pi$ is the duration of one refocusing pulse, and $D'$ is the free evolution duty cycle \cite{connors_charge-noise_2022}. The refocusing pulses are applied about the y axis, and we use $t_\pi=500$~ns for the two single-electron configurations and $t_\pi=1500$~ns for the three-electron configuration, where the longer pulse reduces the heating that appears at large $N_\pi$. The number of refocusing pulses ranges from 1 to 1024 for $Q_R$ in the single-electron configuration and from 1 to 64 for $Q_R$ in the three-electron configuration and for $Q_M$ [Fig.~\ref{S3}].

For sufficiently large $N_\pi$ and duty cycle $D'$, the first peak of $F_{N_\pi}$ is narrow enough to be approximated by a delta function at $f_{\mathrm{CPMG}}$. Using $\int_0^\infty df\,F_{N_\pi}(f,t)=D't/2=\tau_{\mathrm{tot}}/2$, the normalized CPMG amplitude $A_{\mathrm{CPMG}}$ gives the noise power spectral density as
\begin{equation}
S(f_{\mathrm{CPMG}})
\simeq
-\frac{\ln A_{\mathrm{CPMG}}}{\pi^2D't}
=
-\frac{\ln A_{\mathrm{CPMG}}}{\pi^2\tau_{\mathrm{tot}}}.
\label{eq:supp-cpmg-extraction}
\end{equation}

In the noise extraction, we use only measurements with $N_\pi\geq8$ and $D'>0.2$ \cite{connors_charge-noise_2022}. For each $N_\pi$, we measure the normalized $A_{\mathrm{CPMG}}$ as a function of $\tau_{\mathrm{tot}}$ and evaluate Eq.~(\ref{eq:supp-cpmg-extraction}) at every retained point. Since $t$ and $D'$ depend on $\tau_{\mathrm{tot}}$ at fixed $N_\pi$, one value of $N_\pi$ contributes a continuous range of $f_{\mathrm{CPMG}}$ rather than a single frequency. Only points within the window $0.15<A_{\mathrm{CPMG}}<0.85$ are used, where the conversion in Eq.~(\ref{eq:supp-cpmg-extraction}) is least sensitive to the normalization of the amplitude and to the noise floor of the readout. This window is shaded in Fig.~\ref{fig:coherence}(g) of the main text and Fig.~\ref{S3}.

The ranges of $f_{\mathrm{CPMG}}$ obtained from neighboring values of $N_\pi$ overlap, and the values of $S(f_{\mathrm{CPMG}})$ extracted within an overlapping range agree within the scatter of the data. This agreement is consistent with the narrow-band approximation used in Eq.~(\ref{eq:supp-cpmg-extraction}). The points obtained from all $N_\pi\geq8$ form the high-frequency part of the spectra in Fig.~\ref{fig:auto-psd}(a) of the main text.

\section*{S4. DECOMPOSITION OF THE QUBIT SENSOR CORRELATION}
The finite correlation between the estimated $Q_R$ frequency and the charge sensor signal in Fig.~\ref{fig:correlation}(e), measured with $Q_R$ in the three-electron regime using PSB readout, indicates that the two signals share an electrostatic fluctuation. In this readout, the even and odd parity outcomes appear as low and high levels of the sensor signal, and the two levels convert the fluctuation through different local slopes of the Coulomb peak of the sensor dot.

When an electrostatic fluctuation $\xi(t)$ is present in the surroundings, the qubit and the charge sensor respond through their respective transfer functions. At the charge sensor, the fluctuation shifts the electrochemical potential of the sensor dot, which is equivalent to a small horizontal displacement $\delta v=\lambda\xi$ of the operating point along the plunger gate voltage axis $v$, where $\lambda$ converts the fluctuation into an equivalent gate voltage. The recorded signal at readout level $i$ then follows the local slope $k_i$ of the Coulomb peak at that level. At $Q_R$, the same fluctuation produces an electric field at the dot, which displaces the electron within the longitudinal stray field gradient and shifts the qubit frequency $f_R$ by $\eta\xi$, where $\eta$ is the qubit frequency response per unit $\xi$. Including fluctuations specific to each measurement channel,
\begin{equation}
\begin{array}{rcl}
\delta f_R(t) &=& \eta\xi(t)+n_R(t),\\
\delta s_i(t) &=& k_i\lambda\xi(t)+n_i(t),
\end{array}
\label{eq:supp-shared-noise-model}
\end{equation}
where $\delta s_i$ is the recorded charge sensor signal at level $i$. $n_R$ and $n_i$ collect the fluctuations specific to the qubit and charge sensor measurement channels, including noise added by the readout and estimation procedure. 

If we assume that $\xi$, $n_R$, and $n_i$ are mutually uncorrelated, the mixed terms drop out of the time averages that enter Eq.~(\ref{eq:cross_psd}), and
\begin{equation}
\begin{array}{rcl}
\langle\delta f_R(t)\delta f_R(t+\tau)\rangle
&=&\eta^2\langle\xi(t)\xi(t+\tau)\rangle
+\langle n_R(t)n_R(t+\tau)\rangle,\\
\langle\delta s_i(t)\delta s_i(t+\tau)\rangle
&=&k_i^2\lambda^2\langle\xi(t)\xi(t+\tau)\rangle
+\langle n_i(t)n_i(t+\tau)\rangle,\\
\langle\delta f_R(t)\delta s_i(t+\tau)\rangle
&=&\eta k_i\lambda\langle\xi(t)\xi(t+\tau)\rangle.
\end{array}
\label{eq:supp-shared-correlators}
\end{equation}
Taking the Fourier transform of Eq.~(\ref{eq:supp-shared-correlators}) as in Eq.~(\ref{eq:cross_psd}), the auto-PSDs are
\begin{equation}
S_R(f)=\eta^2S_\xi(f)+S_{n_R}(f),
\qquad
S_{s,i}(f)=k_i^2\lambda^2S_\xi(f)+S_{n_i}(f),
\label{eq:supp-shared-auto}
\end{equation}
and the cross-PSD between the qubit frequency and the charge sensor signal is
\begin{equation}
C_i(f)=\eta k_i\lambda S_\xi(f).
\label{eq:supp-shared-cross}
\end{equation}
Under this assumption, the cross-PSD contains only the shared fluctuation. Here $r_i(f)=C_i(f)/[S_R(f)S_{s,i}(f)]^{1/2}$ is the normalized correlation coefficient evaluated for the qubit and the charge sensor at readout level $i$, and its magnitude is the correlation strength. Inserting Eqs.~(\ref{eq:supp-shared-auto}) and (\ref{eq:supp-shared-cross}) gives
\begin{equation}
r_i(f)=
\frac{\eta k_i\lambda S_\xi}
{\left[(\eta^2S_\xi+S_{n_R})(k_i^2\lambda^2S_\xi+S_{n_i})\right]^{1/2}}.
\label{eq:supp-normalized-cross}
\end{equation}
Equation~(\ref{eq:supp-normalized-cross}) shows that the correlation strength is large only when the shared fluctuation carries a large fraction of the total power in both channels. The shared fluctuation reaches the charge sensor signal through the local slope $k_i$, whereas $n_i$ enters after that conversion and we take it to be independent of $k_i$. A steeper local slope then raises the shared contribution while leaving $n_i$ unchanged, so the fraction carried by the shared fluctuation in the charge sensor channel increases and the correlation strength becomes larger.

\begin{figure}[t]
    \centering
    \includegraphics[width=\linewidth]{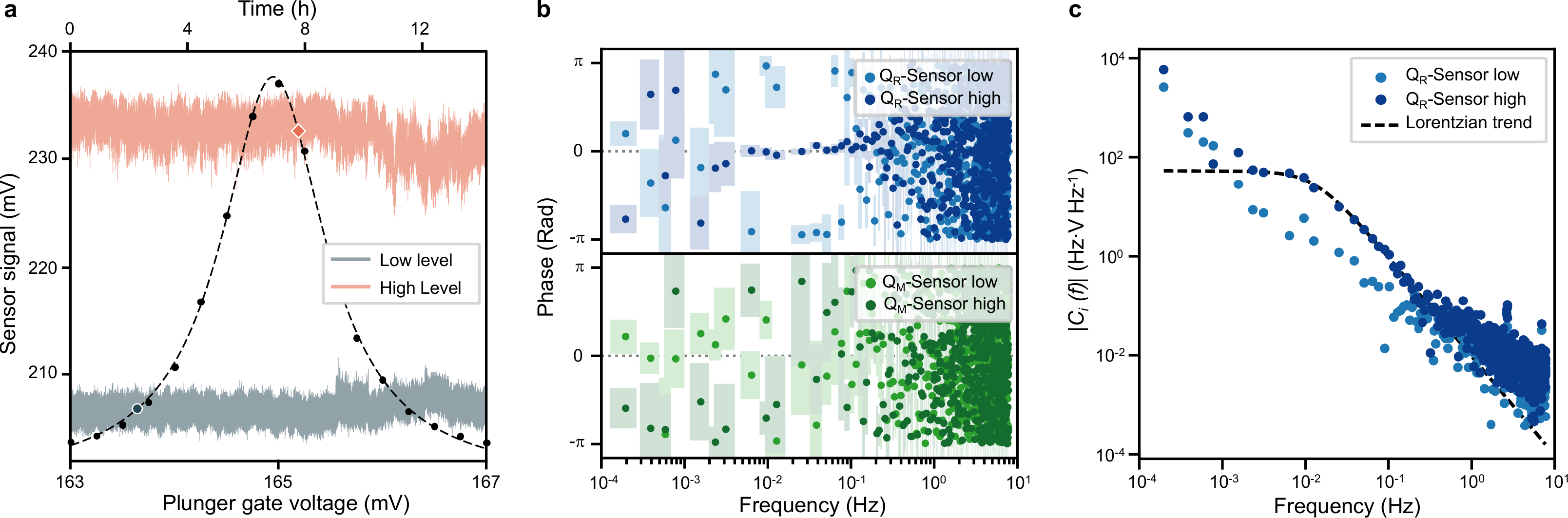}
    \caption{
        Qubit-sensor correlation analysis. (a) Coulomb peak of the sensor dot and the two readout levels used in the correlation analysis. The black circles and dashed curve show the measured sensor response as a function of the plunger gate voltage and its fit. The low and high sensor signals recorded during the correlation measurement are overlaid using the top time axis, and the highlighted markers indicate their mean levels on the fitted curve. (b) Phases of the cross-PSDs between the sensor signal and $Q_R$ (top) and $Q_M$ (bottom), evaluated separately for the low and high readout levels. The shaded regions represent 90~\% confidence intervals. (c) Magnitudes of the $Q_R$--sensor cross-PSDs for the low and high readout levels. The dashed curve shows the Lorentzian trend for the high readout level with $A = 52.5~\mathrm{Hz\cdot V\,Hz^{-1}}$ and $f_c = 13.5~\mathrm{mHz}$.
    }
    \label{S4}
\end{figure}

To determine $k_i$, we fit the Coulomb peak of the sensor dot measured as a function of $v$ and place the two recorded readout levels on the fitted transfer curve [Fig.~\ref{S4}(a)]. This gives $k_{\mathrm{low}}=+8.45$~V/V on the left flank and $k_{\mathrm{high}}=-36.04$~V/V on the right flank, with $|k_{\mathrm{high}}/k_{\mathrm{low}}|=4.27$. The larger magnitude of $k_{\mathrm{high}}$ is consistent with the larger correlation strength measured at the high level [Fig.~\ref{fig:correlation}(e)].

Since $\eta$ and $\lambda$ are independent of the readout level, the opposite signs of $k_{\mathrm{low}}$ and $k_{\mathrm{high}}$ place the phases of $C_i(f)$ at the two levels $\pi$ apart. For the small fluctuations considered here, a horizontal displacement of a peak that is symmetric about its center produces responses with opposite signs on the two flanks, whereas small changes in the peak height or width produce responses with the same sign. At frequencies where Fig.~\ref{fig:correlation}(e) shows appreciable $Q_R$--sensor correlation, the measured cross-PSD phases at the two readout levels differ by approximately $\pi$ [Fig.~\ref{S4}(b)], as expected for a horizontal shift of the operating point.

We take the response coefficients to be frequency independent, so that Eq.~(\ref{eq:supp-shared-cross}) carries all of the frequency dependence of $C_i$ in $S_\xi(f)$ and the shape of the measured cross-PSD follows the spectrum of the shared fluctuation. A charge that switches between two states at random times with a constant probability per unit time has a correlation function $\langle\xi(t)\xi(t+\tau)\rangle$ that decays as $\exp(-|\tau|/\tau_c)$, where $\tau_c$ is the decay time of that correlation function, and its Fourier transform as in Eq.~(\ref{eq:cross_psd}) gives a Lorentzian spectrum, 
\begin{equation}
S(f)=\frac{A}{1+(f/f_c)^2},
\qquad
f_c=\frac{1}{2\pi\tau_c},
\label{eq:supp-tlf-lorentzian}
\end{equation}
where $A$ is a constant. This spectrum is flat below $f_c$ and falls as $1/f^2$ above it \cite{paladino_1_2014}. The dashed curve in Fig.~\ref{S4}(c) shows this spectrum for the high readout level with $A = 52.5~\mathrm{Hz\cdot V\,Hz^{-1}}$ and $f_c = 13.5~\mathrm{mHz}$, corresponding to $\tau_c = 11.8~\mathrm{s}$.

The approximately $\pi$ phase difference between the two readout levels at $Q_R$ identifies the shared quantity as an electrostatic fluctuation acting on both $Q_R$ and the charge sensor, and the plateau and roll-off of the high-level cross-PSD near $10^{-2}$~Hz follow the Lorentzian form of Eq.~(\ref{eq:supp-tlf-lorentzian}). The same fluctuation is not evident at $Q_M$. The $Q_M$--sensor correlation is small at the frequencies where the $Q_R$--sensor correlation is largest, and the phases at the two readout levels are not aligned $\pi$ apart [Figs.~\ref{fig:correlation}(d) and \ref{S4}(b)]. The position of $Q_M$ has the larger frequency susceptibility to dot energy [Fig.~\ref{fig:coherence}(f)], so a fluctuation shared with the charge sensor would appear more strongly at $Q_M$ than at $Q_R$, and the weaker correlation means that the fluctuation couples to $Q_M$ only weakly. Such weak coupling follows from the distance, since $Q_M$ is approximately 100~nm farther from the charge sensor than $Q_R$ and correlations decay with separation \cite{rojas-arias_spatial_2023}. Together with the weak $Q_M$--$Q_R$ correlation [Fig.~\ref{fig:correlation}(c)], the source is therefore on the sensor side of $Q_R$ rather than between the two qubits. A Lorentzian of this kind usually arises from a shared TLF with a constant switching rate, and a device contains many such fluctuators \cite{connors_low-frequency_2019}. The charge sensor also stays on throughout the measurement, and transport through the sensor dot dephases the qubit as long as it is allowed \cite{park_passive_2025,hell_qubit_2016}, so the correlated signal may include a contribution originating in the sensor dot itself. Pulsing the charge sensor into Coulomb blockade during qubit manipulation would help distinguish these possibilities.

Equation~(\ref{eq:supp-normalized-cross}) separates the correlation strength from the absolute frequency noise power. The correlation strength reflects the relative contribution of the shared fluctuation within the two channels, while in the quasi-static Gaussian approximation the frequency variance that determines $T_2^*$ is the integral of $S_R$ above the low-frequency cutoff set by the acquisition time (Appendix B). A finite $Q_R$--sensor correlation therefore does not by itself imply a short $T_2^*$. At frequencies where the correlation is observed, $S_R$ remains about one order of magnitude below $S_M$ [Fig.~\ref{fig:auto-psd}(a)]. Operating at the sweet spot therefore does not require a quiet charge environment, and the coherence it provides comes from the weak conversion of that environment into qubit frequency noise.

\putbib[references_supplement]
\end{bibunit}

\end{document}